\documentclass[12pt]{article}
\usepackage[utf8]{inputenc}
\usepackage[ inner=1cm, outer=1cm, top=1cm, bottom=1.9cm]{geometry}
\usepackage[svgnames]{xcolor}
\usepackage[bottom]{footmisc}
\usepackage{braket}
\usepackage{graphicx}
\usepackage{framed}
\usepackage{csquotes}
\usepackage{tikz}
\usetikzlibrary{decorations.markings}
\usetikzlibrary{decorations.pathmorphing}
\definecolor{shadecolor}{rgb}{0.90,0.90,0.90}
\usepackage{hyperref}
\usepackage{subcaption}
\usepackage{pifont}
\usepackage{setspace}
\usepackage{amsmath, amssymb, amsthm, float, graphicx,amsfonts}
\numberwithin{equation}{section}
\hypersetup{colorlinks=true, linkcolor=DarkRed, citecolor=DarkRed,urlcolor=DarkGreen,linktocpage}
\def\beq{\begin{eqnarray}}\def\eeq{\end{eqnarray}}
\def\be{\begin{equation}}\def\ee{\end{equation}}
\def\a{\alpha}
\def\b{\beta}
\def\c{\chi}
\def\d{\delta}

\def\g{\gamma}

\def\k{\kappa}
\def\l{\lambda}

\def\m{\mu}
\def\n{\nu}
\def\p{\pi}

\def\s{\sigma}

\def\w{\omega}
\def\x{\xi}
\def\z{\zeta}
\def\D{\Delta}
\def\F{\Phi}

\def\G{\Gamma}

\def\Th{\Theta}
\def\pd{\partial}

\def\mf{{\mathcal{F}}}

\def\mi{{\mathcal{I}}}

\def\ml{{\mathcal{L}}}
\def\mm{{\mathcal{M}}}

\def\mr{{\mathcal{R}}}

\def\FM{\mathfrak{M}}

\def\tr{{\rm tr~}}
\def\nn{\nonumber}

\usepackage{bm}

\begin{document}
\thispagestyle{empty}
\hfill
TIFR/TH/20-29
\vspace{2cm}
\begin{center}
{\LARGE\bf Regge amplitudes in Generalized Fishnet and Chiral Fishnet Theories}\\
\bigskip\vspace{1cm}{
{\large Subham Dutta Chowdhury${}^{1, \nu}$, Parthiv Haldar${}^{2, J}$ and Kallol Sen${}^{3, h}$}\let\thefootnote\relax\footnotetext{email: ${}^{\nu}$subham@theory.tifr.res.in,\,${}^{J}$parthivh@iisc.ac.in,\,${}^{h}$kallolmax@gmail.com} 
} \\[7mm]
 {\it${}^{\nu}$ Department of Theoretical Physics, \\ 
 Tata Institute for Fundamental Research, Mumbai 400005}\\
[4mm]
{\it ${}^{J}$Center for High Energy Physics,Indian Institute of Science\\
\it	C.V. Raman Road, Bangalore 560012, India,}\\
[4mm]
{\it ${}^h$ Trinity College Dublin,\\
\it	 The University of Dublin, College Green, Dublin 2, D02 PN40, Ireland.}
 \end{center}
\bigskip
\centerline{\large\bf Abstract}

\begin{quote} \small
We extend the analysis of \cite{Chowdhury:2019hns} to study the Regge trajectories of the Mellin amplitudes of the $0-$ and $1-$ magnon correlators of the generalized Fishnet theory in $d$ dimensions and one type of correlators of chiral fishnet theory in $4$ dimensions. We develop a systematic procedure to perturbatively study the Regge trajectories and subsequently perform the spectral integral. Our perturbative method is very generic and in principle can be applied to correlators whose perturbative Regge trajectories obey some structural conditions which we list down. Our $d$ dimensional results reduce to previously known results in $d=4$ for 0-magnon and 1- magnon. As a non-trivial check, we show that the results for 1-magnon correlator in $d=8$, when evaluated using the exact techniques in \cite{Chowdhury:2019hns, Korchemsky:2018hnb} are in perfect agreement with our $d$ dimensional perturbative results. We also perturbatively compute the Regge trajectories and Regge-Mellin amplitudes of the chiral fishnet correlator $\langle{\rm Tr}[\phi_1(x_1)\phi_1(x_2)]{\rm Tr}[\phi_1^\dagger(x_3)\phi_1^\dagger(x_4)]\rangle$ using the techniques developed in this paper. Since this correlator has two couplings $\k$ and $\omega$, we have obtained closed-form results in the limit $\k \to 0, \omega \to 0$ with $\k/\omega$ held constant. We verify this computation with an independent method of computing the same and obtain perfect agreement. 
\end{quote}

\tableofcontents

\section{Introduction}
Computations in $\mathcal{N}=4$ SYM is in general quite technically challenging. Recently there is a renewed interest in the $\g-$deformed strongly twisted $\mathcal{N}=4$ SYM called the fishnet theory which is far simpler \cite{Gromov:2018hut, Gurdogan:2015csr, Kazakov:2018qez}. The fishnet theory enjoys integrability in the planar limit and also conformal symmetry. In the double scaling limit the scalar, fermions and gauge field decouple thereby breaking the supersymmetry and the $R-$symmetry (for more details refer to \cite{Pittelli:2019ceq, Caetano:2016ydc, Gromov:2019bsj, Gromov:2019aku})\footnote{Also see \cite{Adamo:2019lor, Basso:2019xay, Alfimov:2020obh, Derkachov:2019tzo, Derkachov:2020zvv, Levkovich-Maslyuk:2020rlp, Kazakov:2018ugh} for more related work.}.
This serves as a playground for studying a host of physical phenomenon which would not have been possible in the parent theory.  It is interesting to study the Regge trajectories of correlators in these theories since the exact correlation functions are known \cite{Chowdhury:2019hns, Korchemsky:2018hnb}.
The Regge limit for a scattering process in a theory is defined as a special kinematic limit of $2 \rightarrow 2$ scattering of particles in which the Centre of Mass (COM) momenta is taken to be large. In terms of Mandelstam variables $s ,t$ and $u$, this corresponds to large $s$ at fixed $t$. Regge theory is a study of the analytic properties of the S-matrix, more precisely the classification of singularities of the S-matrix in the complex angular momenta plane \cite{Regge:1959mz}. 
%Regge scattering has important theoretical and phenomenological aspects for which it serves as an important physical quantity to study \cite{Regge:1959mz}. In particular, the Regge limit of the scattering amplitude encodes information about the spectrum of the exchanged particles.
Historically Regge theory has been the organizational tool for studying the theory of strong interactions where it was found that strongly interacting particles had linear Regge trajectories \cite{PhysRevLett.7.394}. There are  quite a few examples attesting to this fact. This can be seen from known examples in QCD \cite{Korchemsky:1994um},  
the Virasoro-Shapiro amplitude in type II superstring theory \cite{PhysRev.177.2309}. In general the Regge amplitude scales with the Mandelstam invariants as 
\be
{\cal A}_{\text Regge} \propto s^{J_{\rm Regge}} 
\ee

For CFTs the meaning of a ``S-matrix" is somewhat ambiguous. However, it was shown that conformal correlators, which naturally obey partial conformal wave decomposition, show similar analogous behaviour leading to the study of ``Conformal Regge theory" \cite{Cornalba:2007fs, Costa2012a, Cornalba:2008qf}. This was initially motivated as studying the implications of the Regge limit of flat space S-matrix in the AdS, for the respective conformal correlators. However, irrespective of the existence of a bulk dual such a study can be undertaken (See \cite{Costa2012a} for a detailed review). In the context of CFTs most notable study of Regge trajectories have been done for $\mathcal{N}=4$ SYM \cite{Costa2012a}.  
For the general dimensional fishnet theories under consideration, we find that for the $0,1-$magnon cases, in the weak coupling, the leading Regge theory is dominated by,
$J_{\rm Regge}=0,-\frac{d}{4}$ respectively while the chiral fishnet theory correlator has $J=0$. In the language of the formal Regge theory, this merits the interpretation that the S-matrix is dominated at this resonance by a particle of negative spin, which we believe is a signature of the inherent non-unitarity of the theory.

Before venturing into the details on the present work, we would like to provide a very brief introduction to the fishnet theories we are considering. In general dimensions \footnote{The $4d$ theory is obtained from the double-scaling limits of the strongly $\gamma$ twisted $\mathcal{N}=4$ SYM. The resulting theory is integrable in the planar limit and non-unitary}, \cite{Sieg:2016vap,Zamolodchikov:1980mb,Chicherin:2017cns,Gromov:2017cja} is given by a proposed lagrangian (in $d-$dimensions) of a bi-scalar theory \cite{Gurdogan:2015csr, Grabner:2017pgm, Kazakov:2018qez} which is a generalization of the four dimensional theory, given by,
\be
\mathcal{L}=N_c\ \tr\left[\bar{X} (-\pd_\m\pd^\m)^\w X+\bar{Z} (-\pd_\m\pd^\m)^{d/2-\w} Z+(4\pi)^{d/2}\xi^2\bar{X}\bar{Z}XZ\right]\,.
\ee
The theory is however not complete at the quantum level without the double trace counterterms   \cite{Fokken:2013aea,Grabner:2017pgm, Kazakov:2018qez}, given by,
\be
\mathcal{L}_\text{dt}=(4\pi)^{d/2}\left(\a_1^2( \tr(X^2)\tr(\bar{X}^2)+ \tr(Z^2)\tr(\bar{Z}^2))-\a_2^2(\tr(XZ)\tr(\bar{X}\bar{Z})+\tr(X\bar{Z})\tr(\bar{X}Z))\right)\,.
\ee
A generalization of the bi-scalar model in four dimensions is chiral fishnet ($\chi-$fishnet) theory which includes three complex scalars and three fermions as explained in \cite{Gurdogan:2015csr, Caetano:2016ydc}\footnote{See also \cite{AuYang:1999hd}.}. This gives the full three coupling generalization of the bi-scalar model, with the lagrangian (see \cite{Kazakov:2018gcy}),
\begin{align}\label{chilag}
	\begin{split}
		\mathcal{L}_{\phi\psi}&=N_c\tr\left(-\frac{1}{2}\pd^\m\phi_j^\dagger\pd_\m\phi^j+i\bar{\psi}^{\dot{\a}}_j(\tilde{\s}^\m)^\a_{\dot{\a}}\pd_\m \psi^j_\a\right)+\mathcal{L}_\text{int}\,,\\
		\mathcal{L}_\text{int}&=N_c\tr\bigg[\xi_1^2\phi_2^\dagger\phi_3^\dagger\phi^2\phi^3+\xi_2^2\phi_3^\dagger\phi_1^\dagger\phi^3\phi^1+\xi_3^2\phi_1^\dagger\phi_2^\dagger\phi^1\phi^2+i\sqrt{\xi_2\xi_3}(\psi^3\phi^1\psi^2+\tilde{\psi}_3\phi_1^\dagger\tilde{\psi}_2)\\
		&+i\sqrt{\xi_1\xi_3}(\psi^1\phi^2\psi^3+\tilde{\psi}_1\phi_2^\dagger\tilde{\psi}_3)+i\sqrt{\xi_1\xi_2}(\psi^2\phi^3\psi^1+\tilde{\psi}_2\phi_3^\dagger\tilde{\psi}_1)\bigg]
	\end{split}
\end{align}
The fishnet theory is conformal and hence we can use the familiar machinery of CFT to determine the dynamics.

In this work we will compute the Regge limit of the fishnet theory in Mellin space using the technique of conformal Regge theory discussed in \cite{Costa2012a} for the general $d$ conformal fishnet theory and the chiral fishnet theory. The corresponding study for $4d$ conformal fishnet theory was taken in \cite{Chowdhury:2019hns}. Let us fiirst define what the mellin amplitude is; the Mellin representation of the connected part of the four-point conformal correlator is, 
\be\label{mellvardef}
\mathcal{G}(u,v)=\frac{1}{(4\pi i)^2}\int_{-i\infty}^{i\infty} ds dt\ u^{t/2}v^{-(s+t)/2}\mu(s,t) \mm(s,t)\,,
\ee
where $\mm(s,t)$ is the Mellin amplitude and
\begin{eqnarray}
	\mu(s,t)&=&\G\bigg(\frac{\D_{34}-s}{2}\bigg)\G\bigg(-\frac{\D_{12}+s}{2}\bigg)\G\bigg(\frac{s+t}{2}\bigg)\G\bigg(\frac{s+t+\D_{12}-\D_{34}}{2}\bigg)\nonumber\\
	&&\G\left(\frac{\D_1+\D_2-t}{2}\right)\G\left(\frac{\D_3+\D_4-t}{2}\right),
\end{eqnarray} 
is the measure with $\D_{ij}=\D_i-\D_j$. The Mellin amplitude admits a partial wave decomposition \cite{Mack:2009mi},
\be\label{mell}
\mm(s,t)=\sum_{J=0}^\infty \int_{-\infty}^{\infty} d\n b_J(\n^2)\g(\n,t)\g(-\n,t)\zeta({\D_i,t})\mathcal{P}_{\n,J}(s,t,\{\D_i\})\,,
\ee
where $\mathcal{P}_{\n,J}(s,t)$ is the Mack polynomial,
\be
\g(\n)=\frac{\G(\frac{\D_1+\D_2+J+i\n-h}{2})\G(\frac{\D_3+\D_4+J+i\n-h}{2})\G\left(\frac{h+i\n-J-t}{2}\right)}{\sqrt{8\pi}\G(i\n)}\,,\ee
and
\be\label{tfactor}
\zeta({\D_i,t})=\frac{1}{\G(\frac{\D_1+\D_2-t}{2})\G(\frac{\D_3+\D_4-t}{2})}.\ee
This will be the focal point of our analysis. We consider the $t$-channel decomposition with $\Delta_1=\Delta_4$ and $\Delta_2=\Delta_3$. In \cite{Costa2012a}, the Regge limit of the mellin amplitude was studied and it was shown that this formalism is equivalent to studying the Regge limit of the position space correlator \cite{Cornalba:2007fs}. In Mellin space the Regge limit is defined as large $s$ and fixed $t$ behaviour of the Mellin amplitude \cite{Costa2012a},
\be
\lim_{s\rightarrow\infty}\mathcal{P}_{\n,J}(s,t)=s^J a_J\,, \ \text{where}\ a_J=\frac{ (2-h-i\n+J)_J(2-h+i\n+J)_J}{(h+i\n-1)_J(h-i\n-1)_J}\,. \label{aJ}
\ee
The factor $a_J$ becomes $1$ for general $\n$ and integer $J$. In Regge theory of scattering amplitudes, the partial waves of the highest spin exchange dominates- this is apparent in the mellin language as the mack polynomial contributes $s^J$ for each $J$. Naively the sum over spins seems divergent but this can be taken care of by the ``Sommerfeld-Watson" transform. This step is technical and we refer the readers to section 3 of \cite{Chowdhury:2019hns} for a self-contained review of the summation procedure. The basic procedure involves, converting the sum over $J$ in \eqref{mell} to a contour integral over $J$ and then deforming the contour to analytically continue the spin to the complex plane. For convenience we write down the Regge Mellin amplitude for general $d$ Fishnet correlators,

\begin{align}\label{melfn}
	\begin{split}
		\mm^{(n)}_\pm(s,t)&=\frac{1}{2\pi i} \oint dJ \frac{\pi}{\sin\pi J} \int_{-\infty}^\infty d\n \left(\frac{s}{4}\right)^J e^{i\pi J/2}\nu \sinh \pi\nu \,\zeta_n({\D_i,t})\\
		&\times
		\frac
		{
			(-1)^{-J}
			\Gamma (h+J)
			\Gamma (h+J-i \nu )
			\Gamma (h+J+i \nu )
			\Gamma \left(\frac{h-J-t-i \nu }{2}\right) 
			\Gamma \left(\frac{h-J-t+i \nu }{2}\right)
		}
		{2\pi ^{ 2(h+1)}
			\Gamma (J+1)
			\Gamma \left(\frac{h+J-\D_{12}-i \nu }{2}\right)
			\Gamma \left(\frac{h+J+\D_{12}-i \nu }{2}\right)
			\Gamma \left(\frac{h+J-\D_{34}+i \nu }{2}\right) 
			\Gamma \left(\frac{h+J+\D_{34}+i \nu }{2}\right)
		}
		\, b_J^{n}(\n) P_J^{\pm}, 
	\end{split}
\end{align} 
%\frac{\left(E_{\Delta,J}^{(n)}\right)^p}{1-\chi_n E^{(n)}_{\Delta,J}}	

where,
\be
\z_n(\D_i,t)=\frac{1}{\G\left(\frac{\D_1+\D_2-t}{2}\right)\G\left(\frac{\D_3+\D_4-t}{2}\right)},\qquad P_J^{\pm}=\begin{cases} \cos \pi J/2\,, +(\text{even spin})\\ - i\sin \pi J/2\,, -(\text{odd spin})\end{cases}\,.
\ee
$ b_J^{n}(\n)$ is the spectral function for respective correlators of the theories we are considering and contains the dynamical information, $\D_{ij}=\D_i-\D_j$ and for the exchanged operator $\D=h+i\n$ with $h=\frac{d}{2}$. %Further $p=1,2,1$ for  $n=0,1,2$ magnon graphs respectively. Also we have $\c_0=\c_2=((4\p)^h\x^2)^2$ and $\c_1=(4\p)^h\x^2$.
The Regge trajectories are given by the poles of the spectral function, 
\be \label{specdef}
b_J^{n}(\n).
\ee 
That the term $e^{J\p/2}P_{J}^{\pm}$ takes care of $(s\rightarrow -s)$ in the Sommerfeld-Watson transform and from now on we will dispense with this term by writing out the $(s\rightarrow -s)$ term separately. Further, the factor $(-1)^J$ can be taken care of by overall sign for even and odd spins separately. %Now we move on to investigate the Mellin amplitude below case by case.

\subsection*{Summary of results:} 
\paragraph{ }Following \cite{Chowdhury:2019hns}, we extend the analysis to few more examples including $d-$dimensional extensions of the $0,1-$magnon cases in \cite{Gromov:2018hut} and chiral fishnet theories in four dimension in \cite{Kazakov:2018gcy}. We observe that in general dimensions as for the chiral fishnet theory, obtaining exact analytical solutions $J(\n,\xi)$ for the Regge trajectories as a function of the coupling$-\xi$ is rendered almost impossible due to complexity of the spectral function. Nevertheless one can obtain the Regge trajectories depending on whether $\n>\xi$ or $\n<\xi$. The precise dependence of $\n$ on coupling depends on the details of the theory being considered. These two regimes offer tractable solutions to the spectral function which can be shown to be analytic continuations of each other in certain cases where closed form polynomial solutions in principle can be found. Further, we can use the different regimes of the solutions to reduce the Mellin amplitudes. From here on, we will discuss the leading Regge trajectories for the theories under consideration. Without further ado, we will dive into the main findings of the work. \\
\subsubsection*{General $d$ fishnet theory: 0 - magnon }
\paragraph{}For $0-$magnon in generalised $d$ dimensional fishnet theory, we are dealing with the four point correlators $\langle{\rm Tr} \left(X(x_1)X(x_2)\right){\rm Tr}\left(\bar{X}(x_3)\bar{X}(x_4)\right)\rangle$ where the operator $X$ has dimension $\D=d/4$. The definitions ``inner" and ``outer" solutions have been defined in section \ref{genRegge}. In short, since we are working in weak coupling, there are broadly two regimes of the coupling $\xi$, relative to the integrand $\nu$, where we perturbatively evaluate the Regge trajectories. The inner solution corresponds to the regime 
$\nu<\xi^\alpha$ while the outer region corresponds to $\nu>\xi^\alpha$ (where the exponent $\alpha$ depends on the type of correlator we are looking at). A more precise definition is given below \eqref{Jpert1}.    

\begin{itemize}
	\item {\bf Regge trajectories: Outer solution}   The outer solution for $\n>\xi^2$ is,
\begin{shaded}
	\be
	J_{\text{o}\pm}^{(0)}(\nu)=\pm i\n+\a_{1\pm}\xi^4+\sum_{k=2}^\infty \a_k(\pm\n) \xi^{4k}\,.
	\ee
\end{shaded}	
	with the coefficients are dimension dependent and the first few are given by,
\begin{shaded}	
	\begin{align}
	\begin{split}
	\a_2(\pm\n)&=-\frac{F_1(\pm i\n)}{2}\a_{1\pm}^2\,,\ \a_3(\pm\n)=\frac{F_2(\pm i\n)+3F_1(\pm i\n)^2}{8}\a_{1\pm}^3\,,\\
	~ \a_4(\pm\n)&=-\frac{F_3(\pm i\n)+12F_1(\pm i\n)F_2(\pm i\n)+16F_1(\pm i\n)^3}{48}\a_{1\pm}^4\,.
	\end{split}
	\end{align}
\end{shaded}	
where $F_r(\nu)$ is given explicitly in \eqref{FaG} and  $\a_{1\pm} =\G(h)\G(i \n)/(2\G(h \pm i\n))$. 
	\item {\bf Regge trajectories: Inner solution} The inner solution for $\n\leq\xi^2$ is,
	\begin{shaded}
	\be
	J_{\text{i}\pm}^\text{(0)}(x)=\tilde{\a}_{1\pm}\xi^2+\sum_{k=2}^\infty \tilde{\a}_k(x)\xi^{2k}\,, \ \n=x\xi^2\,, x<1\,,
	\ee
	\end{shaded}
	with the coefficients given by,
	\begin{shaded}
	\begin{align}
	\begin{split}
	\tilde{\a}_2(x)&=-\frac{H_{h-1}}{2}(\tilde{\a}_{1\pm}^2+x^2)\,, \ \tilde{\a}_3(x)=\frac{(\tilde{\a}_{1\pm}^2+x^2)}{8\tilde{\a}_{1\pm}}\left[(H_{h-1})^2(x^2+3\tilde{\a}_{1\pm}^2)+H_{h-1,2}(\tilde{\a}_{1\pm}^2-x^2)\right]\,,\\
	\tilde{\a}_4(x)&=-\frac{\tilde{\a}_{1\pm}^2+x^2}{24}\left[6H_{h-1} H_{h-1,2} \tilde{\a}_{1\pm}^2+H_{h-1,3} (\tilde{\a}_{1\pm}^2-3x^2)+2(H_{h-1})^3(3x^2+4\tilde{\a}_{1\pm}^2)\right]\,.\end{split}
	\end{align}
	\end{shaded}
where $\tilde{\a}_{1\pm}=\pm\sqrt{1-x^2}$ and  $H_{n,r}$ is the generalized harmonic number of order $r$ of $n$. The outer solution serves as a reference to determine the correct radius of convergence for the inner solution which determines the limits of the integral to be computed for the Mellin amplitude.
	
	\item {\bf Regge-Mellin Amplitude:} 
	Finally, from conformal Regge theory, the final Mellin amplitude for the leading Regge trajectory can be written as a perturbative expansion in coupling,
	\begin{shaded}
	\be
	\mm_+^{(0)}=2\int_{-1}^1 dy \sum_{k=0}^\infty A_k(L,y)\frac{e^{Ly}\sqrt{1-y^2}}{y^{k+1}}\xi^{2k+2}\,,\ L=\xi^2\log\frac{s}{4}\,.
	\ee
	\end{shaded}
this is a perturbative expansion of the Mellin amplitude obtained for $\xi\rightarrow0$, $s\rightarrow\infty$ and $L=\xi^2\log(s/4)=\text{constant}$. The functions $A_k(L,y)$ are given by,
	\begin{shaded}
	\begin{align}
	\begin{split}
	A_0(L,y)&= \frac{  \Gamma^3 (h)  }{(2\p)^{1+4 h} \Gamma^4 \left(\frac{h}{2}\right)}\\
	A_1(L,y)&=\frac{A_0(L,y)}{2}
	\left[H_{h-1}-y \left(\psi ^{(0)}(h) (L-4 y)+2 y \psi ^{(0)}\left(\frac{h-t}{2}\right)+4 y \psi ^{(0)}\left(\frac{h}{2}\right)+\gamma  L\right)\right]\,.
	\end{split}
	\end{align}
	\end{shaded}
\end{itemize}
where $\gamma$ is Euler–Mascheroni constant.
\vspace*{0.5cm}
\subsubsection*{General $d$ fishnet theory: 1 - magnon }
\paragraph{}For $1-$magnon case, we consider the four point correlator $\langle{\rm Tr}(X(x_1)Z(x_1)X(x_2)){\rm Tr}(\bar{Z}(x_3)\bar{X}(x_4)\bar{Z}(x_4))\rangle$. The dimensions are $\D_1=d/2$ and $\D_2=d/4$. We have two distinct cases for even and odd spin. For {\bf even spin}, the solution for the leading Regge trajectory are,
\begin{itemize}

\item{\bf Regge trajectories: Outer solution} The outer solution for $\n>\xi$ is,
\begin{shaded}
\be
J_{o \pm}^{(1e)}(\n)=-\frac{h}{2}\pm i \nu + \b_{1\pm} \x^2 +\sum _{k=2}^\infty \b_k( \pm\n) \xi ^{2 k}\,.
\ee
\end{shaded}
with the coefficients given by
\begin{shaded}
\begin{align}
\begin{split}
\b_2(\pm\n)&=-\frac{\b_{1\pm}^2}{2}G_{1}(\pm i\n),~~~
	\b_3(\pm\n)=\frac{\b_{1\pm}^3}{8}\left[G_{2}(\pm i\n)+3G_{1}(\pm i\n)^3\right],\\
	\b_4(\pm\n)&=-\frac{\b_{1\pm}^4}{48}\left[ G_{3}(\pm i\n)+12 G_{2}(\pm i\n)G_{1}(\pm i\n)+16 G_{1}(\pm i\n)^3\right]
\end{split}
\end{align}
\end{shaded}	
The functions $G_r(\n)$ are given by \eqref{FaG} and $\b_{1\pm} =\G(h/2)\G(i \n)/(2\G(h/2 \pm i\n))$.	

\item {\bf Regge trajectories: Inner solution} The inner solution for $\n< \xi$ 
\begin{shaded}
\be
J_{i\pm}^{(1e)}(x)= -\frac{h}{2}+\tilde{\b}_{1\pm}\x+\sum_{k=2}^{\infty}\tilde{\b}_k( x)\x^{k}\,, \ \n=x\xi^2\,, x<1\,,
\ee
\end{shaded}
where the coefficients are given respectively by,
\begin{shaded}
\begin{align}
\begin{split}
\tilde{\b}_2(x)&=-\frac{1}{2}(\tilde{\b} _{1\pm}^2+x^2)H_{\frac{h}{2}-1},\,\,\,\tilde{\b}_3(x)=\frac{(\tilde{\b}_{1\pm}^2+x^2)}{8 \tilde{\b}_{1\pm}} \left[\left(H_{\frac{h}{2}-1}\right)^2 (3 \tilde{\b}_{1\pm}^2+x^2)+H_{\frac{h}{2}-1,2} (\tilde{\b}_{1\pm}^2-x^2)\right]\\
	\tilde{\b}_4(x)&=-\frac{1}{24} (\tilde{\b}_{1\pm}^2+x^2) 
	\left[
	\left(H_{\frac{h}{2}-1}\right)^3 (8 \tilde{\b}_{1\pm}^2+6 x^2)
	+6  H_{\frac{h}{2}-1} \,H_{\frac{h}{2}-1,2}\,\tilde{\b}_{1\pm}^2
	+H_{\frac{h}{2}-1,3} (\tilde{\b}_{1\pm}^2-3 x^2)\right].
	\end{split}
	\end{align}
\end{shaded}	
where $\tilde{\b}_{1\pm}=\pm\sqrt{1-x^2}$ and  $H_{n,r}$ is the generalized harmonic number of order $r$ of $n$.
\item{\bf Regge-Mellin Amplitude:} The reduced Mellin amplitude is then,
\begin{shaded}
\be
\mm_+^{(1)}=\int_{-1}^1 dy s^{-\frac{h}{2}}\sqrt{1-y^2}e^{Q y} \sum_{k=0}^\infty B_k(Q,y)\xi^{2+k}+(s\rightarrow-s)\,, \ Q=\xi\log\frac{s}{4}\,.
\ee	
\end{shaded}
this is a perturbative expansion of the Mellin amplitude obtained for $\xi\rightarrow0$, $s\rightarrow\infty$ and $L=\xi\log(s/4)=\text{constant}$. The functions $B_k(Q,y)$ given by,
\begin{shaded}
\begin{align}\label{1magevenresult2}
	\begin{split}
	B_0(Q,y)&=-\frac{  \csc \left(\frac{\pi  h}{2}\right) \Gamma \left(\frac{h}{2}\right)  \Gamma \left(\frac{3 h}{4}-\frac{t}{2}\right)^2}{2^{(4h+3)} \pi ^{4 h}\Gamma \left(1-\frac{h}{2}\right)},\\
	B_1(Q,y)&=B_0(Q,y)\left(-y \left(H_{-\frac{h}{2}}+\psi ^{(0)}\left(\frac{3 h}{4}-\frac{t}{2}\right)\right)-\frac{1}{2} Q \left(\psi ^{(0)}\left(\frac{h}{2}\right)+\gamma \right)+\pi  y \cot \left(\frac{\pi  h}{2}\right)\right)
	\end{split}
	\end{align}
\end{shaded}	
	
\end{itemize}
	
Similarly for $1-$magnon {\bf odd spin} case,

\begin{itemize}
\item{\bf Regge trajectories: Outer solution} For $\n>\xi$,
\begin{shaded}
\be
J^{(1o)}_{o\pm}(\n)=-\frac{h}{2}+\l_\pm\xi^2+\sum_{k=2}^\infty \l_k(\n)\xi^{2k}\,,
\ee
\end{shaded}
with the coefficients given by,
\begin{shaded}
\begin{align}
\begin{split}
\l_2(\n)&=-\frac{\l_{1\pm}^2}{2}G_{1}(i\n),~~~
\l_3(\n)=\frac{\l_{1\pm}^3}{8}\left[G_{2}(i\n)+3G_{1}(i\n)^3\right],\\
\l_4(\n)&=-\frac{\l_{1\pm}^4}{48}\left[ G_{3}(i\n)+12 G_{2}(i\n)G_{1}(i\n)+16 G_{1}(i\n)^3\right]\,,
\end{split}
\end{align} 
\end{shaded}
where the functions $G_r(\n)$ are given in \eqref{FaG} and $\l_{1\pm} =-\G(h/2)\G(i \n)/2\G(h/2 \pm i\n)$.	

\item{\bf Regge trajectories: Inner solution} The inner solution is given by,
\begin{shaded}
\be
J_{i\pm}^{(1o)}(x)=-\frac{h}{2}+\tilde{\l}_\pm\xi+\sum_{k=2}^\infty \tilde{\l}_k(x)\xi^k\,, \ \ \n=x\xi\,, x<1,
\ee	
\end{shaded}
with the first few coefficients given by,
\begin{shaded}
\begin{align}
\begin{split}
\tilde{\l}_2(x)&=-\frac{1}{2}(\tilde{\l}_{1\pm}^2+x^2)H_{\frac{h}{2}-1},\,\,\,\tilde{\l}_3(x)=\frac{(\tilde{\l}_{1\pm}^2+x^2)}{8 \tilde{\l}_{1\pm}} \left[\left(H_{\frac{h}{2}-1}\right)^2 (3 \tilde{\l}_{1\pm}^2+x^2)+H_{\frac{h}{2}-1,2} (\tilde{\l}_{1\pm}^2-x^2)\right]\\
\tilde{\l}_4(x)&=-\frac{1}{24} (\tilde{\l}_{1\pm}^2+x^2) 
\left[\left(H_{\frac{h}{2}-1}\right)^3 (8 \tilde{\l}_{1\pm}^2+6 x^2)
+6  H_{\frac{h}{2}-1} \,H_{\frac{h}{2}-1,2}\,\tilde{\l}_{1\pm}^2
+H_{\frac{h}{2}-1,3} (\tilde{\l}_{1\pm}^2-3 x^2)\right].
\end{split}
\end{align} 
\end{shaded}
where $\tilde{\l}_{1\pm}=\pm i\sqrt{1+x^2}$ and  $H_{n,r}$ is the generalized harmonic number of order $r$ of $n$.
\item{\bf Regge-Mellin amplitude} Finally, the Mellin amplitude for odd spin can be shown to be the analytic continuation of the even spin result for $\xi\rightarrow i\xi$. 
\end{itemize}
The functions $F_r$ and $G_r$ are given by,
\begin{align}\label{FaG}
\begin{split}
F_{r}(z)&:=(r-1)! H_{h-1,r}+(-1)^{r-1} \left[\psi^{(r-1)}(h+z)-\psi^{(r-1)}(z)\right]\,,\\
G_{r}(z)&:=(r-1)! H_{\frac{h}{2}-1,r}+(-1)^{r-1} \left[\psi ^{(r-1)}\left(\frac{h}{2}+z \right)-\psi ^{(r-1)}(z )\right]\,.
\end{split}
\end{align}
 \\
 \subsubsection*{Chiral Fishnet theory in $d=4$} 
\paragraph{ } For the {\bf chiral fishnet theory}, we consider correlation functions $\langle{\rm Tr}[\phi_1(x_1)\phi_1(x_2)]{\rm Tr}[\phi_1^\dagger(x_3)\phi_1^\dagger(x_4)]\rangle$ %$\tr[\phi_j(x)^2]$.
 The dimension of these operators are $\D=1$. There are two coupling constants $\k$ and $\omega$. We work in the limit $\k \to 0, \omega\to 0$ with $\k/\omega$ held constant. 
%\begin{align}
%\begin{split}
%b^{{\rm CI}}_J(\n)=&1/\left(-\omega ^4+\frac{\left(J^2+\nu ^2\right) \left((J+2)^2+\nu ^2\right)}{16 (J+1) \nu } \left((J+1) \nu +i \kappa ^4 \left(-\psi ^{(1)}\left(\frac{1}{4} (J+i \nu +2)\right)+\psi ^{(1)}\left(\frac{1}{4} (J+i \nu +4)\right)\right.\right.\right.\\
%&\left.\left.\left. +\psi ^{(1)}\left(\frac{1}{4} (J-i \nu +2)\right)-\psi ^{(1)}\left(\frac{1}{4} (J-i \nu +4)\right)\right)\right)\right)\,,
%\end{split}
%\end{align}
%and expand around the $0-$magnon case,
%\be
%b_J^{{\rm Cl}}(\n)=\sum_{k=0}^\infty \frac{c_k(J,\n)}{a_0(J,\n)^{k+1}}\,.
%\ee
\begin{itemize}
\item{\bf Regge trajectories: Outer solution} The outer solution is,
\begin{shaded}
\be
J_{\text{o}\pm}^{(c)}(\nu)=\pm i\n+\g_{1\pm}\omega^4+\sum_{k=2}^\infty \g_k(\pm\n) \omega^{4k}\,.
\ee
\end{shaded}
the first few coefficients in the expansion are given by
\begin{shaded}
\begin{align}
\begin{split}
\g_2(\pm\n)&=\frac{\g_{1\pm} \left(\g_{1\pm} (3-3  (\nu^2 \pm 3 i\n))-2 z^4 \left(3 \psi ^{(1)}\left(\pm\frac{i \nu }{2}+1\right)-3 \psi ^{(1)}\left(\pm\frac{i \nu }{2}+\frac{1}{2}\right)+\pi ^2\right)\right)}{6  (\nu^2 \pm i\n)}\\
\g_3(\pm\n) &=\frac{\g_{1\pm} }{36 \nu ^2 (\nu \mp i)^2}
\Bigg[3 \g_{1\pm} \Bigg\{6 \g_{1\pm} (1+\nu  (\nu  (-9+\nu  (\nu \mp 5 i))+5 i))+z^4 \Bigg(4 \pi ^2 (-1+\nu  (\nu \mp 4 i))\\
& \hspace{3.5 cm}+12 (-1+\nu  (\nu \mp4 i)) \psi ^{(1)}\left(\pm \frac{i \nu }{2}+1\right)-12 (-1+\nu  ( \n\mp 4 i)) \psi ^{(1)}\left(\pm\frac{i \nu }{2}+\frac{1}{2}\right)\\
&\hspace{3.5 cm}+3 \nu  (\n\mp i ) \left(-\psi ^{(2)}\left(\pm\frac{i \nu }{2} +1\right)+\psi ^{(2)}\left(\pm\frac{i \nu }{2}+\frac{1}{2}\right)+12 \zeta (3)\right)\Bigg)\Bigg\}\\
&\hspace{3.5 cm}+4 z^8 \left(3 \psi ^{(1)}\left(\pm\frac{i \nu }{2}+1\right)-3 \psi ^{(1)}\left(\pm \frac{i \nu }{2}+\frac{1}{2}\right)+\pi ^2\right)^2\Bigg],
\end{split}
\end{align}
\end{shaded}
where $\g_{1\pm}=-\frac{2}{(\nu^2 \pm i\nu)}$ and  we have used $\kappa= z \omega$ and $z\geq 0$.

\item{\bf Regge trajectories: Inner solution} The inner solution is,
    \begin{shaded}
    	\be 
    	J_{\text{i}\pm}^{(c)}(x)=\tilde{\d}_{1\pm}\omega^2 + \sum_{k=2}^{\infty} \tilde{\d}_k(x)\omega^{2k}
    	\ee 
    \end{shaded}	
    with first few coefficients $\tilde{\d}_k,\, k\ge 2$ given by
    \begin{shaded}
        	\begin{align}
    	\begin{split}
    	\tilde{\d}_2(x)&=-\frac{\tilde{\d}_{1}^2}{2}-2 x^2, \qquad
    	\tilde{\d}_3(x)= \frac{\left(\tilde{\d}_{1}^2+4 x^2\right) \left(\tilde{\d}_{1}^2+6 z^4 \zeta (3)\right)}{2 \tilde{\d}_{1}},\\
    	\tilde{\d}_4(x)&= -\frac{1}{120} \left(\tilde{\d}_{1}^2+4 x^2\right) \left[75 \tilde{\d}_{1}^2+60 x^2+z^4 \left\{30 \left(12 \zeta (3)+\psi ^{(2)}(1)-\psi ^{(2)}\left(\frac{1}{2}\right)\right)+7 \pi ^4\right\}\right] ,
    	\end{split} 
    	\end{align}
    \end{shaded}
where $\tilde{\d}_{1\pm}=\pm 2\sqrt{1-x^2}$ and we have used $\kappa= z \omega$ and $z\geq 0$.

\item{\bf Regge-Mellin amplitude}
%\begin{equation}
%\begin{split}
%I_k=\int d\n\oint dJ \frac{l_0(J,\n)^k M(J,\n)}{a_0(J,\n)^{k+1}}=2\omega^2\int_{-1}^1 dx \frac{x}{\sqrt{1-x^2}}\left(\frac{l_0(J(x),x)^k M(J(x),x)}{((J-J_1^+)(J-J_1^-)(J-J_2^+)(J-J_2^-))^k}\right)_{Res J=J_+}\,.
%\end{split}
%\end{equation}
We obtain the perturbative expansion of the Mellin amplitude for $\w\rightarrow0$, $s\rightarrow\infty$ and $Q=\w^2\log(s/4)=\text{constant}$, 
\begin{shaded}
	\be
		\begin{split}
		&\mm^{(c)}_+(s,t)\\
		&=2\int_{-1}^1 dx~\Gamma \left(1-\frac{t}{2}\right)^2\sqrt{1-x^2}e^{2 q x} \left( \frac{ \omega ^2 }{128 \pi ^9 x} + \frac{ \omega ^4}{128 \pi ^9 x^2}  \left(-2 q x-2 x^2 \psi ^{(0)}\left(1-\frac{t}{2}\right)+4 x^2+1\right)\right.\\
		&\left. +\frac{\omega ^2 }{384 \pi ^9 x^3} \left(8 \omega ^4 \left(x \left(6 q^2 x-6 q \left(2 x^2+1\right)+x \left(\pi ^2 \left(2 x^2+1\right)+6\right)\right)+3 x^2 \left(2 x \psi ^{(0)}\left(1-\frac{t}{2}\right)\right.\right.\right.\right.\\
		&\left.\left.\left.\left. \left(2 q+x \psi ^{(0)}\left(1-\frac{t}{2}\right)-4 x\right)+\left(2 x^2-1\right) \psi ^{(1)}\left(1-\frac{t}{2}\right)\right)+3\right)+9 \kappa ^4 \zeta (3) (x (q+x)-1)\right)\right.\\
		&\left. \frac{\omega ^4}{7680 \pi ^9 x^4}  \left(-\kappa ^4 \left(180 \zeta (3) \left(-2 x \left(-2 q^2 x+q \left(3-2 x^2\right)+2 x^3+x\right)+2 x^2 (2 x (q+x)-1) \psi ^{(0)}\left(1-\frac{t}{2}\right)\right.\right.\right.\right.\\
		&\left.\left.\left.\left. +3\right)+7 \pi ^4 x^2 (2 x (q+x)-1)\right)-320 \omega ^4 \left(2 q \left(2 q^2+\pi ^2+3\right) x^3-\left(6 q^2+\pi ^2+3\right) x^2+6 x^4 (2 x (q-2 x)+1)\right.\right.\right.\\
		&\left.\left.\left. \psi ^{(0)}\left(1-\frac{t}{2}\right)^2+3 x^2 \left(2 x \left(2 x^2-1\right) (q-2 x)+1\right) \psi ^{(1)}\left(1-\frac{t}{2}\right)+2 x^4 \psi ^{(0)}\left(1-\frac{t}{2}\right)\right.\right.\right.\\
		&\left.\left.\left. \left(6 q (q-2 x)+\left(6 x^2-3\right) \psi ^{(1)}\left(1-\frac{t}{2}\right)+\pi ^2 \left(2 x^2+1\right)\right)+4 \left(\pi ^2-6\right) q x^5+6 q x\right.\right.\right.\\
		&\left.\left.\left. +4 x^6 \psi ^{(0)}\left(1-\frac{t}{2}\right)^3+\left(4 x^2-3\right) x^4 \left(\psi ^{(2)}\left(1-\frac{t}{2}\right)+12 \zeta (3)\right)-8 \pi ^2 x^6+12 x^4-3\right)\right)\right)\\
		& +(s \rightarrow -s) + {\cal O}(\omega^{10}, \omega^6\kappa^4, \omega^2\kappa^8) 
	\end{split}
	\ee
\end{shaded}    
\end{itemize}    
The remainder of the paper is organized as follows. In section \ref{genRegge}, we discuss  general principles for  perturbative evaluation of the Regge trajectories in generalized fishnet theories in $d-$dimensions. Specifically, we focus on the issues regarding the convergence of the solutions and solutions appropriate for different perturbative regimes. In section \ref{0mag} we apply the general principles to the $d-$dimensional $0-$magnon case including details of the Regge trajectories, the Mellin amplitudes and further analysis. In section \ref{1mag} we perform the same analysis for the $1-$magnon case separately for even and odd spin. In section \ref{chm1}, we extend the logic to chiral fishnet theory in four dimensions. This is a bit involved due to multiple couplings but give rise to interesting observations. We finish the work with conclusions in section \ref{concl} and point out some future directions. In the appendix \ref{iaoi0mgd}, we point out the integrals required for the inner and outer solutions. Specifically, we point out the subtleties involved in the choice of contour manipulations, such as Wick rotation and analytic continuation. In appendix \ref{chfema1int}, we do the same for the chiral fishnet theory. Finally in appendix \ref{intweak}, we provide some details of the integrals needed for the perturbative computations of the Mellin amplitudes.

%\section{Regge Theory of $d$-dimensional Fishnet Theory}\label{regfish}

\section{Perturbative Regge Trajectories: General principles}\label{genRegge}
In this section we give a  general outline of perturbative analysis for obtaining Regge trajectories in the $d-$dimensional conformal fishnet theory.  Generically, the spectral functions \eqref{specdef} are very complicated functions of spin $J$, the spectral representation parameter $\nu$ and the coupling constant $\xi$ of the theory. The Regge trajectories are yielded by the  poles in spin $J$ of the spectral function $b_J^n(\n)$ which, for $d-$dimensional fishnet theories, has the generic structure
\be \label{gendspec}
\frac{\left(E_{h+i\n,J}^{(n)}\right)^p}{1-\c_nE_{h+i\n,J}^{(n)}}
\ee where $E_{h+i\n,J}^{(n)}$ is the eigenvalue of the graph-building operator $\hat{H}^{(n)}$ for n-magnon graph. One has $p=1,2$ for  $0$ and $1$ magnon graphs respectively and $\c_0=((4\p)^h\x^2)^2,\,\c_1= (4\p)^h\x^2$. For a self-contained review of these structures the readers are referred to section 3 of \cite{Chowdhury:2019hns}.
However, often $E_{h+i\n,J}^{(n)}$ contains transcendental or hypertranscendental functions like gamma function, polygamma function etc., making it almost always impossible to obtain exact expression of the Regge spin $J$ as a function of the scaling dimension ($\Delta =h+ i \nu$) as well as the coupling $\x$. To circumvent this, we will  work perturbatively and obtain an expression for the Regge spin perturbative in $\x$ with expansion coefficients as functions of $\n$. This kind of perturbative analysis was first initiated in \cite{Chowdhury:2019hns} for studying Regge theory of two magnon correlators in $4d$ conformal fishnet theory. We formalize the argument presented there in this section. \par
Essentially there are two extreme regions of couplings where we can consider the perturbative expansion in $\x$. One is the weak coupling region, $\x\to0$, and another is the strong coupling region $\x\to\infty$. In this work, we will focus on the weak coupling limit only. In the following, we will chalk out the basic principles of perturbative evaluation for the Regge trajectories using abstract arguments. Explicit evaluations based on this abstract analysis will be presented for various cases in subsequent sections.

In the weak coupling region we can consider  following generic perturbation series about $\x=0$%\footnote{Observe that, the Regge trajectory is a function of $i\n$ rather than just $\n$ i.e., $J(\n)\equiv J(i\n)$. This is because, ultimately, $J$ is a function of $\D$ and we have expressed $\D$ as $h+i\n$.}
\be \label{Jpert1}
J^{Regge}=\sum_{k=0}^\infty f_k(\n)\,\x^{ \a k}
\ee where $\a>0$ is non-universal and depends on the spectral function of the correlator we are looking at. Due to the range of integration of $\n$, $(\infty,+\infty)$, there is a slight subtlety in the perturbative expansion.  We can distinguish two distinct regions for $\n$: $|\n|\gg h(\x)$ and  $\,|\n|< h(\x)$ for some $h(\x)$ ( which is, again, non-universal and depends on the correlator), giving rise to two distinct perturbative representations of $J$ in the regions above.  
\paragraph{} 
First we consider the region of $\n$ where $\n\gg\xi$. We observe that, the Regge spin $J^{Regge}$ satisfies
\begin{align} \label{ReggeEq}
	\left(E_{\Delta,J^{Regge}}^{(n)}\right)^{1-p}\left[\left(E^{(n)}_{\Delta,J^{Regge}}\right)^{-1}-\c_n\right]=0,\,~~~p\ge1.
\end{align} Now, this has two solutions: one is provided by the pole(s) of $E^{(n)}_{\Delta,J^{Regge}}$. This solution is independent of the coupling $\x$. Let us call this Regge trajectory to be $J^{(n)}_f(\n)$ for later reference. The subscript $f$ is to denote the fact that, in the free limit i.e., $\x\to0$, this is the only solution that is there. Therefore, we will call $J_f^{(n)}$  \emph{free} Regge trajectory.
The other solution which, clearly, depends upon the coupling  is given by the solution of the equation\footnote{Note that, for this solution we are assuming quite justifiably $(E^{(n)}_{\Delta,J^{Regge}})^{-1}\neq 0$.}
\be \label{Reggesol2}
\c_n\,E_{\D,J^{Regge}}^{(n)}=1.
\ee 

We have $\c_n\sim \x^{2q}$, $q$  being either $1$ or $2$. Then, for \eqref{Reggesol2} to have a solution for $J^{Regge}$ in the limit $\x\to0$ one must have that, in the perturbative expansion of $E_{\D,J^{Regge}}^{(n)}$\footnote{We use the ansatz in \eqref{Jpert1} into $E_{\D,J^{Regge}}^{(n)}$ to obtain a weak coupling perturbative expansion of the latter.} in $\x$ there is a term $\sim \x^{-2q}$ as $\x\to 0$.   Essentially, we are looking for singular behavior of $E_{\D,J^{Regge}}^{(n)}$ in the limit $\x\to 0$. This can be achieved by considering the ansatz in \eqref{Jpert1} to be a weak coupling perturbation about the free Regge trajectory $J^{(n)}_f$: 
\be \label{Jpert2}
J^{(n)}_{\text{o}}(\n):=J_f^{(n)}(\n)+\sum_{k=1}^{\infty}f_k(\n)\x^{\a k}.
\ee 
Now, $\a$ can be determined by the pole structure of $E^{(n)}_{\D,J^{Regge}}$. Assuming $J=J_f^{(n)}$ to be a simple pole (which is true for all of our cases), it is straightforward to obtain
\be \label{alphaval}
\a=2q\,.
\ee 
Using this ansatz
into the left hand side of \eqref{Reggesol2} and  expanding the same about $\x=0$ \emph{assuming $\n$ to be non-perturbative}, one solves for $\{f_p(\n)\}$ by solving \eqref{Reggesol2} order by order in $\x$. % It turns out that %\textbf{\textcolor{red}{(PH: I am  using the word \enquote{turns out} because I could not find a mathematical reason, even a hand wavy one, for why this is happening!)}},
%,$f_p(\n),\,p\ge 2$ can be expressed in terms of $f_1(\n)$ such that $f_p(\n)=0$ \emph{for all} $p\ge2$ when $f_1(\n)=0$.  Further, the fact that
The fact that $J_\x^{Regge}$ is a pole of the spectral function  manifests itself as a pole in $f_1(\n)$ with residue proportional to $\x^{-2pq}$ i.e., one has the following structure for the spectral function \eqref{specdef},
\be \label{polestructo}
\frac{\left(E_{\Delta,J_{\text{o}}^{(n)}}^{(n)}\right)^p}{1-\chi_n E^{(n)}_{\Delta,J_{\text{o}}^{(n)}}}= \x^{-2pq}\frac{\mathfrak{B}_{\text{o}}(\n)}{[f_1(\n)]^{p-1}\left[f_1(\n)-F(\n)\right]}
\ee where, it turns out that, $\mathfrak{B}_{\text{o}}(\n)$ can be \emph{solely} expressed in terms of  to the \emph{residue} of $E_{\Delta,J}^{(n)}$ at $J=J_f^{(n)}$. Also, $F(\n)$ is actually the same residue but some $\n$-independent numerical factor! The readers are referred to appendix \cite{sec2det} for the detailed analysis. Also  Note that, this pole structure, \eqref{polestructo}, also account for the Regge spin $J_f^{(n)}(\n)$ by
virtue of the pole at $f_1(\n)=0$. This is expected by construction of $J_{\text{o}}^{(n)}$.  \paragraph{}
The perturbative expression \eqref{Reggesol2} can't be valid over the entire range of $\n\in(-\infty,\infty)$. Especially, near $\n\to0$ this perturbation series can't hold good simply because now $\n$ and $\x$ are comparable. Therefore, the assumption of $\n$ being non-perturbative does not hold good in this domain. In order to estimate the radius of convergence, we perform a ratio test. The perturbative expansion converges as long as 
\be 
\lim_{k\to\infty}\left|\frac{f_{k+1}(\n)}{f_k(\n)}\right|\x^{2q}< 1.
\ee This is true for any $|\n|\sim O(1)$. However, as $|\n|$ becomes less than unity there will come a point where this condition will fail to hold good. This is determined by investigating the ratio $|f_{k+1}(i\n)/f_{k}(i\n)|$ in the limit $\n\to0$. In particular, say, we have
\be 
\lim_{\substack{\n\to0\\k\to\infty}}\left|\frac{f_{k+1}(\n)}{f_k(\n)}\right|\sim |\n|^{-b},~~~b>0.
\ee 
Then the perturbative expansion will break down if 
\be |\n|^{-b}\,\x^{2q}\gtrsim1.\ee
Thus, we can't use the perturbative expansion in \eqref{Jpert2}  for $|\n|\sim O\left(\x^{2q-b}\right)$ . To proceed further, let us introduce the scaling relation
\be \label{innerscaling}
\n=x\,\x^{2q-b}
\ee so that now the entire region of $\n$ is labelled by that of $x$ and we are interested into two regions: $|x|<1$ and $|x|>1$. We will call the former to be \emph{inner region} and the latter \emph{outer region}. In the \emph{outer region} we can use \eqref{Jpert2} with $\a=2q$ but in the inner  region we will need to use a new perturbation expansion. In general with the scaling behaviour \eqref{innerscaling} we have , % To develop a perturbation series in the inner region, first, we will investigate the equation
\be \label{innereq}
1-\c_n E^{(n)}_{h+ix\,\x^{2q-b},J^{Regge}}=0
\ee
% in the limit $x\to0$. In this limit the equation becomes $1-\c_n E^{(n)}_{h,J^{Regge}}=0$. Now, in this limit we can construct a perturbative (perturbative in $\x$) ansatz for $J^{Regge}$ around the pole of $E^{(n)}_{h,J^{Regge}}$ which is clearly at $J^{Regge}=J_f^{(n)}(\n=0)$. Note that, this ansatz is independent of $x$. 
For $x\neq 0,\, |x|<1$ we can consider the perturbative ansatz for $J^{Regge}$ solving \eqref{innereq} above to be 
\be \label{Jifinal}
J_{\text{i}}^{(n)}(x)=J_f^{(n)}(0)+\sum_{k=1}h_k(x)\x^{k\left(2q-b\right)}.
\ee Here, the subscript $\text{i}$ denotes that this perturbative ansatz is valid in the \emph{inner region}. We will, hereafter, call this solution by the name \emph{inner solution} and the perturbative expression \eqref{Jpert2} by the name \emph{outer solution}\footnote{Hence the extra $o$ in the subscript!}.As before, the Regge pole structure of the spectral function manifests itself into a pole of the spectral function in $h_1(x)$. Thus, in the inner region the spectral function has the structure
\be \label{polestructi}
\frac{\left(E_{h+ix\,\x^{2q-b},J_{\text{i}}^{(n)}}^{(n)}\right)^p}{1-\chi_n E^{(n)}_{h+ix\,\x^{2q-b},J_{\text{i}}^{(n)}}}\equiv
\x^{-2pq}
\frac
{
	\mathfrak{B}_{\text{i}}(x)
}
{
	[h_1-H_f(x)]^{p-1}\left[h_1-H_\x(x)\right]
}
\ee Here, the pole at $h_1=H_f(x)$ corresponds to the  $J_f^{(n)}(x)$ while, $h_1=H_\x(x)$ corresponds to the inner solution $J_{\text{i}}^{(n)}(x)$.  We will now consider zero magnon and one magnon correlators in the $d-$dimensional Fishnet theory for our explicit calculation. It is also worth mentioning that, abstract arguments above have been laid out keeping in mind the structure of the spectral function, \eqref{gendspec}, for $d-$dimensional Fishnet theories. However, similar arguments can also be put into use, if required, for other structures of the spectral functions. For instance, this has been done for our analysis of Regge trajectories in chiral fishnet theories.  

\section{Zero Magnon correlator in $d-$dimensional fishnet theory}\label{0mag}
In the zero-magnon case we have $\D_1=\D_2=\D_3=\D_4=\frac{h}{2}$. Putting these into  \eqref{melfn} we have for zero-magnon correlator,
\begin{align}\label{Mellin0}
\begin{split}
\mm^{(0)}_{\pm}(s,t)=\left[\frac{\pm1}{2\pi i}\right. &\left.\oint dJ \frac{\pi}{\sin\pi J} \int_{-\infty}^\infty d\n \left(\frac{s}{4}\right)^J \nu \sinh \pi\nu \,\zeta_0({\D_i,t})\right.\\
&\times\left. \frac
{  
	\Gamma (h+J) 
	\Gamma (h+J-i \nu ) 
	\Gamma (h+J+i \nu ) 
	\Gamma \left(\frac{h-J-t-i \nu }{2}\right) 
	\Gamma \left(\frac{h-J-t+i \nu }{2}\right)
}
{
	2\pi ^{ 2(h+1)} 
	\Gamma (J+1) 
	\Gamma \left(\frac{h+J-i \nu }{2}\right)^2 
	\Gamma \left(\frac{h+J+i \nu }{2}\right)^2
} \frac{E_{\Delta,J}^{(0)}}{1-\chi_0 E^{(0)}_{\Delta,J}}\right]\\
& \pm(s\rightarrow -s),
\end{split}
\end{align} 
with $\z_0(\D_i,t)=\G\left(\frac{h-t}{2}\right)^{-2}$. In general $d$, the eigenvalue of the zero-magnon graph building operator is given by eq.(C.7) of \cite{Gromov:2018hut}
\be \label{E0}
E^{(0)}_{h+i\n,J}=c^4 \p^{2h}\frac
{
	\Gamma \left(\frac{J-i\n }{2}\right) \Gamma \left(\frac{J+i\n }{2}\right)
}
{
	\Gamma \left(\frac{2 h+J-i\n }{2}\right) \Gamma \left(\frac{2h+J+i\n }{2}\right)
}
\ee
and the spectral function is given by \eqref{gendspec} with $\c_0=(4\p)^{2h}\x^4,\,p=1$,
\be \label{spec0}
\frac{E_{h+i\n,J}^{(0)}}{1-\c_0 E^{(0)}_{h+i\n,J}}\,.
\ee Note that, $E^{(0)}_{h+i\n,J}$ as well as the spectral function is invariant under $\n\to-\n$. Thus, if $J^{Regge}(\n)$ is a Regge trajectory so is $J^{Regge}(-\n)$. Now, we see that the spectral function is in terms of Gamma functions thereby, making the exact evaluation of the Regge trajectories difficult. So we will follow the analysis chalked out in section \ref{genRegge} to investigate the Regge theory in the weak coupling limit $\x\to0$. 
\subsection{Evaluation of Regge Trajectories}
We initiate determination of  Regge trajectory in the weak coupling limit following section \ref{genRegge}. First, observe that the free Regge trajectory $J_f^{(0)}$ is given in this case by

\be
J^\text{(0)}_{f\pm}(p)=\pm i\n-2p,\,~~~\text{where}~
\begin{cases} 
	0\leq p\leq h-1,~~ h\in\mathbb{Z}_{\ge0},\\ 
	p\geq0,~~~~~~~~~~~h\in\mathbb{Z}_{\ge 0}+\frac{1}{2}.
\end{cases}
\ee 
Clearly, $J^\text{(0)}_{f\pm}(0)$ corresponds to the leading Regge trajectory for all $h$. 
We will concentrate our attention upon the leading Regge trajectory from hereon. 
	\vspace*{0.1cm}
\paragraph{Outer Solution:} For perturbative evaluation of the leading Regge trajectory, we will follow the analysis chalked out in section \ref{genRegge}. First, we will consider the \emph{outer solution}. 
Observing that $J^\text{(0)}_{f\pm}(0)$ is a simple pole of $E_{h+i\n,J}^{(0)}$, we consider the perturbative solutions
\begin{shaded}
\be\label{J0out}
J_{\text{o}\pm}^{(0)}(\nu)=\pm i\n+\a_{1\pm}\xi^4+\sum_{k=2}^\infty \a_k(\pm\n) \xi^{4k}\,.
\ee
\end{shaded}
the first few coefficients in the expansion \eqref{J0out} are given by\footnote{To clear notational cluttering, we have used the a rescaling of $c$ 
	\be 
	(2 \pi )^{h} \sqrt{\frac{2}{\Gamma (h)}}~c\to 1.
	\ee  },
\begin{shaded}
\begin{align}\label{ocoeff0}
\a_2(\pm\n)&=-\frac{F_1(\pm i\n)}{2}\a_{1\pm}^2\,,\ \a_3(\pm\n)=\frac{F_2(\pm i\n)+3F_1(\pm i\n)^2}{8}\a_{1\pm}^3\,,\\
~ \a_4(\pm\n)&=-\frac{F_3(\pm i\n)+12F_1(\pm i\n)F_2(\pm i\n)+16F_1(\pm i\n)^3}{48}\a_{1\pm}^4\,.
\end{align}
\end{shaded}
with the definition
\be
F_{r}(z)=(r-1)! H_{h-1,r}+(-1)^{r-1} \left[\psi^{(r-1)}(h+z)-\psi^{(r-1)}(z)\right]\,,
\ee 
where $\psi^{(n)}(z)$ is the usual polygamma function and $H_{n,r}$ is the \emph{generalized harmonic number of order $r$ of $n$}\footnote{\label{harmonic} The generalized harmonic number of order $r$ of $n$ is defined by
\be 
H_{n,r}:=\sum_{k=1}^{n}\frac{1}{k^r}.
\ee The special case $r=0$ gives $H_{n,0}=n$ and the special case $r=1$ corresponds to the usual harmonic number $H_n$ i.e.,
\be 
H_n:=\sum_{k=1}^{n}\frac{1}{k}
\ee  }. Note that we have two leading outer solutions. Using the solutions \eqref{J0out}, the pole structure of the spectral function manifests as
\be\label{0pole}
\frac{E_{h+i\n,J_{\text{o}\x}^{(0)}}}{1-\c_0E_{h+i\n,J_{\text{o}\x}^{(0)}}}=\x^{-4}\frac{(4\p)^{-2 h}B(h,\pm i\n)/2}{[\a_{1\pm}-B(h,\pm i\n)/2]}\,.
\ee
where $B(a,b)=\G(a)\G(b)/\G(a+b)$. This has a simple pole at $\a_{1\pm} =B(h, \pm i\n)/2$.   While evaluating the Mellin amplitude we have to take into account \emph{both} $J_{\text{o}\x+}^\text{(0)}(\n)$ and $J_{\text{o}\x-}^\text{(0)}(\n)$. 
	\vspace*{0.1cm}
\paragraph{Inner Solution:} Now, we will move onto evaluation of the inner solution. Following the analysis of section \ref{genRegge}, we find out that  the \emph{inner region} corresponds to $|x|<1$ with $\n=x\x^2$.
Thus the perturbative expression for the inner solution turns out to be
\begin{shaded}
\be\label{J0in}
J_{\text{i}}^\text{(0)}(x)=\sum_{k=1}^\infty \tilde{\a}_k(x)\xi^{2k}\,,
\ee
\end{shaded}
where first few coefficients $\tilde{\a}_k(x),\,k\ge2$ are expressed  as
\begin{shaded}
\begin{align}\label{icoef0}
\begin{split}
\tilde{\a}_2(x)&=-\frac{H_{h-1}}{2}(\tilde{\a}_{1}^2+x^2)\,, \ \tilde{\a}_3(x)=\frac{(\tilde{\a}_{1}^2+x^2)}{8\tilde{\a}_{1}}\left[(H_{h-1})^2(x^2+3\tilde{\a}_{1}^2)+H_{h-1,2}(\tilde{\a}_{1}^2-x^2)\right]\,,\\
\tilde{\a}_4(x)&=-\frac{\tilde{\a}_{1}^2+x^2}{24}\left[6H_{h-1} H_{h-1,2} \tilde{\a}_{1}^2+H_{h-1,3} (\tilde{\a}_{1}^2-3x^2)+2(H_{h-1})^3(3x^2+4\tilde{\a}_{1}^2)\right]\,.\end{split}
\end{align}
\end{shaded}
 The spectral function takes the form,
\be\label{0inspec}
\frac{E_{h+ix\x^2, J_{\text{i}\pm}^\text{(0)}(x)}}{1-\c_0 E_{h+ix\x^2, J_{\text{i}\pm}^\text{(0)}(x)}}=\x^{-4}\frac{(4\p)^{-2h}}{\left(\tilde{\a}_{1}-\sqrt{1-x^2}\right)\left(\tilde{\a}_{1}+\sqrt{1-x^2}\right)}\,.
\
\ee
The poles are at $\tilde{\a}_1=\tilde{\a}_{1\pm}:=\pm\sqrt{1-x^2}$. Thus, analogous to the outer region, we have contribution of two Regge trajectories in the region $\nu < \xi^2$. They are given by 
$J_{\text{i}+}^\text{(0)}(x)$ for $\tilde{\a}_{1}=\tilde{\a}_{1+}$ and $J_{\text{i}-}^\text{(0)}(x)$ for $\tilde{\a}_{1}=\tilde{\a}_{1+}$

\subsection{Evaluation of Mellin Amplitude}\label{Mell0}
Now, we turn to evaluating the zero-magnon Mellin amplitude as given in \eqref{Mellin0}. Let us start by rewriting the expression for the Mellin amplitude as following:

\be\label{melr}
\mathcal{M}^{(0)}_\pm(s,t)=\pm\zeta_0(\D_i,t)\int_{-\infty}^\infty d\n  \oint \frac{dJ}{2\p i} \,M^{(0)}(J,\n) \left(\frac{s}{4}\right)^J
\frac{E^{(0)}_{h+i\n,J}}{1-\c_0 E^{(0)}_{h+i\n,J}}
\pm (s\rightarrow -s)\,
\ee
with
\be\label{Mdef}
M^{(0)}(J,\n):=\p\n\sinh(\pi\n) \frac{\G(h+J)\G(h+J+i\n)\G(h+J-i\n)\G(\frac{h-J-t+i\n}{2})\G(\frac{h-J-t-i\n}{2})}
{2\p^{2(h+1)}\sin(\p J)\G(1+J)\G(\frac{h+J+i\n}{2})^2\G(\frac{h+J-i\n}{2})^2}\,.
\ee

The contour integral over $J$ is supposed to pick up the contribution of Regge pole by virtue of Cauchy's integral formula. But we have already seen that,  the pole structure of the spectral function manifests itself as a pole in  the expansion coefficient of the lowest power of the $\x$ in the perturbative expression for the Regge trajectory. Thus, we can change the integration variable from $J$ to this coefficient, let us call the coefficient generically $\d$, so that now we consider a contour integral over this expansion coefficient $\d$ rather than $J$ itself. This change of variable will naturally introduce a Jacoboian of  transformation. Thus,
the integral over $J(\d)$ can be recast as,
\be
\oint \frac{dJ}{2\p i} ~M^{(0)}(J,\n)\left(\frac{s}{4}\right)^J\frac{E^{(0)}_{h+i\n,J}}{1-\c_0 E^{(0)}_{h+i\n,J}}
\rightarrow
\oint \frac{d\d}{2\p i} 
~\Theta(\d)
M^{(0)}(J(\d),\n) \left(\frac{s}{4}\right)^{J(\d)}\frac{E^{(0)}_{h+i\n,J}}{1-\c_0 E^{(0)}_{h+i\n,J}}
\ee where $\d$ is the said coefficient which, in the present case, is $\a_1$ for outer solution and $\tilde{\a}_1$ for inner solution. Further, $\Theta(\d)$ is the Jacobian of transformation given by
\be\label{Jacobdef} 
\Th(\d):=\frac{\pd J}{\pd\d}.
\ee 
 Now, we turn our attention towards the $\n$ integral. We will divide the $\n$ integration domain into inner and  outer regions and perform the integrals accordingly followed by combining them at the end. Accordingly, we will call these \emph{inner integral} and \emph{outer integral} respectively. Thus, the $\n$ integral can be expressed,
 \be \label{0maggendint}
 \mi^{(0)}(\x,s,t)=\int_{-\infty}^\infty d\n \mf(J,\,\n)=
 \left(\int_{-\infty}^{-\x^{2}}+\int_{\x^{2}}^\infty\right) d\n 
 \mf_\text{outer}
 +
 \int^{\x^{2}}_{-\x^{2}} d\n
 \mf_\text{inner}\,.
 \ee 
 where
 \be
 \mf(J,\,\n)= \zeta_0(\D_i,t)\oint \frac{d\d}{2\p i} 
 ~\Theta(\d)
 M^{(0)}(J(\d),\n) \left(\frac{s}{4}\right)^{J(\d)}\frac{E^{(0)}_{h+i\n,J}}{1-\c_0 E^{(0)}_{h+i\n,J}}
 \ee 
 Here, $\mf_\text{inner}$ is obtained by using $J_{\text{i}\pm}^\text{(0)}$  for $\mf(J,\n)$ and $\mf_\text{outer}$ by putting $J_{\text{o}\pm}^{(0)}$ into the same. In can be shown that the full integral \eqref{0maggendint} can be reduced to the following (based on some unproven but not improbable assumptions regarding analytic continuation of inner and outer solution as argued in Appendix \ref{iaoi0mgd}), 
 \begin{shaded}
 \begin{align}\label{I0maggendfinal}
 \begin{split}
  \mi^{(0)}(\x,s,t)&=2\x^2\zeta_0(\D_i,t)\int_{-1}^{1} \frac{y dy}{\sqrt{1-y^2}}\left[\F^{(0)}_+\left(\sqrt{1-y^2}\right)\right]\,.\\
 \end{split}
 \end{align}
 \end{shaded}
  Where $\F^{(0)}_{+}(\sqrt{1-y^2})$, $\Theta^{(0)}_{\text{i}}(\tilde{\a}_{1+})$ and $M^{(0)}(J,\n)$  has been defined in the \eqref{phii0}, \eqref{Jacoi0} and \eqref{Mdef}  respectively. For convenience of the reader we reproduce  the expressions from the appendix here. 
  \begin{align} 
  \begin{split}
  &\F^{(0)}_{+}(\sqrt{1-y^2}):=\Theta^{(0)}_{\text{i}}(\tilde{\a}_{1+})\,M^{(0)}\left(J_{\text{i}+}^\text{(0)}(\sqrt{1-y^2}), \sqrt{1-y^2}\right)\,\left(\frac{s}{4}\right)^{J_{\text{i}+}^\text{(0)}(\sqrt{1-y^2})}\,\left[\pm\frac{(4\p)^{-2h}}{2\x^4y}\right],\\
 &\Theta^{(0)}_{\text{i}}(\tilde{\a}_{1+}):=\frac{\pd J_{\text{i}+}^{(0)}( \sqrt{1-y^2})}{\pd\tilde{\a}_{1+}}=\xi^2\left[1-H_{h-1}^1\tilde{\a}_{1+}\xi^2+\frac{\x^4}{8\tilde{\a}_{1+}^2} \right.\\
 &\left.\phantom{\Theta^{(0)}_{\text{i}}(\tilde{\a}_{1+}):=abcgdehhhjj}+H_{h-1}^2 ((1-y^2)^2+3\tilde{\a}_{1+}^4)+(H_{h-1}^1)^2(9\tilde{\a}_{1+}^4+4(1-y^2)\tilde{\a}_{1+}^2-(1-y^2)^2) + {\cal O}(\xi^6)\right],
  \end{split}
   \end{align}
 where $\tilde{\a}_{1+}= y$. The final Mellin amplitude is given by
\be 
\mm_{+}^{(0)}(s,t)=\zeta_0(\D_i,t)\mi^{(0)}(\x,s,t)\,+\,(s\to-s).
\ee  
Note that $\mm_{-}^{(0)}(s,t)$ is zero because of the fact that the correlator is symmetric under $s \to -s$. In the limit $$\xi \rightarrow 0, ~~ s \rightarrow \infty,~~L=\xi^2\log \frac{s}{4} \to {\rm Constant}$$ one finds the generic structure
\begin{shaded}
\be \label{integrand0}
\mm_{+}^{(0)}(s,t)=\int_{-1}^1 dy \sum_{k=0}^\infty A_k(L,y)~\frac{e^{L y}\sqrt{1-y^2}}{y^{k+1}}\x^{2k+2} +(s \to -s)
\ee 
\end{shaded}
where, we have introduced $L=\x^2\log(s/4)$. The first three coefficients $A_0, A_1$ and $A_2$ are given as following,
\begin{shaded}
\begin{align}\label{coef}
\begin{split}
A_0(L,y)&= \frac{  \Gamma^3 (h)  }{(2\p)^{1+4 h} \Gamma^4 \left(\frac{h}{2}\right)}\\
A_1(L,y)&=\frac{A_0(L,y)}{2}
\left[H_{h-1}-y \left(\psi ^{(0)}(h) (L-4 y)+2 y \psi ^{(0)}\left(\frac{h-t}{2}\right)+4 y \psi ^{(0)}\left(\frac{h}{2}\right)+\gamma  L\right)\right]\\
A_2(L,y)&=\frac{A_0(L,y)}{48} \left(6 y \left(\psi ^{(0)}(h)^2 \left(L^2 y-L \left(6 y^2+1\right)+16 y^3\right)+4 \gamma  L y^2 \psi ^{(0)}\left(\frac{h-t}{2}\right)+2 y \left(-\psi ^{(1)}(h) (y (L-8 y)\right.\right.\right.\\
&\left.\left.\left. +2)+\left(2 y^2-1\right) \psi ^{(1)}\left(\frac{h-t}{2}\right)+\left(2-4 y^2\right) \psi ^{(1)}\left(\frac{h}{2}\right)\right)+2 \psi ^{(0)}(h) \left(2 y^2 (L-4 y) \psi ^{(0)}\left(\frac{h-t}{2}\right)\right.\right.\right.\\
&\left.\left.\left. +\gamma  L (y (L-2 y)-1)\right)+L \psi ^{(1)}(h)+4 y^3 \psi ^{(0)}\left(\frac{h-t}{2}\right)^2\right)+48 y^3 \psi ^{(0)}\left(\frac{h}{2}\right) \left(\psi ^{(0)}(h) (L-4 y)\right.\right.\\
&\left.\left. +2 y \psi ^{(0)}\left(\frac{h-t}{2}\right)+\gamma  L\right)+96 y^4 \psi ^{(0)}\left(\frac{h}{2}\right)^2-12 \psi ^{(1)}(h)+\pi ^2 \left(L \left(2 y^2-1\right) y+4 y^2+2\right)\right.\\
&\left.+6 \gamma ^2 L y (y (L+2 y)-1)\right)
\end{split}
\end{align}
\end{shaded}
One can easily determine the higher order terms. Now we would like to point out that, the integrand \eqref{integrand0} has  singularity at $y=0$. Thus the integral \eqref{finalint} has to be considered in the sense of Cauchy principal value integral. The integrals to be done are 
\be 
\int_{-1}^{1} dy\,y^{-n}\sqrt{1-y^2}e^{Ly},~~~ n\in\mathbb{Z}_{\ge}.
\ee These integrals can be expressed in terms of Bessel function $J_\m(L)$ and modified Struve functions $\mathbf{L}_\n(L)$\footnote{ See appendix \ref{intweak}.}.

\subsection{Agreement with $d=4$ results for 0-magnon}\label{awd4r0m}
                      In this section we show explicitly that our $d$ dimensional results in the limit $d \to 4$ match with results obtained in \cite{Chowdhury:2019hns} and thereby \cite{Korchemsky:2018hnb}.
 Since we have used a rescaling here of the factor $c$ for convenience, it is better to reproduce $d=4$ answer for clarity. For $d=4$, spectral function takes the values 
  \be \left.\frac{E_{\Delta,J}^{(0)}}{1-\chi_0 E^{(0)}_{\Delta,J}}\right|_{d=4}=\frac{1}{64 \pi ^4 \left(\left(J^2+\nu ^2\right) \left((J+2)^2+\nu ^2\right)-4 \xi ^4\right)}
  \ee                      
The leading Regge poles are at \be J_{\pm}=-1+\sqrt{2 \sqrt{\xi ^4-\nu ^2}-\nu ^2+1} \ee The Regge poles are exact in coupling and hence the Mellin amplitude in the limit in the limit $$\xi \rightarrow 0, ~~ s \rightarrow \infty,~~\xi^2\log \frac{s}{4} \to {\rm Constant}$$ is evaluated along the same lines as \cite{Chowdhury:2019hns}. For brevity we just reproduce the answer here, 
\begin{eqnarray} \label{0mag4d}
\mm_{+,~d=4}^{(0)}(s,t)&=&2\int_{-1}^{1} dy\left(\frac{\xi ^2 \sqrt{1-y^2} e^{L y} }{512 \pi ^9 y} -\frac{\xi ^4 \sqrt{1-y^2} e^{L y}  \left(y (L-4 y)+2 y^2 \psi ^{(0)}\left(1-\frac{t}{2}\right)-1\right)}{1024 \pi ^9 y^2}\right.\nonumber\\
&&\left.+ \frac{\xi ^6 \sqrt{1-y^2} e^{L y} }{12288 \pi ^9 y^3} \left(3 L^2 y^2+6 y^2 \left(2 y \psi ^{(0)}\left(1-\frac{t}{2}\right) \left(L+y \psi ^{(0)}\left(1-\frac{t}{2}\right)-4 y\right)\right.\right.\right.\nonumber\\
&&\left.\left.\left. +\left(2 y^2-1\right) \psi ^{(1)}\left(1-\frac{t}{2}\right)\right)-6 L \left(2 y^3+y\right)+2 \left(2 y^2+1\right) \left(\pi ^2 y^2+3\right)\right)\right)
\end{eqnarray} 
where we have done the scaling $\nu =\xi^2\sqrt{1-y^2}$. 
In the limit $d \rightarrow 4$ \eqref{integrand0} exactly matches with \eqref{0mag4d}. Note that apriori this was not clear that this can happen since the integral in \eqref{0mag4d} is an exact manipulation without breaking up into ``inner" and \enquote{outer} regions. Let us take a moment to emphasise the non-triviality of this check in the sense that integral over the $d$-dimensional case involved jacobian of a non-trivial transformation due to perturbative evaluation of the Regge pole where as such things were completely absent for the the non-perturbative analysis done in \cite{Chowdhury:2019hns} and \cite{Korchemsky:2018hnb}.
This also serves as a non-trivial check of our assumption about analytic continuation made in the appendix \ref{iaoi0mgd}.

\section{One Magnon correlator in general $d$ fishnet theory}\label{1mag}
For the one magnon case we have $\D_1=\D_4=h,\,\D_2=\D_3=\frac{h}{2}.$ Putting these into \eqref{melfn}, one obtains
\begin{align}
\begin{split}
\mm^{(1)}_{\pm}(s,t)=\Bigg[\frac{\pm 1}{2\p i}&\zeta_1(\D_1,t)\oint dJ\frac{\p}{\sin \p J}\int_{-\infty}^{\infty}d\n\,\frac{s}{4}^{J}\, e^{i\p J/2}\, \n\sinh\p\n \\
&\times 
\frac{
	\G(h+J)
	\G\left(\frac{h+J-i\n}{2}\right)
	\G\left(\frac{h+J+i\n}{2}\right)
	\G\left(\frac{h-J-t-i\n}{2}\right)
	\G\left(\frac{h-J-t+i\n}{2}\right)
}
{
	2\pi ^{2(h+1)}
	\G(J+1)
	\Gamma \left(\frac{h+2 J-2 i \nu }{4}\right) 
	\Gamma \left(\frac{3 h+2 J-2 i \nu }{4}\right) 
	\Gamma \left(\frac{h+2 J+2 i \nu }{4}\right) 
	\Gamma \left(\frac{3 h+2 J+2 i \nu }{4}\right)
}
\frac{
	\left(E_{\D,J}^{(1)}\right)^2
}
{
	1-\c_1 E_{\D,J}^{(1)}
}\Bigg]\\
&\hspace{12 cm}\,\pm (s\to-s)
\end{split}\label{melmeasure1}
\end{align} with $\zeta_1(\D_i,t)=\G\left(\frac{3h-2t}{2}\right)^{-2}$ and $\c_1=(4\p)^{h}\x^2$. $\mm^{(1)}_{+}(s,t)$ is the one relevant to even spin while $\mm^{(1)}_{-}(s,t)$ for odd spin. In general $d$, the eigenvalue of the one-magnon graph building operator is given by (C.14) of \cite{Gromov:2018hut}
\be \label{E1}
E_{h+i\n,J}^{(1)}=c^2\p^{h}(-1)^J\,\frac{\G\left(\frac{h+2J-2i\n}{4}\right)\G\left(\frac{h+2J+2i\n}{4}\right)}{\G\left(\frac{3h+2J-2i\n}{4}\right)\G\left(\frac{3h+2J+2i\n}{4}\right)}\,.
\ee Regge trajectories are given by the poles of the spectral function
\be \label{spec1}
\frac{\left(E_{h+i\n,J}^{(1)}\right)^2}{1-(4\p)^h\x^2\,E_{h+i\n,J}^{(1)}}\,.
\ee As in the zero magnon case, the spectral function is in terms of Gamma functions thus making the exact evaluation of the Regge trajectories difficult. Therefore the analysis of section \ref{genRegge} will be followed as usual.

\subsection{Evaluation of the Regge trajectories}\label{1Regge}
First, the free Regge trajectory is given by
\be 
J_f^{(1)}(p;\pm i\n)=-2p-\frac{h}{2}\pm i\n, ~~~p\in\mathbb{Z}_{\ge},~~\text{s.t.}
\begin{cases} 
	0\leq p\leq \frac{h}{2},~~ h\in\mathbb{Z}_{\ge0}\,;\\ 
	p\geq0,~~~~~~~~h\in\mathbb{Z}_{\ge 0}+\frac{1}{2}.
\end{cases}
\ee   $E_{h+i\n}^{(1)}$ has simple poles in these locations. The leading trajectory in this family is given by $p=0$. The free Regge trajectories are same for even and odd spins. Now, we turn to evaluation of  the coupling dependent Regge trajectory. As in the zero magnon case we will consider only the leading trajectory in the following analysis.
\subsubsection*{{\bf I}. {\bf Even Spin}}

	For even spin we have\footnote{The superscript \enquote{$1e$} is to label that we are considering even spin case. Similarly, we will use \enquote{$1o$} to denote the odd spin case.}
	\be \label{1even}
	E^{(1e)}_{h+i\n,J}=c^2 \pi ^h\frac{ 
		\Gamma \left(\frac{h+2 J-2 i \nu }{4}\right) 
		\Gamma \left(\frac{h+2 J+2 i \nu }{4}\right)}
	{\Gamma \left(\frac{3 h+2 J-2 i \nu }{4}\right)
		\Gamma \left(\frac{3 h+2 J+2 i \nu }{4}\right)}.
	\ee  Using the fact that  $J_f^{(1)}$ is simple pole of $E_{h+i\n,J}^{(1)}$ we reach the following expressions for the inner and outer solutions.\\
	
	\textbf{Outer Solution:} The outer solution is given by the perturbative expression
	\begin{shaded}
	\be\label{J1eout}
	J_{\text{o}\pm}^{(1e)}( \n)=-\frac{h}{2}\pm i \nu + \b_{1\pm} \x^2 +\sum _{k=2}^\infty \b_k( \pm\n) \xi ^{2 k}
	\ee  
	\end{shaded}
	where the first few members of the sequence  $\{\b_k(\pm\n):k\ge2\}$ can be expressed as\footnote{We have used the rescaling \be \frac{2 (2\p)^h}{\G\left(\frac{h}{2}\right)}c\to 1.\ee We will stick to this scaling for the rest of the analysis of one magnon case.}
	\begin{shaded}
	\begin{align}\label{coeffe1}
	\begin{split}
	&\b_2(\pm\n)=-\frac{\b_{1\pm}^2}{2}G_{1}(\pm i\n),~~~
	\b_3(\pm\n)=\frac{\b_{1\pm}^3}{8}\left[G_{2}(\pm i\n)+3G_{1}(\pm i\n)^3\right],\\
	&\b_4(\pm\n)=-\frac{\b_{1\pm}^4}{48}\left[ G_{3}(\pm i\n)+12 G_{2}(\pm i\n)G_{1}(\pm i\n)+16 G_{1}(\pm i\n)^3\right].
	\end{split}
	\end{align}
	\end{shaded}
	where we have defined the function
	\be \label{Gdef}
	G_{r}(z):=(r-1)! H_{\frac{h}{2}-1,r}+(-1)^{r-1} \left[\psi ^{(r-1)}\left(\frac{h}{2}+z \right)-\psi ^{(r-1)}(z )\right]\,,
	\ee  
	Using this, the pole structure of the spectral function manifests as
	\begin{align}\label{pole1e}
	\begin{split}
	\frac{\left(E_{h+i\n,J_{\text{o}\pm}^{(1e)}(\n)}^{(1e)}\right)^2}{1-\c_1 E^{(1e)}_{h+i\n,J_{\text{o}\pm}^{(1e)}(\n)}}
	=\frac{2^{-4 h-1} \pi ^{-2 h} \Gamma \left(\frac{h}{2}\right)^2 \Gamma (\pm i \nu )^2}{\xi ^4 \b_{1\pm}
		\left( 2 \b_{1\pm} \Gamma \left(\frac{h}{2}\pm i \nu \right)^2-\Gamma \left(\frac{h}{2}\right) \Gamma (\pm i \nu ) \Gamma \left(\frac{h}{2}\pm i \nu \right)\right)}.
	\end{split}
	\end{align} Note that, the pole at $\b_{1\pm}=0$ corresponds to the free theory poles (it is immediately seen that $\b_{1\pm}=0$ leads to vanishing of \eqref{coeffe1}). The other non-trivial pole is given at
	\be 
	\b_{1\pm}=\frac{1}{2 }B\left(\frac{h}{2},\pm i\n\right),
	\ee   
where $B(a,b)=\G(a)\G(b)/\G(a+b)$.	In summary we have two sets of Regge poles depending on $\b_{1\pm}$. 
	
	\vspace*{1cm}
	
	 \par

	\textbf{Inner Solution:} Next, we turn toward the inner solution. Following the argument of section \ref{genRegge}, it is straightforward to obtain that, the inner region corresponds to $|x|<1$ with $\n=x\x$. Then, the perturbative expression for the inner solution is given by
	\begin{shaded}
	\be \label{1mageveninner}
	J_{\text{i}}^{(1e)}(  x)=-\frac{h}{2}+\sum_{k=1}^{\infty}\tilde{\b}_k( x)\x^{k}
	\ee 
	\end{shaded}
	with first few coefficients $\tilde{\b}_k,\,k\ge2$ given by
	
	\begin{shaded}
	\begin{align}\label{1mei}
	\begin{split}
	\tilde{\b}_2(x)&=-\frac{1}{2}(\tilde{\b} _{1}^2+x^2)H_{\frac{h}{2}-1},\,\,\,\tilde{\b}_3(x)=\frac{(\tilde{\b}_{1}^2+x^2)}{8 \tilde{\b}_{1}} \left[\left(H_{\frac{h}{2}-1}\right)^2 (3 \tilde{\b}_{1}^2+x^2)+H_{\frac{h}{2}-1,2} (\tilde{\b}_{1}^2-x^2)\right]\\
	\tilde{\b}_4(x)&=-\frac{1}{24} (\tilde{\b}_{1}^2+x^2) 
	\left[
	\left(H_{\frac{h}{2}-1}\right)^3 (8 \tilde{\b}_{1}^2+6 x^2)
	+6  H_{\frac{h}{2}-1} \,H_{\frac{h}{2}-1,2}\,\tilde{\b}_{1}^2
	+H_{\frac{h}{2}-1,3} (\tilde{\b}_{1}^2-3 x^2)\right].
	\end{split} 
	\end{align}
	\end{shaded}
	$x$-dependence appears in the form of functional dependence upon $x^2$  which is in perfect harmony with  the symmetry of the spectral function under $i\n\to-i\n$.
	The pole structure is manifested as ,
	\be 
	\frac{\left(E_{h+ix \x,\,J_{\text{i}\pm}^{(1e)}(  x)}^{(1e)}\right)^2}{1-\c_0 E^{(1e)}_{h+ix\x,\,J_{\text{i}\pm}^{(1e)}(x)}}
	=\frac{(4\pi) ^{-2 h}}{\xi ^4 \left(\tilde{\beta}_{1}-\sqrt{1-x^2}\right)\left(\tilde{\beta}_{1}+\sqrt{1-x^2}\right) \left(\tilde{\beta}_{1}-ix\right)\left(\tilde{\beta}_{1}+ix\right)}.
	\ee  
	The $\tilde{\b}_1=\pm ix$ poles correspond to the free Regge trajectory, which can be easily seen by observing that in \eqref{1mei}, all $\tilde{\b}_k(x) \to 0$ for this choice. The nontrivial pole is now given at 
	\be \tilde{\b}_1=\tilde{\beta}_{1\pm}:=\pm\sqrt{1-x^2}\,.\ee
	  We will denote the inner solution with $\tilde{\b}_1=\tilde{\b}_{1\pm}$ by $J^{1e}_{\text{i}\pm}$.
\subsubsection*{{\bf II}. \textbf{Odd Spin}}
	For odd spin we have
	\be \label{1odd}
	E^{(1o)}_{h+i\n,J}=-c^2 \pi ^h\frac{ 
		\Gamma \left(\frac{h+2 J-2 i \nu }{4}\right) 
		\Gamma \left(\frac{h+2 J+2 i \nu }{4}\right)}
	{\Gamma \left(\frac{3 h+2 J-2 i \nu }{4}\right)
		\Gamma \left(\frac{3 h+2 J+2 i \nu }{4}\right)}.
	\ee
	Now, we can follow the same procedure as in the even spin case to obtain the inner and outer solutions.\par 
	
	\textbf{Outer Solution:} The outer solution is given by
	\begin{shaded}
	\be \label{J1oout}
	J_{\text{o}\pm}^{(1o)}(\n)=-\frac{h}{2}+\l_{1\pm}\x^2+\sum_{k=2}^{\infty}\l_k(\pm\n)\x^{2k}
	\ee 
	\end{shaded}
	where the first few members of $\{\l_k(\pm\n):\,k\ge 2\}$ is given by,
	\begin{shaded}
	\begin{align}\label{coeffo1}
	\begin{split}
	&\l_2(\pm\n)=-\frac{\l_{1\pm}^2}{2}G_{1}(\pm i\n),~~~
	\l_3(\pm\n)=\frac{\l_{1\pm}^3}{8}\left[G_{2}(\pm i\n)+3G_{1}(\pm i\n)^3\right],\\
	&\l_4(\pm\n)=-\frac{\l_{1\pm}^4}{48}\left[ G_{3}(\pm i\n)+12 G_{2}(\pm i\n)G_{1}(\pm i\n)+16 G_{1}(\pm i\n)^3\right],
	\end{split}
	\end{align} 
	\end{shaded}
$G_i(i\n)$ being defined in \eqref{Gdef}. The pole structure of the spectral function manifests itself  as 
	\begin{align}
	\begin{split}
	\frac{\left(E_{h+i\n,J_{\text{o}\pm}^{(1o)}( \n)}^{(1o)}\right)^2}{1-\c_1 E^{(1o)}_{h+i\n,J_{\text{o}\pm}^{(1o)}(\n)}}
	=\frac{2^{-4 h-1} \pi ^{-2 h} \Gamma \left(\frac{h}{2}\right)^2 \Gamma (\pm i \nu )^2}{\xi ^4\l_{1\pm} \left(2 \l_{1\pm} \Gamma \left(\frac{h}{2}\pm i \nu \right)^2+ \Gamma \left(\frac{h}{2}\right) \Gamma (\pm i \nu ) \Gamma \left(\frac{h}{2}\pm i \nu \right)\right)}
	\end{split}
	\end{align}
	As in the even spin case, the pole at $\l_{1\pm}=0$ corresponds to the free Regge trajectory $J_f^{(1)}$. The non-trivial pole is at
	\be 
	\l_{1\pm}=-\frac{1}{2}B\left(\frac{h}{2},\pm i\n\right).
	\ee 
	where  $B(a,b)=\G(a)\G(b)/\G(a+b)$.

	\textbf{Inner Solution:} Similar to the even spin case, one obtains that, the inner solution region corresponds to $|x|<1$ with $\n=x\x$. Thereby, one has the perturbative expression for the inner solution
\begin{shaded}
	\be\label{J1oin} 
	J_{\text{i}}^{(1o)}(x)=-\frac{h}{2}+\sum_{k=1}^{\infty} \tilde{\l}_k(x)\x^{k}
	\ee 
\end{shaded}	
	with first few coefficients $\tilde{\l}_k,\,k\ge2$ given by
	\begin{shaded}
	\begin{align}\label{coeffi1}
	\begin{split}
	\tilde{\l}_2(x)&=-\frac{1}{2}(\tilde{\l}_{1}^2+x^2)H_{\frac{h}{2}-1},\,\,\,\tilde{\l}_3(x)=\frac{(\tilde{\l}_{1}^2+x^2)}{8 \tilde{\l}_{1}} \left[\left(H_{\frac{h}{2}-1}\right)^2 (3 \tilde{\l}_{1}^2+x^2)+H_{\frac{h}{2}-1,2} (\tilde{\l}_{1}^2-x^2)\right]\\
	\tilde{\l}_4(x)&=-\frac{1}{24} (\tilde{\l}_{1}^2+x^2) 
	\left[
	\left(H_{\frac{h}{2}-1}\right)^3 (8 \tilde{\l}_{1}^2+6 x^2)
	+6  H_{\frac{h}{2}-1} \,H_{\frac{h}{2}-1,2}\,\tilde{\l}_{1}^2
	+H_{\frac{h}{2}-1,3} (\tilde{\l}_{1}^2-3 x^2)\right].
	\end{split} 
	\end{align}
	\end{shaded}
	Again, as in the even spin case, $x$-dependence appears in the form of functional dependence upon $x^2$  which is in perfect harmony with  the symmetry of the spectral function under $i\n\to-i\n$.
	The pole structure is now  manifested as,
	\be 
	\frac{\left(E_{h+ix \x,\,J_{\text{i}}^{(1o)}(  x)}^{(1o)}\right)^2}{1-\c_0 E^{(1o)}_{h+ix\x,\,J_{\text{i}}^{(1o)}(x)}}=\frac{ (4\pi) ^{-2 h}}{\xi ^4 \left(\tilde{\l}_{1}-i\sqrt{1+x^2}\right)\left(\tilde{\l}_{1}+i\sqrt{1+x^2}\right) \left(\tilde{\l}_{1}-ix\right)\left(\tilde{\l}_{1}+ix\right)}
	\ee 
	The poles  at $\tilde{\l}_{1}(x)=\pm ix$ correspond to the free Regge trajectory. The non-trivial poles are at 
	\be 
	\tilde{\l}_1=\tilde{\l}_{1\pm}(x):=\pm i\sqrt{1+x^2}\,.
	\ee 
	We will denote the inner solution with $\tilde{\l}_1=\tilde{\l}_{1\pm}$ by $J^{1o}_{\text{i}\pm}$.
Before proceeding further we would like to make a few comments.
\begin{enumerate}
	\item While evaluating the Mellin amplitude, we will \emph{always} consider the non-trivial poles. Specifically, We will \emph{not consider} the contribution of the free Regge trajectories $J_f^{(1)}(0;\pm i\n)$.
	\item There is an important connection between the even spin Regge trajectory and the odd spin Regge trajectory. From expressions of the Regge trajectories above, it is straightforward to observe that,
	\begin{align}\label{eorel}
	\begin{split} 
	J_{\text{o}\pm}^{(1e)}(\n) \xrightarrow[\n\, \text{fixed}]{\x\to i\x}J_{\text{o}\pm}^{(1o)}(\n),\qquad
	J_{\text{i}\pm}^{(1e)}(x) \xrightarrow[x\to -ix]{\x\to i\x}J_{\text{i}\pm}^{(1o)}(x).
	\end{split}
	\end{align}
	where, $J_{\text{i}\pm}^{(1e)}(x):=J_{\text{i}}^{(1e)}\left(\tilde{\b}_{1\pm}=\pm\sqrt{1-x^2}\right),\,\text{and}\,J_{\text{i}\pm}^{(1o)}(x):=J_{\text{i}}^{(1o)}\left(\tilde{\l}_{1\pm}=\pm i\sqrt{1+x^2}\right).$ 
	 This connection can be explained as following. The eigenvalue of the graph building operator  can be written as 
	\be 
	E_{h+i\n,J}^{(1e)}=+ \mathcal{E}(i\n,J),~~~E_{h+i\n,J}^{(1o)}=- \mathcal{E}(i\n,J)
	\ee  with
	\be 
	\mathcal{E}(i\n,J):=c^2 \pi ^h\frac{ 
		\Gamma \left(\frac{h+2 J-2 i \nu }{4}\right) 
		\Gamma \left(\frac{h+2 J+2 i \nu }{4}\right)}
	{\Gamma \left(\frac{3 h+2 J-2 i \nu }{4}\right)
		\Gamma \left(\frac{3 h+2 J+2 i \nu }{4}\right)}.
	\ee The corresponding spectral functions thus turn out to be
	\begin{align} 
	\begin{split}
	\frac{\left(E_{h+i\n,J}^{(1e)}\right)^2}{1-\c_1 E_{h+i\n,J}^{(1e)}}=\frac{\mathcal{E}(i\n,J)^2}{1-\x^2(4\p)^h\mathcal{E}(i\n,J)},\qquad
	\frac{\left(E_{h+i\n,J}^{(1o)}\right)^2}{1-\c_1 E_{h+i\n,J}^{(1o)}}=\frac{\mathcal{E}(i\n,J)^2}{1+\x^2(4\p)^h\mathcal{E}(i\n,J)}.
	\end{split}
	\end{align}
	Clearly,
	\be 
	\frac{\left(E_{h+i\n,J}^{(1e)}\right)^2}{1-\c_1 E_{h+i\n,J}^{(1e)}}\xrightarrow[\n\, \text{fixed}]{\x\to i\x} \frac{\left(E_{h+i\n,J}^{(1o)}\right)^2}{1-\c_1 E_{h+i\n,J}^{(1o)}}.
	\ee This  can be seen as an \enquote{analytic continuation} in complex $\x$ plane. From this, the relations in \eqref{eorel} follow at once. Since the outer solution is expressed in terms of $\n$ and $\x$, only analytic continuation in $\x$ suffices. On the other hand, in the inner solution the relevant variables are $\x$ and $x$. Since $x=\n/\x$, $x$ needs to be analytically continued in the \emph{opposite sense} of that of $\x$ to keep $\n$ \emph{fixed}. 
	The \emph{choice of phase} is very important for defining the analytic continuation between the inner solutions. To be specific, under this choice of phase $J^{(1e)}_{\text{i}+}$ analytically continues to $J^{(1o)}_{\text{i}+}$ and $J^{(1e)}_{\text{i}-}$ to $J^{(1o)}_{\text{i}-}$. If we want to define the opposite analytic continuation i.e., from $J^{(1o)}_{\text{i}\pm}$ to $J^{(1e)}_{\text{i}\pm}$ then, we need to consider the opposite phase i.e., under the choice $\x\to-i\x,\,x\to ix$ one obtains $J^{(1o)}_{\text{i}\pm}\to J^{(1e)}_{\text{i}\pm}$. 
\end{enumerate}
\subsection{Evaluation of Mellin Amplitude}
For the evaluation of the Mellin amplitude, we will follow the procedure laid out in details for zero magnon case in section \ref{Mell0}. Main arguments of the analysis does not alter; Only changes are in explicit functional forms. This is almost immediately obvious since for the even case, the inner and outer solutions are structurally similar to that of the 0-magnon and satisfy the same crucial transformation properties (see \eqref{transinint} and \eqref{transoutint} ). Thus, we just give the final expressions here. For future reference let us express the expression for the Mellin amplitude as

\begin{align}
\begin{split}
\mm^{(1)}_{\pm}(s,t)=\Bigg[\frac{\pm 1}{2\p i}&\zeta_1(\D_1,t)\oint dJ\frac{\p}{\sin \p J}\int_{-\infty}^{\infty}d\n\,\left(\frac{s}{4}\right)^{J}\, M^{(1)}(J, \nu)\frac{
	\left(E_{\D,J}^{(1)}\right)^2
}
{
	1-\c_1 E_{\D,J}^{(1)}
}\Bigg]\pm (s\to-s)
\end{split}
\end{align}

where
\begin{align}\label{1mel}
\begin{split}
M^{(1)}(J, \n)= \n\sinh\p\n  
\frac{
	\G(h+J)
	\G\left(\frac{h+J-i\n}{2}\right)
	\G\left(\frac{h+J+i\n}{2}\right)
	\G\left(\frac{h-J-t-i\n}{2}\right)
	\G\left(\frac{h-J-t+i\n}{2}\right)
}
{
	2\pi ^{2 (h+1)}
	\G(J+1)
	\Gamma \left(\frac{h+2 J-2 i \nu }{4}\right) 
	\Gamma \left(\frac{3 h+2 J-2 i \nu }{4}\right) 
	\Gamma \left(\frac{h+2 J+2 i \nu }{4}\right) 
	\Gamma \left(\frac{3 h+2 J+2 i \nu }{4}\right)
}
\end{split}
\end{align}
\subsubsection*{{\bf I}. \textbf{Even Spin}}
	
	For even spin, the Mellin amplitude is given by 
	\be \label{Melev}
	\mm_{+}^{(1)}(s,t)=2\zeta_1(\D_i,t)\x\int_{-1}^{1}dy\,\frac{y}{\sqrt{1-y^2}}\,\F_{+}^{(1)}(\sqrt{1-y^2})\,+\,(s\to-s)
	\ee where, $\F_{+}^{(1)}(\sqrt{1-y^2})$ is obtained analogous to $\F_{+}^{(0)}(\sqrt{1-y^2})$ and is given by 
\begin{align} 
	\begin{split}
	&\F^{(1)}_{+}(\sqrt{1-y^2}):=\Theta^{(1)}_{\text{i}}(\tilde{\b}_{1+})\,M^{(1)}\left(J_{\text{i}+}^\text{(1)}(\sqrt{1-y^2}), \sqrt{1-y^2}\right)\,\left(\frac{s}{4}\right)^{J_{\text{i}+}^\text{(1)}(\sqrt{1-y^2})}\,\left[\frac{(4\p)^{-2h}}{2\x^4y}\right],\\
	&\Theta^{(1)}_{\text{i}}(\tilde{\b}_{1+}):=\frac{\pd J_{\text{i}+}^{(1)}( \sqrt{1-y^2})}{\pd\tilde{\b}_{1+}}=\xi\Bigg[1-\tilde{\b}_{1+} \xi H_{\frac{h}{2}-1}+\frac{\xi^2}{8 \tilde{\b}_{1+}^2} \left\{\left(H_{\frac{h}{2}-1}\right){}^2 \left[9 \tilde{\b}_{1+}^4+4 \tilde{\b}_{1+}^2 \left(1-y^2\right)\right.\right.\\
	&\left.\left.\hspace{8.5 cm}-\left(1-y^2\right)^2\right]+H_{\frac{h}{2}-1}^{(2)} \left[3 \tilde{\b}_{1+}^4+\left(1-y^2\right)^2\right]\right\}+O(\x^3)\Bigg],
	\end{split}
	\end{align}
	where $\tilde{\b}_{1+}= y$, $M^{(1)}(J,\n)$ and $J_{\text{i}+}^\text{(1)}(x)$ are given in \eqref{1mel} and \eqref{1mageveninner} respectively. The integrand is then evaluated in the limit 
	$$\xi \rightarrow 0, ~~ s \rightarrow \infty,~~Q=\xi\log \frac{s}{4} \to {\rm Constant}$$ 
	to give, 
	\begin{shaded}
	\be \label{1magevenresult}
	\mm_{+}^{(1)}(s, t)=\int_{-1}^1 dy ~s^{-\frac{h}{2}}\sqrt{1-y^2}e^{Q y}\sum_{k=0}^{\infty} B_k(Q,y)\x^{2+k} + (s \to -s)
	\ee
	\end{shaded}
	 where,
	 \begin{shaded}
	\begin{align}\label{1mageve2}
	\begin{split}
	B_0(Q,y)&=-\frac{  \csc \left(\frac{\pi  h}{2}\right) \Gamma \left(\frac{h}{2}\right)  \Gamma \left(\frac{3 h}{4}-\frac{t}{2}\right)^2}{2^{(4h+3)} \pi ^{4 h}\Gamma \left(1-\frac{h}{2}\right)},\\
	B_1(Q,y)&=B_0(Q,y)\left[-y \left(H_{-\frac{h}{2}}+\psi ^{(0)}\left(\frac{3 h}{4}-\frac{t}{2}\right)\right)-\frac{1}{2} Q \left(\psi ^{(0)}\left(\frac{h}{2}\right)+\gamma \right)+\pi  y \cot \left(\frac{\pi  h}{2}\right)\right]
	\end{split}
	\end{align}
	\end{shaded}
 and so on. One can  systematically obtain higher order terms in $\x$ expansion above. Thus, one is left with doing integrals of the form
	\be 
	\int_{-1}^1 dx\,x^n\sqrt{1-x^2}e^{Qx},~~~n\in\mathbb{Z}_{\ge}\,.
	\ee These integrals can be expressed in terms of modified Bessel functions of first kind, $I_{\m}(Q)$\footnote{ See appendix \ref{intweak}.}.
\subsubsection*{{\bf II}. \textbf{Odd spin}}\label{oddspin1maggend}

    To deal the odd spin case, we can exploit  the analytic continuation between the even spin Regge trajectory and the odd spin Regge trajectory that we delineated towards the end of section \ref{1Regge}. So, what one needs to do is to consider the monodromies $\x\to i\x, x\to-ix$ to the even spin Mellin amplitude integrals at the very first following up with the usual analysis. The final effect of this is to just take the analytic continuation of the final $\x$ dependent answer under $\x\to i\x$. To be little bit more explicit, if the even spin Mellin amplitude from \eqref{Melev} is given, generically, by
    \be 
    \mm_{+}^{(1)}=\FM(\x,s,t)\,+\,(s\to -s)    \ee then the odd spin amplitude is given by
    \be \label{1magoddresult}
    \mm_{-}^{(1)}=\FM(i\x,s,t)\,-\,(s\to-s).
    \ee

    We conjecture this and verify this with an explicit calculation for $d=4$ and $d=8$ in the next subsection where the Regge poles are exactly known.\footnote{One might think that \eqref{1magoddresult} should come with a -ve sign because the odd Mellin measure also has a $(-1)^J$. But note that given the integrated result, we can only do $\xi \to i \xi$.} 
	
\subsection{Agreement with $d=4$ and $d=8$ results for 1-magnon}\label{awd4d8r1m}

 In this subsection we explicitly compute $d=4$ and $d=8$ exact Regge trajectories and compute perturbative Regge Mellin amplitudes (both odd and even) of one magnon correlators following \cite{Chowdhury:2019hns} and match it explicitly with the higher dimensional result.
 
 \subsubsection*{{\bf I}. Even spin}
  First we consider the even spin. The spectral function relevant for our discussion is given by \eqref{1even}, 
 \begin{eqnarray}
 	\frac{(E^{(1e)}_{d=4})^2}{1-16\pi^2\xi^2 E^{(1e)}_{d=4}}&=& \frac{1}{256 \pi ^4 \left((J+1)^2+\nu ^2\right) \left(J (J+2)+\nu ^2-\xi ^2+1\right)}\nonumber\\
 	\frac{(E^{(1e)}_{d=8})^2}{1-256\pi^4\xi^2 E^{(1e)}_{d=8}}&=& \frac{1}{4096 \pi ^8 (J-i \nu +2) (J-i \nu +4) (J+i \nu +2) (J+i \nu +4) }\nonumber\\
 	&&\times \frac{1}{\left(2 J (J+6) \nu ^2+J (J+6) (J (J+6)+16)+\nu ^4+20 \nu ^2-4 \xi ^2+64\right)} 
 \end{eqnarray}
 The exact Regge poles can be listed as follows. 
  \begin{eqnarray}
  J^{d=4,e}_\pm= -1\pm\sqrt{\xi ^2-\nu ^2},\qquad  J^{d=8,e}_\pm=-3+\sqrt{1-\nu ^2\pm 2 \sqrt{\xi ^2-\nu ^2}}
  \end{eqnarray}
We note that, the Regge poles for the $d=8$ case is essentially that of the type we encounter in $d=4$ 0-magnon. The Mellin amplitude can be calculated according to the techniques outlined in \cite{Chowdhury:2019hns}. In the limit 
$$\xi \rightarrow 0, ~~ s \rightarrow \infty,~~Q=\xi\log \frac{s}{4} \to {\rm Constant}$$ 
We get the Mellin amplitudes as follows 
\begin{eqnarray}
\mm_{+,~d=4}^{(1)}(s,t)&=&\int_{-1}^{1} dy~\sqrt{1-y^2} e^{Q y} \left(-\frac{\xi ^2 }{2048 \pi ^9} +\frac{\xi ^3}{2048 \pi ^9} y  \psi ^{(0)}\left(\frac{3}{2}-\frac{t}{2}\right) -\frac{\xi ^4 }{24576 \pi ^9}\left(6 y^2 \psi ^{(0)}\left(\frac{3}{2}-\frac{t}{2}\right)^2+\left(6 y^2\right. \right.\right.\nonumber\\
&&\left.\left.\left. -3\right)  \psi ^{(1)}\left(\frac{3}{2}-\frac{t}{2}\right)+\pi ^2 \left(2 y^2+1\right)\right) \right) + (s \to -s)\nonumber\\
\mm_{+,~d=8}^{(1)}(s,t)&=& \int_{-1}^{1} dy~\sqrt{1-y^2} e^{Q y}\left(-\frac{\xi ^2 }{524288 \pi ^{17} s^2}+\frac{\xi ^3 \left(Q+2 y \psi ^{(0)}\left(3-\frac{t}{2}\right)+2 y\right)}{1048576 \pi ^{17} s^2}-\frac{\xi ^4 }{12582912 \pi ^{17} s^2}\left(3 Q^2 \right.\right.\nonumber\\
&&\left.\left. +12 (y (Q+2 y)+1) \psi ^{(0)}\left(3-\frac{t}{2}\right)+24 Q y+12 y^2 \psi ^{(0)}\left(3-\frac{t}{2}\right)^2+6 \left(2 y^2-1\right) \psi ^{(1)}\left(3-\frac{t}{2}\right)\right.\right.\nonumber\\
&&\left.\left. +4 \left(\pi ^2-6\right) y^2+2 \pi ^2+24\right) \right)  + (s \to -s)
\end{eqnarray}
This is in perfect agreement with  \eqref{1magevenresult} and \eqref{1magevenresult2} in the limit $d=4$ and $d=8$.

\subsubsection*{{\bf II}. Odd spin}
 For odd spin the spectral functions become,   
\begin{eqnarray}
\frac{(E^{(1e)}_{d=4})^2}{1-16\pi^2\xi^2 E^{(1o)}_{d=4}}&=& \frac{1}{256 \pi ^4 \left((J+1)^2+\nu ^2\right) \left(J (J+2)+\nu ^2+\xi ^2+1\right)}\nonumber\\
\frac{(E^{(1e)}_{d=8})^2}{1-256\pi^4\xi^2 E^{(1o)}_{d=8}}&=& \frac{1}{4096 \pi ^8 (J-i \nu +2) (J-i \nu +4) (J+i \nu +2) (J+i \nu +4) }\nonumber\\
&&\times \frac{1}{\left(2 J (J+6) \nu ^2+J (J+6) (J (J+6)+16)+\nu ^4+20 \nu ^2+4 \xi ^2+64\right)} 
\end{eqnarray}
The Regge poles are given by, 
\begin{eqnarray}
J^{d=4,o}_\pm= -1\pm i\sqrt{\xi ^2+\nu ^2},\qquad  J^{d=8,o}_\pm=-3+\sqrt{1-\nu ^2\pm 2i \sqrt{\xi ^2+\nu ^2}}
\end{eqnarray}
Following \cite{Chowdhury:2019hns}, the Mellin amplitude in the Regge limit can be evaluated. 
The integrals can also be done following \cite{Chowdhury:2019hns}. The integrated amplitude is given by \footnote{Note that one has to be careful about the -ve sign coming from $(-1)^{J_{{\rm odd}}}$ in the Mellin measure \eqref{melfn} (which translates to the overall -ve sign of the Mellin amplitude in \eqref{melmeasure1}.}
\begin{eqnarray}
\mm_{-,~d=4}^{(1)}(s,t)&=&\frac{\xi ^2}{8192 \pi ^8 Q^2 s}  \left(2 \xi  J_2(Q) \left(2 Q \psi ^{(0)}\left(\frac{3}{2}-\frac{t}{2}\right)+3 \xi  \psi ^{(0)}\left(\frac{3}{2}-\frac{t}{2}\right)^2+\xi  \left(3 \psi ^{(1)}\left(\frac{3}{2}-\frac{t}{2}\right)+\pi ^2\right)\right) \right. \nonumber\\
&&\left.-Q J_1(Q)\left(\pi ^2 \xi ^2+\xi ^2 \left(2 \psi ^{(0)}\left(\frac{3}{2}-\frac{t}{2}\right)^2+\psi ^{(1)}\left(\frac{3}{2}-\frac{t}{2}\right)\right)-4\right)\right) - (s \to -s)\nonumber\\
\mm_{-,~d=8}^{(1)}(s,t)&=&\frac{\xi ^2}{4194304 \pi ^{16} Q^2 s^2} \left(4 \xi  J_2(Q) \left(\xi  \left(2 Q^2+\pi ^2-6\right)+\psi ^{(0)}\left(3-\frac{t}{2}\right) \left(\xi  \left(Q^2+6\right)+2 Q \right.\right.\right.\nonumber\\
&&\left.\left.\left. +3 \xi  \psi ^{(0)}\left(3-\frac{t}{2}\right)\right)+2 Q  +3 \xi  \psi ^{(1)}\left(3-\frac{t}{2}\right)\right)+Q J_1(Q) \left(\xi  \left(Q (\xi  Q+4)-2 \pi ^2 \xi \right) \right.\right.\nonumber\\
&&\left.\left.-2 \xi ^2 \left(2 \psi ^{(0)}\left(3-\frac{t}{2}\right) \left(\psi ^{(0)}\left(3-\frac{t}{2}\right) +3\right)+\psi ^{(1)}\left(3-\frac{t}{2}\right)\right)+8\right)\right)- (s \to -s)\nonumber\\
\end{eqnarray}

%\begin{comment}
%Here I had not taken into account the -ve sign for $(-1)^J$ from eq 2.1: Subham 
%\end{comment}

The integrals of \eqref{1magevenresult} and \eqref{1magevenresult2}, can be done using the integrals in Appendix \ref{intweak} and, subsequently, using the identity
\be 
I_\nu(i z) = i^\nu J_\nu (z),\,\,\n\in\mathbb{Z}_{\ge}
\ee  
we obtain perfect agreement with $d=4$ and $d=8$ results using the proposal for odd spin Mellin amplitude \eqref{1magoddresult}. 

%%Chiral starts here.%%%
\section{Regge theory of Chiral fishnet theories in $d=4$}\label{chm1}
                        In the second part of our work, we will study Regge amplitudes of one of the chiral fishnet theories first studied in 
\cite{Kazakov:2018gcy}. More precisely we study the correlator $\langle{\rm Tr}[\phi_1(x_1)\phi_1(x_2)]{\rm Tr}[\phi_1^\dagger(x_3)\phi_1^\dagger(x_4)]\rangle$. We first show that the perturbative analysis of the previous section holds and then reproduce the answer via an independent perturbative calculation. These serves as a concrete check for our formalism. We begin by quoting the Mellin amplitude  relevant for our purpose \eqref{melfn} with $\Delta_i=1$ and $h=2$ but instead of the general $d$ spectral function for 0 and 1 magnon, we have the relevant spectral function for the chiral fishnet correlator.
\begin{align}\label{Mellinc}
\begin{split}
\mm^{(c)}_{\pm}(s,t)=\left[\frac{\pm1}{2\pi i}\right. &\left.\oint dJ \int_{-\infty}^\infty d\n \left(\frac{s}{4}\right)^J e^{i\pi J/2} M^{(c)}(J,\nu) \,\zeta_c({\D_i,t}) b^{c}_J(\n)\right] \pm(s\rightarrow -s),
\end{split}
\end{align} 

where the Mellin measure and the spectral function (\cite{Kazakov:2018gcy}) are given by, 
\be\label{melc}
M^{(c)}(J,\nu)=\frac{\pi}{\sin \pi J}\nu \sinh \pi\nu \,\zeta_c({\D_i,t}) \frac
{  
	\Gamma (h+J) 
	\Gamma (h+J-i \nu ) 
	\Gamma (h+J+i \nu ) 
	\Gamma \left(\frac{h-J-t-i \nu }{2}\right) 
	\Gamma \left(\frac{h-J-t+i \nu }{2}\right)
}
{
	2\pi ^{2 (h+1)} 
	\Gamma (J+1) 
	\Gamma \left(\frac{h+J-i \nu }{2}\right)^2 
	\Gamma \left(\frac{h+J+i \nu }{2}\right)^2
},
\ee 
\begin{align}\label{chiralspec}
\begin{split}
b^{c}_J(\n)=&1/\left(256\pi^4\left(-\omega ^4+\frac{\left(J^2+\nu ^2\right) \left((J+2)^2+\nu ^2\right)}{16 (J+1) \nu } \left((J+1) \nu +i \kappa ^4 \left(-\psi ^{(1)}\left(\frac{1}{4} (J+i \nu +2)\right)\right.\right.\right.\right.\\
&\left.\left.\left.\left.+\psi ^{(1)}\left(\frac{1}{4} (J+i \nu +4)\right) +\psi ^{(1)}\left(\frac{1}{4} (J-i \nu +2)\right)-\psi ^{(1)}\left(\frac{1}{4} (J-i \nu +4)\right)\right)\right)\right)\right)
\end{split}
\end{align}
with $\z_c(\D_i,t)=\G\left(\frac{2-t}{2}\right)^{-2}$ and the coupling parameters $\k,\,\w$ are defined by the relations
\be 
\k^4=\x_2^2\x_3^2\,,\qquad\w^4=(\x_2^2-\x_3^2)^2.
\ee 
Here, $\x_2,\,\x_3$ are the couplings used to write the $\c-$fishnet Lagrangian in  \eqref{chilag}. 
%\subsection{Possible scenarios}
%The spectral function has two couplings $(\k,\w)$. It is evident that for $\k=\w=0$, there are two sets of poles,
%\be\label{trivpoles}
%J=\pm2i\n\,, \ -2\pm2i\n\,.
%\ee
%This is the free theory scenario. For non-zero coupling, there are various scenarios which we list below, starting with the simplest ones. 
%\begin{itemize}
%\item \underline{$\w=0,\k\neq0$} For this case, we retrieve corrections to \eqref{trivpoles}. The leading trajectory is $J=\pm2i\n$, with corrections in terms of $\k-$coupling.
%\item \underline{$\k=0,\w\neq0$} This reduces to the $0-$magnon case with a leading Regge trajectory at $J=\pm2i\n$.
%\item\underline{$(\w,\k)\neq0$} For $\w\neq0$, we obtain corrections to the $0-$magnon  trajectory due to $\k$. 

\subsection{Evaluation of Regge trajectories}
       The exact evaluation of Regge trajectories is not possible for this case and we proceed to evaluate the trajectories perturbatively as in the general $d$ fishnet theory. Since there are two coupling constants $\kappa$ and $\omega$, we work in the limit 
$\kappa \to 0, \omega \to 0$ with $\frac{\kappa}{\omega}$ held constant. We note that the leading free Regge trajectories are given by 
\be 
J_{f\pm}^{(c)}=\pm i\n
\ee

\paragraph{Outer Solution:} We begin by evaluating the outer solution 
which we evaluate perturbatively with the following ansatz \begin{shaded}
\be\label{Jcout}
J_{\text{o}\pm}^{(c)}(\nu)=\pm i\n+\g_{1\pm}\omega^4+\sum_{k=2}^\infty \g_k(\pm\n) \omega^{4k}\,.
\ee
\end{shaded}
the first few coefficients in the expansion \eqref{Jcout} are given by\begin{shaded}
\begin{align}\label{ocoeffc}
\begin{split}
\g_2(\pm\n)&=\frac{\g_{1\pm} \left(\g_{1\pm} (3-3  (\nu^2 \pm 3 i\n))-2 z^4 \left(3 \psi ^{(1)}\left(\pm\frac{i \nu }{2}+1\right)-3 \psi ^{(1)}\left(\pm\frac{i \nu }{2}+\frac{1}{2}\right)+\pi ^2\right)\right)}{6  (\nu^2 \pm i\n)}\\
\g_3(\pm\n) &=\frac{\g_{1\pm} }{36 \nu ^2 (\nu \mp i)^2}
\Bigg[3 \g_{1\pm} \Bigg\{6 \g_{1\pm} (1+\nu  (\nu  (-9+\nu  (\nu \mp 5 i))+5 i))+z^4 \Bigg(4 \pi ^2 (-1+\nu  (\nu \mp 4 i))\\
& \hspace{3.5 cm}+12 (-1+\nu  (\nu \mp4 i)) \psi ^{(1)}\left(\pm \frac{i \nu }{2}+1\right)-12 (-1+\nu  ( \n\mp 4 i)) \psi ^{(1)}\left(\pm\frac{i \nu }{2}+\frac{1}{2}\right)\\
&\hspace{5.5 cm}+3 \nu  (\n\mp i ) \left(-\psi ^{(2)}\left(\pm\frac{i \nu }{2} +1\right)+\psi ^{(2)}\left(\pm\frac{i \nu }{2}+\frac{1}{2}\right)+12 \zeta (3)\right)\Bigg)\Bigg\}\\
&\hspace{7.7 cm}+4 z^8 \left(3 \psi ^{(1)}\left(\pm\frac{i \nu }{2}+1\right)-3 \psi ^{(1)}\left(\pm \frac{i \nu }{2}+\frac{1}{2}\right)+\pi ^2\right)^2\Bigg],
\end{split}
\end{align}
\end{shaded}
where we have used $\kappa= z \omega$ and $z\geq 0$.
Using this, the pole structure of the spectral function \eqref{chiralspec} becomes
\be\label{specresc}
b^{{\rm CI}}_{J_{\text{o}\pm}^{(c)}(\nu)}=-\frac{1}{128\pi^4\omega ^4 \left(2+\g_{1\pm}   (\nu^2 \pm i\nu)\right)}
\ee
The poles are at $\g_{1\pm}=-\frac{2}{(\nu^2 \pm i\nu)}$. Analogous to previous cases, we consider $J_{\text{o}+}^{(c)}(\nu)$ corresponding to $\g_{1+}$ and $J_{\text{o}-}^{(c)}(\nu)$ corresponding to $\g_{1-}$.
\\

\textbf{Inner Solution:}
    The inner solution region corresponds to $|x|<1$ with $\n=2x\omega^2$. Thereby, one has the perturbative expression for the inner solution
    \begin{shaded}
    	\be \label{cmelinnerj}
    	J_{\text{i}\pm}^{(c)}(x)=\sum_{k=1}^{\infty} \tilde{\d}_k(x)\omega^{2k}
    	\ee 
    \end{shaded}	
    with first few coefficients $\tilde{\d}_k,\, k\ge 2$ given by
    \begin{shaded}\label{cmelinnerj2}
    	\begin{align}\label{coeffic}
    	\begin{split}
    	\tilde{\d}_2(x)&=-\frac{\tilde{\d}_{1}^2}{2}-2 x^2, \qquad
    	\tilde{\d}_3(x)= \frac{\left(\tilde{\d}_{1}^2+4 x^2\right) \left(\tilde{\d}_{1}^2+6 z^4 \zeta (3)\right)}{2 \tilde{\d}_{1}},\\
    	\tilde{\d}_4(x)&= -\frac{1}{120} \left(\tilde{\d}_{1}^2+4 x^2\right) \left[75 \tilde{\d}_{1}^2+60 x^2+z^4 \left\{30 \left(12 \zeta (3)+\psi ^{(2)}(1)-\psi ^{(2)}\left(\frac{1}{2}\right)\right)+7 \pi ^4\right\}\right] ,
    	\end{split} 
    	\end{align}
    \end{shaded}
where we have used $\kappa= z \omega$ and $z\geq 0$.
Using this, the pole structure of the spectral function \eqref{chiralspec} becomes
\be\label{specrescin}
b^{{\rm CI}}_{J_{\text{i}\pm}^{(c)}(x)}=\frac{1}{64\pi^4\omega ^4 \left(\d_{1}-2 \sqrt{1-x^2}\right)\left(\d_{1}+2 \sqrt{1-x^2}\right)}
\ee
The poles are at $\tilde{\d}_1=\tilde{\d}_{1\pm}:=\pm 2\sqrt{1-x^2}$. Analogous to previous cases, we consider $J_{\text{i}+}^{(c)}(x)$ corresponding to $\tilde{\d}_1=\tilde{\d}_{1+}$ and $J_{\text{i}-}^{(c)}(x)$ corresponding to $\tilde{\d}_1=\tilde{\d}_{1-}$.

\subsection{Evaluation of Mellin amplitude}
For the evaluation of the Mellin amplitude, we will follow the procedure laid out in details for zero magnon case in section \ref{Mell0}. This is again due to the fact that the inner and outer solutions satisfy the transformation properties as that of 0-magnon (see \eqref{transinint} and \eqref{transoutint} ). Thus, we just give the final expressions here.
 
\be \label{Melc}
\mm_{+}^{(c)}=4\zeta_1(\D_i,t)\omega^2\int_{-1}^{1}dy\,\frac{y}{\sqrt{1-y^2}}\,\F_{c}^{(1)}(\sqrt{1-y^2})\,+\,(s\to-s)
\ee where\footnote{The extra factor of 2 is due to the fact that we had chosen $\nu = 2 x \omega^2$ for the inner solution compared to $\nu =  x \x^2$ for the 0-magnon case. }, $\F_{+}^{(c)}(\sqrt{1-y^2})$ i is given by 
\begin{align} 
\begin{split}
&\F^{(c)}_{+}(\sqrt{1-y^2}):=\Theta^{(c)}_{\text{i}}(\tilde{\d}_{1+})\,M^{(c)}\left(J_{\text{i}+}^\text{(c)}(\sqrt{1-y^2}), \sqrt{1-y^2}\right)\,\left(\frac{s}{4}\right)^{J_{\text{i}+}^\text{(c)}(\sqrt{1-y^2})}\,\left[ \frac{1}{256\pi^4 \omega ^4 y} \right],\\
&\Theta^{(c)}_{\text{i}}(\tilde{\d}_{1+}):=\left|\frac{\pd J_{\text{i}+}^{(c)}(\tilde{\d}_{1+}, \sqrt{1-y^2})}{\pd\tilde{\d}_{1+}}\right|=\omega ^2-y \omega ^4 +\omega ^6 \left(\frac{3 \left(5 y^2-4\right) z^4 \zeta (3)}{y^2}-\frac{y^2}{2}+2\right)
\\
&\phantom{\left|\frac{\pd J_{\text{i}+}^{(c)}(\tilde{\d}_{1+}, \sqrt{1-y^2})}{\pd\tilde{\d}_{1+}}\right|dsdsdsdsd44}+ \omega ^8 \left(\frac{7 y^3}{2}+y \left(-12 z^4 \zeta (3)-\frac{7 \pi ^4 z^4}{60}-6\right)\right)\\
\end{split}
\end{align}
where $\tilde{\d}_{1+}= y$, $M^{(c)}(J,\n)$ and $J_{\text{i}+}^\text{(c)}(x)$ are given in \eqref{melc} and \eqref{cmelinnerj} respectively. The integrand is then evaluated in the limit 
$$\omega \rightarrow 0, ~~ s \rightarrow \infty,~~Q=\omega^2 \log \frac{s}{4} \to {\rm Constant}$$ 
to give,
 
\begin{shaded}
	\be\label{chiralpert}
	\begin{split}
		\mm^{(c)}_+(s,t)&=2\int_{-1}^1 dx~\Gamma \left(1-\frac{t}{2}\right)^2\sqrt{1-x^2}e^{2 q x} \left( \frac{ \omega ^2 }{128 \pi ^9 x} + \frac{ \omega ^4}{128 \pi ^9 x^2}  \left(-2 q x-2 x^2 \psi ^{(0)}\left(1-\frac{t}{2}\right)+4 x^2+1\right)\right.\\
		&\left. +\frac{\omega ^2 }{384 \pi ^9 x^3} \left(8 \omega ^4 \left(x \left(6 q^2 x-6 q \left(2 x^2+1\right)+x \left(\pi ^2 \left(2 x^2+1\right)+6\right)\right)+3 x^2 \left(2 x \psi ^{(0)}\left(1-\frac{t}{2}\right)\right.\right.\right.\right.\\
		&\left.\left.\left.\left. \left(2 q+x \psi ^{(0)}\left(1-\frac{t}{2}\right)-4 x\right)+\left(2 x^2-1\right) \psi ^{(1)}\left(1-\frac{t}{2}\right)\right)+3\right)+9 \kappa ^4 \zeta (3) (x (q+x)-1)\right)\right.\\
		&\left. \frac{\omega ^4}{7680 \pi ^9 x^4}  \left(-\kappa ^4 \left(180 \zeta (3) \left(-2 x \left(-2 q^2 x+q \left(3-2 x^2\right)+2 x^3+x\right)+2 x^2 (2 x (q+x)-1) \psi ^{(0)}\left(1-\frac{t}{2}\right)\right.\right.\right.\right.\\
		&\left.\left.\left.\left. +3\right)+7 \pi ^4 x^2 (2 x (q+x)-1)\right)-320 \omega ^4 \left(2 q \left(2 q^2+\pi ^2+3\right) x^3-\left(6 q^2+\pi ^2+3\right) x^2+6 x^4 (2 x (q-2 x)+1)\right.\right.\right.\\
		&\left.\left.\left. \psi ^{(0)}\left(1-\frac{t}{2}\right)^2+3 x^2 \left(2 x \left(2 x^2-1\right) (q-2 x)+1\right) \psi ^{(1)}\left(1-\frac{t}{2}\right)+2 x^4 \psi ^{(0)}\left(1-\frac{t}{2}\right)\right.\right.\right.\\
		&\left.\left.\left. \left(6 q (q-2 x)+\left(6 x^2-3\right) \psi ^{(1)}\left(1-\frac{t}{2}\right)+\pi ^2 \left(2 x^2+1\right)\right)+4 \left(\pi ^2-6\right) q x^5+6 q x\right.\right.\right.\\
		&\left.\left.\left. +4 x^6 \psi ^{(0)}\left(1-\frac{t}{2}\right)^3+\left(4 x^2-3\right) x^4 \left(\psi ^{(2)}\left(1-\frac{t}{2}\right)+12 \zeta (3)\right)-8 \pi ^2 x^6+12 x^4-3\right)\right)\right)\\
		& +(s \rightarrow -s) + {\cal O}(\omega^{10}, \omega^6\kappa^4, \omega^2\kappa^8) 
	\end{split}
	\ee
\end{shaded}                                
Although we donot explicitly list the integrated results here, it can be obtained using integrals listed in Appendix \ref{intweak}

\subsection{Alternative perturbative method for Regge amplitude}\label{apmra}
 In this section we evaluate the Mellin amplitude \eqref{Mellinc} in a different perturbative approach. Note that we can rewrite the spectral function \eqref{chiralspec} as follows
\begin{align}\label{chf1sf}
\begin{split}
b^{c}_J(\n)=&\frac{1}{256\pi^4\left(a_0(\nu, J) +l_0(\nu, J)\right)}
\end{split}
\end{align}
where $a_0(\nu, J)$ is the spectral function corresponding to the 0-magnon like spectral function and $l_0(\nu, J)$ are given by 
\begin{equation}\label{0magspecc}
\begin{split}
a_0(\nu, J)&=-\omega ^4+\left(J^2+\nu ^2\right) \left((J+2)^2+\nu ^2\right)\\
l_0(\nu, J)&=\frac{i \kappa ^4 }{16  (J+1) \nu }\left(J^2+\nu ^2\right) \left((J+2)^2+\nu ^2\right) \left(-\psi ^{(1)}\left(\frac{1}{4} (J+i \nu +2)\right)+\psi ^{(1)}\left(\frac{1}{4} (J+i \nu +4)\right)\right.\\
&\left.+\psi ^{(1)}\left(\frac{1}{4} (J-i \nu +2)\right)-\psi ^{(1)}\left(\frac{1}{4} (J-i \nu +4)\right)\right)
\end{split}
\end{equation}

Therefore we try to solve the Mellin amplitude as perturbation expansion in $\kappa$ about 0-magnon. To be precise let us first write down the relevant Mellin amplitude.

\be\label{chfema1}
\begin{split}
\mm^{(c)}_{\pm}(s,t)=\left[\frac{\pm1}{2\pi i}\right. &\left.\oint dJ \int_{-\infty}^\infty d\n \left(\frac{s}{4}\right)^J e^{i\pi J/2} M^{(c)}(J,\nu) \,\zeta_c({\D_i,t}) \left(\sum_{k=0}^\infty\frac{c_k(J,\n)}{a_0(J,\n)^{k+1}}\right)\right] \pm(s\rightarrow -s)
\end{split}	 
\ee
where $M^{(c)}(J,\n)$ is given by \eqref{melc}. We have also expanded the spectral function 
$b^{c}_J(\n)$ about zero-magnon as follows

\be
\begin{split}
	b^{c}_J(\n) &=\sum_{k=0}^\infty\frac{c_k(J,\n)}{a_0(J,\n)^{k+1}},\qquad	c_k(J, \n)=l_0(J, \n)^k 
\end{split} 
\ee 
For the rest of the section let us focus on $\mm^{(c)}_{+}(s,t)$.
\subsubsection{k=0}
For $k=0$, this is the usual $0-$magnon case. The spectral function corresponding to  $0-$magnon ($a_0(\nu, J)$), given by \eqref{0magspecc}, has poles at,
\be\label{0magpolesc}
J_{1,2}^\pm=-1\pm\sqrt{1-\n^2\pm 2\sqrt{4\omega^4-\n^2}}\,.
\ee
The leading Regge poles are,
\be
J_1^\pm=-1+\sqrt{1-\n^2\pm 2\sqrt{4\omega^4-\n^2}}\,.
\ee
For the zeroth order case ,\eqref{chfema1} reduces to the $0-$magnon case. Defining  $\n=2\omega^2\sqrt{1-x^2}$, the amplitude becomes (This manipulation has been outlined in \cite{Chowdhury:2019hns} and for convenience worked out in Appendix \ref{chfema1int} corresponding to higher orders of $k$),
\be
I_0=4\omega^2\int_{-1}^1 dx\ \left(\frac{x}{256\pi^4 \sqrt{1-x^2}}\frac{16 M(J(x),x)}{\left(4\left(J+1\right) \left(J \left(J+2\right)+4\omega^4(1-x^2)\right)\right)}\,\right)|_{{\rm Res }~ J=J^1_+} + (s \to -s) 
\ee
with the explicit form of $M(J,\n)$ given by $M(J,\n)=\left(\frac{s}{4}\right)^J e^{i\pi J/2} M^{(c)}(J,\nu) \,\zeta_c({\D_i,t})$. The contribution from the $0-$th order case is denoted by $I_0$ and for convenience and its explicit expression is given in \eqref{chm1i0}. \footnote{Note that the factors of $\psi^{(i)}\left(1-\frac{t}{2}\right)^j$ should be dropped to get the momentum space amplitude of \cite{Korchemsky:2018hnb}},
These integrals can be easily done (see \cite{Chowdhury:2019hns},\cite{Korchemsky:2018hnb}). We will present the integrated results shortly. 
\subsubsection{k $\geq$ 1}

For $k \geq 1$, the evaluation of this integal becomes difficult since we have multiplicity of poles as is evident from \eqref{chfema1}. Nevertheless, in Appendix \ref{chfema1int}, we argue that the general form for the integral can be brought to the following form 

\begin{equation}\label{chfema1k1} 
\begin{split}
I_k&=\left[\frac{1}{2\pi i}\oint dJ \int_{-\infty}^\infty d\n \left(\frac{s}{4}\right)^J e^{i\pi J/2} M^{(c)}(J,\nu) \,\zeta_c({\D_i,t}) \left(\frac{c_k(J,\n)}{a_0(J,\n)^{k+1}}\right)\right] + (s\rightarrow -s)\\
&=4\omega^2\int_{-1}^1 dx \left(\frac{x}{256 \pi^4 \sqrt{1-x^2}} \frac{16^{k+1}(l_0(J(x),x))^k M(J(x),x)}{((J-J_1^+)(J-J_1^-)(J-J_2^+)(J-J_2^-))^{k+1}}\right)|_{{\rm Res }~ J=J^1_+}\, + (s\rightarrow -s)
\end{split}
\end{equation}
where for $k=1, 2$ {\it i.e.} the first two corrections given by $I_1$ and $I_2$ are given in \eqref{chm1i1} and \eqref{chm1i2} respectively. Adding up $I_0, I_1$ and $I_2$ (from \eqref{chm1i0}, \eqref{chm1i1}, \eqref{chm1i1} respectively), we explicitly obtain perfect agreement with \eqref{chiralpert}. This is a non-trivial check of our formalism.

%\subsection{Chiral fishnet: Type II}
%For this case the spectral function is given by, 
%\be\label{chspectfn2}
%b_J(\n)^{-1}=(J^2+4\n^2)((J+2)^2+4\n^2)+\frac{128(J(J+2)-4\n^2)\l^2\m^2}{J(J+2)(J(J+2)-4\l^2)+8(2+J(J+2)+2\l^2)\n^2+16\n^4}-16\l^4\,.
%\ee

\section{Conclusions}\label{concl}

In this work, we have extended the analysis of Regge trajectories of Mellin amplitudes for the connected part of  4-$d$ 0, and 1-magnon fishnet correlators initially studied in \cite{Korchemsky:2018hnb} and \cite{Chowdhury:2019hns}, to general dimensional 0 and 1-magnon fishnet correlators (\cite{Kazakov:2018qez}) and also one of the chiral fishnet correlators studied in \cite{Kazakov:2018gcy}. We note the salient features of our computation.

\begin{itemize}
	\item Unlike $d=4$ biscalar fishnet theory, the Regge poles for these cases cannot be solved exactly and we needed to resort to perturbative evaluation of the Regge trajectories at weak coupling. The perturbative regge poles are given in \eqref{J0in} and \eqref{J0out} for 0-magnon correlator, \eqref{J1eout} and \eqref{1mageveninner} for the even spin Regge trajectory for 1-magnon correlator,  \eqref{J1oout} and \eqref{J1oin} for the odd spin Regge trajectory for 1-magnon correlator, \eqref{Jcout} and \eqref{cmelinnerj} for the chiral fishnet correlator. 
	
	\item We have also been able to find closed-form expressions (perturbative in respective couplings) for the Mellin-Regge amplitudes for all these cases after doing the spectral integral. To be precise, we find that for the general dimensional 0-magnon and 1- magnon correlators, the Regge trajectories scale as 1 and $s^{-\frac{d}{4}}$ respectively at weak coupling while for the chiral fishnet correlator it scales as 1. 
	
	\item In order to do the respective spectral integrals over $\nu$, we extend the machinery used in \cite{Chowdhury:2019hns} to study 2-magnon Regge trajectories perturbatively in $d=4$. The method turns out to be quite general and just dependent on the generic features of the Regge poles and is one of the main results of this paper (see \eqref{integrand0}, \eqref{coef} for 0-magnon amplitude, \eqref{1magevenresult}, \eqref{1mageve2} and section \ref{oddspin1maggend} for 1-magnon amplitude and finally \eqref{chiralpert} for the Mellin-Regge amplitude for chiral fishnet theory correlator).  
	
	\item We have several crosschecks of our results also; for 0 and 1-magnon correlators we can explicitly compute closed-form expressions for Regge poles in $d=4$ and $d=4, 8$ respectively and can explicitly derive an exact expression for the spectral integral following the techniques in \cite{Chowdhury:2019hns} and \cite{Korchemsky:2018hnb}. These exact integrals, when evaluated perturbatively, are in perfect agreement with our $d$-dimensional results. The explicit comparisons have been made in subsection \ref{awd4r0m} for the 0-magnon and subsection \ref{awd4d8r1m} for 1-magnon correlator. We want to emphasize that this is a non-trivial check since the perturbative technique is structurally different from the exact analysis.
	
	\item For chiral fishnet correlator, this analysis is complicated by the fact that there are two couplings ($\k$ and $\omega$) in the theory. We can obtain perturbative Regge trajectories in the special limit $\k \to 0,~ \omega \to 0$ with $\k/\omega$ held constant. From observing the generic structure of the Regge poles, we are able to use our general method (used to compute the 0 and 1-magnon correlators, in general, $d$ fishnet theory) to compute the spectral integral. We also note that the spectral function for the chiral correlator can be rewritten as a perturbation expansion in $\kappa$ about a 4-$d$ 0-magnon spectral function with coupling constant $\omega$. This resulting spectral integral (perturbative in $\kappa$) can be exactly evaluated in $\omega$. This analysis, done in subsection \ref{apmra}, facilitates an entirely different way of evaluating the spectral integral for the chiral fishnet correlator. In the limit, $\k \to 0,~ \omega \to 0$ with $\k/\omega$ held constant, we obtain perfect agreement between our two methods.

\end{itemize}

The computation of the general fishnet theories in d dimensions and for multiple couplings reveal some features which are worth exploring in the future.
\begin{itemize}
	\item The computation makes it clear that we do not need to solve the spectral function exactly in order for the perturbative computation of leading Regge trajectory. We can start by computing the solutions of the regge trajectory perturbatively from the start regardless of dimensions and other complexities. This begs the question, how can our technology be used to find the sub-leading Regge trajectory and amplitude. In the context of $d=4$ fishnet theory, this has been explored in a nice paper very recently \cite{Caron-Huot:2020nem}. Their analysis is restricted to $d=4$ where closed-form expression for the conformal blocks, used in their computation, is known. For general d dimensions, it remains an open question to find sub-leading Regge trajectory in general and in particular for the Fishnet theory.   
	
	\item This computation forms the basis against which the computations of diagrammatic perturbation theory can be compared. The presence of transcendental functions of Mellin variables translates into purely transcendental contributions of the cross ratios.
	
	\item  An $n-$magnon case is suppressed by $s^{-\alpha}$ contribution, followed by a series expansion in Log[s] in the Regge limit. For the leading order in $s$, one can ask the following question: whether the contributions can be re-summed into something which can be interpreted in terms of the cross channel contribution as happens in conformal Regge bootstrap. 
	
	\item In principle, this study of Regge trajectories can be extended to 2-magnon correlators in general $d$ fishnet theory\cite{Gromov:2018hut, Kazakov:2018qez} and also the 2nd type of chiral fishnet correlator\cite{Kazakov:2018gcy}. The immediate challenge for extending this to the 2 magnon correlator in the general dimension is the fact that closed-form expressions for the spectral function is not known and would involve extending the $4d$ analysis in Appendix C.4 of \cite{Gromov:2018hut} to general $d$ dimensions.
	
	\item For the chiral correlator of the type $\langle{\rm Tr}[\phi_1(x_1)\phi_2^\dagger (x_2)]{\rm Tr}[\phi_1^\dagger(x_3)\phi_2(x_4)]\rangle$  studied in \cite{Kazakov:2018gcy}, we find that our techniques do not go through. Explicitly, the general method developed here involved evaluating the integral at weak coupling $\xi$ say. For this purpose we found perturbative Regge trajectories corresponding to the regions $\nu < \xi^\alpha$ (inner solution) and $\nu > \xi^\alpha$ (outer solution). The coefficient $\alpha$ is non-universal and depends on the theory. It has to be estimated by a ratio test of the outer perturbative solution in order to determine where exactly the perturbation series breaks down. In particular, we were unable to find such a region or the ``$\alpha$" for the chiral correlator of the second type. We hope to address these issues in the future.
\end{itemize}

\section*{Acknowledgements}
We thank Vladimir Kazakov for suggesting the problem to us and Gregory Korchemsky for discussions. The work of SDC is supported by the Infosys Endowment for the study of the Quantum Structure of Spacetime. K.S. is partially supported by the Laureate Award IRCLA/2017/82 from the Irish Research Council.

%%%%%%%%%%%%%%%%%%%%%%%%%%%%%%%%%%%%%%%%%%%%%%%%%%
%%%%%%%%%%%%%%%%%%%%%%%%%%%%%%%%%%%%%%%%%%%%%%%%%%
%appendix starts here
\appendix
%\section{Connection between Inner Solution and the Outer Solution }\label{inout}
\section{Details of Perturbative evaluation of Regge Trajectories }\label{sec2det}
In this appendix, we offer some detailed analysis of the general principles for obtaining the outer solution, which we chalked out in section \ref{genRegge}. In particular, we detail out one possible abstract argument for  how one can  reach \eqref{polestructo} starting with the perturbative ansatz  \eqref{Jpert2} which we rewrite here for convenience
\be 
J_{\text{o}}^{(n)}(\n)=J_f^{(n)}(\n)+\sum_{k=1}^{\infty}\,f_k(\n)\,\x^{\a k},\,\,\,\a>0
\ee where, recall that, $J_f^{(n)}(\n)$ is assumed to be a  simple pole (this is satisfied by both $0$ and $1-$magnon correlators in generalized fishnet theory) in $J$ of $E_{\D,J}^{(n)}$ with $\D=h+i\n$. In doing so, we will work under certain assumptions. One can try to see that, to what extent these assumptions can be relaxed and generalized. However, we do not engage ourselves with that task! Further, consider the series
\be \label{P1ser}
P_1(\n):=\sum_{k=1}^{\infty}\,f_k(\n)\,\x^{\a k},\,\,\,\a>0
\ee  converging pointwise to $L_1(\n)$ on a domain $\mathcal{D}$ over real $\n$   by ratio test
\be\label{ratio} 
\lim_{k\to\infty}\left|\frac{f_{k+1}(\n)}{f_{k}(\n)}\right|\x^{\a }<1~~~\forall~\n\in\mathcal{D}.
\ee The implication of this convergence criterion has been discussed in section \ref{genRegge} itself. We will not go into that. We will only \emph{use the fact} that this series converges. Now, consider the  series 
\be\label{P2ser}
P_2(\n):=\sum_{k=1}^{\infty}A_k(\n)\,\x^{\a k}
\ee where,
\be 
A_k(\n):=\frac{f_{k+1}(\n)}{f_1(\n)}.
\ee A straightforward application of the ratio test \eqref{ratio} implies that, if $P_1(\n)$ converges pointwise on $\mathcal{D}$ so does $P_2(\n)$. Let, $P_2(\n)$ converges to $L_2(\n)$. We assume that,
\be 
L_2(\n)<1~~~ \forall~\n\in\mathcal{D}.
\ee 
\paragraph{} First, we will evaluate $\a$. To do so, we need to consider the pole structure of $E_{\D,J}^{(n)}\equiv E_{\n}^{(n)}(J)$ in $J$. Since $J_f^{(n)}(\n)$ is a simple pole of $E_{\n}^{(n)}(J)$,  there exists an \emph{annular} region\footnote{All regions are defined to be open connected sets in the usual metric topology of $\mathbb{C}$.} in complex $J$-plane where $E_{\n}^{(n)}(J)$ admits a Laurent series representation of the form
\be \label{Laurent}
E_{\n}^{(n)}(J)=\frac{\mathcal{R}(\n)}{J-J_f^{(n)}(\n)}+\sum_{k=0}^{\infty}c_k (\n)\,\left(J-J_f^{(n)}(\n)\right)^k
\ee where, $\mr(\n)$ is residue of $J_f^{(n)}(\n)$ at $J=J_f^{(n)}(\n)$. Now, in this representation we will use the ansatz $J_{\text{o}}^{(n)}(\n)$ in place of $J$. Under what condition one can do this is tightly connected to the convergence property of the power series $P_1$. For the purpose of evaluation of $\a$, we will fix our attention to the singular piece of the Laurent expansion. In particular, this piece can now be expressed as 
\be 
\frac{\mathcal{R}(\n)}{J-J_f^{(n)}(\n)}=\x^{-\a}\frac{\mr(\n)}{f_1(\n)}\,\left[1+P_2\right]^{-1}
\ee   Now, since we have assumed that $P_2$ converges to $L<1$ we can expand $[1+P_2]^{-1}$ as 
\be 
\left[1+P_2\right]^{-1}=1-P_2+2P_2^2+\dots
\ee  resulting in the following expression for the singular piece:
\be\label{polepiece} 
\frac{\mathcal{R}(\n)}{J-J_f^{(n)}(\n)}=\x^{-\a}\frac{\mr(\n)}{f_1(\n)}\,\left[1-P_2+2P_2^2+\dots\right].
\ee Then, it is straightforward to observe that, 
\be 
E_\n^{(n)}\left(J_{\text{o}}^{(n)}(\n)\right)=\x^{-\a}\frac{\mr(\n)}{f_1(\n)}+O(\x^0).\ee
Next, we need to solve \eqref{Reggesol2} which we reproduce here (with new notations) for convenience
\be \label{Reggesol2'}
\c_n E_\n^{(n)}\left(J_{\text{o}}^{(n)}(\n)\right)=1
.\ee 
Putting $\c_n= (4\p)^{qh}\x^{2q}$ into this, we have
\be \label{ReggeSOl3}
(4\p)^{qh}\x^{2q-\a}\frac{\mr(\n)}{f_1(\n)}+O(\x^{2q})=1
\ee Since $q>0$, it is evident that, one will have a solution to above if and  only if 
\be\label{alpha} 
\a=2q.
\ee 

\paragraph{} Next, we look into \eqref{ReggeSOl3} more closely. Let us write down the $O(\x^{2q})$ terms explicitly. From the Laurent expansion \eqref{Laurent} and \eqref{polepiece}, it is straightforward to obtain
\be \label{EJLaurent}
E_\n^{(n)}\left(J_{\text{o}}^{(n)}(\n)\right)=\x^{-2q}\frac{\mr(\n)}{f_1(\n)}[1-P_2+2P_2^2+\dots]+\sum_{k=0}^{\infty}c_k(\n)\, P_1^k.
\ee  Using this  into \eqref{Reggesol2'} above along with the expression for $\c_n$ one has
\be 
\frac{\mr(\n)}{f_1(\n)}[1-P_2+2P_2^2+\dots]+\x^{2q}\sum_{k=0}^{\infty}c_k(\n)\, P_1^k=(4\p)^{-qh}
\ee  This equation  above breaks effectively into two equations
\begin{eqnarray} 
	\frac{\mr(\n)}{f_1(\n)}&=&(4\p)^{-qh}\label{xi0}\\\label{xi2q}
	\frac{\mr(\n)}{f_1(\n)}[-P_2+2P_2^2+\dots]&+&\x^{2q}\sum_{k=0}^{\infty}c_k(\n)\, P_1^k=0.
\end{eqnarray} Solving \eqref{xi0} one gets
\be 
f_1(\n)=(4\p)^{qh}\mr(\n)
\ee and
using \eqref{P1ser} and \eqref{P2ser}, one can solve  \eqref{xi2q} order by order in $\x$ to obtain
\begin{eqnarray}\label{coeff}
	\x^{2q}&:&\qquad \qquad f_2(\n)=f_1(\n)^2\,\frac{c_0(\n)}{\mr(\n)},\nn\\
	\x^{4q}&:&\qquad\qquad f_3(\n)=-f_1(\n)^3\left[\frac{c_1(\n)}{\mr(\n)}+2\left(\frac{c_0(\n)}{\mr(\n)}\right)^2\right] .
\end{eqnarray}One can proceed further in this manner. The important structural point coming out of this is that, the coefficient function $f_{k\ge2}(\n)$ is proportional to $f_1(\n)^k$. At this point, it is worth the mention that, this structural simplicity is consequence of the simplicity of the equation \eqref{Reggesol2'}. To be specific, this kind of equation arises in the Regge study of $n-$magnon correlator of normal Fishnet theories. But, the same is not true, for example, for chiral fishnet theories in $d=4$ which has been studied in section \ref{chm1}. In particular, we would like to draw attention of the reader to the \eqref{ocoeffc} which clearly doesn't conform to the simplistic structure of \eqref{coeff}.

\paragraph{}Finally, we have all the ingredients to reach \eqref{polestructo}. In particular, we want to show that the Regge pole structure of the spectral function \eqref{gendspec} is manifested as a pole in $f_1(\n)$. To do so, we substitute \eqref{EJLaurent} into the spectral function to obtain
\be 
\frac{\left[E_\n^{(n)}\left(J_{\text{o}}^{(n)}(\n)\right)\right]^p}{1-\c_nE_\n^{(n)}\left(J_{\text{o}}^{(n)}(\n)\right)}=\x^{-2pq}\frac{[\mr(\n)]^p}{[f_1(\n)]^{p-1}\left[f_1(\n)-(4\p)^{qh}\mr(\n)\right]}
\ee 
where we have used  the coefficients $\{f_k:k\ge2\}$ expressed in terms of $f_1(\n)$ (thus, using \eqref{xi2q}) but leaving $f_1(\n)$ as it is. Now we have reached \eqref{polestructo} with 
\be 
\mathfrak{B}_{\text{o}}(\n)=[\mr(\n)]^p,\qquad F(\n)=(4\p)^{qh}\mr(\n).
\ee 
\section{Inner and outer integral 0-magnon general $d$}\label{iaoi0mgd}
                 In this section we outline in detail how to arrive at \eqref{I0maggendfinal}.
Recall from \eqref{0maggendint} that, 
\be 
\mi^{(0)}(\x,s,t)=\int_{-\infty}^\infty d\n \mf(J,\,\n)=
\left(\int_{-\infty}^{-\x^{2}}+\int_{\x^{2}}^\infty\right) d\n 
\mf_\text{outer}
+
\int^{\x^{2}}_{-\x^{2}} d\n
\mf_\text{inner}\,.
\ee 
where
\be
\mf(J,\,\n)= \zeta_0(\D_i,t)\oint \frac{d\d}{2\p i} 
~\Theta(\d)
M^{(0)}(J(\d),\n) \left(\frac{s}{4}\right)^{J(\d)}\frac{E^{(0)}_{h+i\n,J}}{1-\c_0 E^{(0)}_{h+i\n,J}},
\ee 
and we have assumed that we have perturbatively solved for the Regge pole $J$ as a functions of $\d$ (hence the jacobian and the integration over $\delta$). For convenience, $M(J, \n)$ and $\Th(\d)$ are given by \eqref{Mdef} and \eqref{Jacobdef},
\begin{eqnarray}\label{appM}
M^{(0)}(J,\n)&=&\p\n\sinh(\pi\n) \frac{\G(h+J)\G(h+J+i\n)\G(h+J-i\n)\G(\frac{h-J-t+i\n}{2})\G(\frac{h-J-t-i\n}{2})}
{2\p^{2(h+1)}\sin(\p J)\G(1+J)\G(\frac{h+J+i\n}{2})^2\G(\frac{h+J-i\n}{2})^2}\,\nonumber\\
 \Th(\d)&=&\frac{\pd J}{\pd\d}.
\end{eqnarray}
Here, $\mf_\text{inner}$ is obtained by using $J_{\text{i}\pm}^\text{(0)}(x)$ \eqref{J0in} into $\mf(J,\n)$ and $\mf_\text{outer}$ by putting $J_{\text{o}\pm}^{(0)}(\nu)$ \eqref{J0out} into the same. Now, we will focus upon the inner and outer integrals individually. Let us define the inner and outer intgralas as follows 
\be
\mi_{\text{i}}^{(0)}=\int^{\x^{2}}_{-\x^{2}} d\n
\mf_\text{inner}\,\qquad \mi_{\text{o}}^{(0)}:=
\left(\int_{-\infty}^{-\x^{2}}+\int_{\x^{2}}^\infty\right) d\n 
\mf_\text{outer}
\ee

\subsubsection*{I. Inner Integral}
Let us start by investigating the inner integral. In the integral we make the change of variable $\n\to x \xi^2$ and write the integral as 
\begin{align}
\begin{split}
\mi_{\text{i}}^{(0)}:=&\int^{\x^{2}}_{-\x^{2}} d\n
\mf_\text{inner}\,=\zeta_0(\D_i,t)\x^2\int_{-1}^{1}dx \oint \frac{dJ}{2\p i}~M^{(0)}(J,x)\left(\frac{s}{4}\right)^J\frac{E^{(0)}_{h+i x\x^2,J}}{1-\c_0 E^{(0)}_{h+ix\x^2,J}}
\end{split}
\end{align}
where the solution for $J_{\text{i}\pm}^\text{(0)}(x)$ is given by \eqref{J0in}
\be
J_{\text{i}}^\text{(0)}(x)=\sum_{k=1}^\infty \tilde{\a}_k(x)\xi^{2k}\,,
\ee
where first few coefficients $\tilde{\a}_k(x),\,k\ge2$ are expressed  as
\begin{align}
\begin{split}
\tilde{\a}_2(x)&=-\frac{H_{h-1}}{2}(\tilde{\a}_{1}^2+x^2)\,, \ \tilde{\a}_3(x)=\frac{(\tilde{\a}_{1}^2+x^2)}{8\tilde{\a}_{1}}\left[(H_{h-1})^2(x^2+3\tilde{\a}_{1}^2)+H_{h-1,2}(\tilde{\a}_{1}^2-x^2)\right]\,,\\
\tilde{\a}_4(x)&=-\frac{\tilde{\a}_{1}^2+x^2}{24}\left[6H_{h-1} H_{h-1,2} \tilde{\a}_{1}^2+H_{h-1,3} (\tilde{\a}_{1}^2-3x^2)+2(H_{h-1})^3(3x^2+4\tilde{\a}_{1}^2)\right]\,.\end{split}
\end{align}
Next, we make the change of variable $J\to \tilde{\a}_1$ and introduce the required Jacobian 
\begin{align}\label{Jacoi0}
\begin{split}
\Theta^{(0)}_{\text{i}}(\tilde{\a}_{1\pm}):=\frac{\pd J_{\text{i}\x}^{(0)}(\tilde{\a}_{1\pm},x)}{\pd\tilde{\a}_{1\pm}}=&\xi^2\left[1-H_{h-1}^1\tilde{\a}_{1}\xi^2+\frac{H_{h-1}^2 (x^4+3\tilde{\a}_{1}^4)+(H_{h-1}^1)^2(9\tilde{\a}_{1}^4+4x^2\tilde{\a}_{1}^2-x^4)}{8\tilde{\a}_{1}^2}\xi^4\right.\\
&\left.-\frac{\tilde{\a}_{1}}{6}(H_{h-1}^3(\tilde{\a}_{1}^2-x^2)+3H_{h-1}^1 H_{h-1}^2(2\tilde{\a}_{1}^2+x^2)+(H_{h-1}^1)^3(7x^2+8\tilde{\a}_{1}^2))\xi^6+\dots\right].
\end{split}\end{align}
We also use the pole structure of the spectral function in terms of $\tilde{\a}_{1\pm}$, given by \eqref{0inspec}, 
\be
\frac{E_{h+ix\x^2, J_{\text{i}\pm}^\text{(0)}(x)}}{1-\c_0 E_{h+ix\x^2, J_{\text{i}\pm}^\text{(0)}(x)}}=\x^{-4}\frac{(4\p)^{-2h}}{\left[(\tilde{\a}_{1}-\sqrt{1-x^2})(\tilde{\a}_{1}+\sqrt{1-x^2})\right]}\,.
\
\ee
 Putting everything together we have,
\begin{align}
\begin{split} 
\mi_{\text{i}}^{(0)}=\zeta_0(\D_i,t)\,\x^{2}
\int_{-1}^{1}dx  \oint \frac{d\tilde{\a}_{1}}{2\p i}\Theta^{(0)}_{\text{i}}(\tilde{\a}_{1},x)~M\left(J_{\text{i}}^{(0)}(\tilde{\a}_1,x), x\right)\left(\frac{s}{4}\right)^{J_{\text{i}}^{(0)}(\tilde{\a}_1,x)}\,\left[\frac{\x^{-4}(4\p)^{-2h}}{\left[\tilde{\a}_{1}(x)^2-(1-x^2)\right]}\right].
\end{split}
\end{align} The contour integration over $\tilde{\a}_1(x)$ will pickup contributions from two poles at $\tilde{\a}_1=\tilde{\a}_{1\pm}:=\pm\sqrt{1-x^2}$. Recall that, we defined $J_{\text{i}\pm}^{(0)}(x):=J_{\text{i}}^{(0)}(\tilde{\a}_1=\tilde{\a}_{\pm},x)$. Further, we note that, $J_{\text{i}\pm}^{(0)}(x)$ as well as $\Theta(\tilde{\a}_1=\tilde{\a}_{1\pm},x)$ are even under $x\to-x$. Also,
 from \eqref{appM}, it is straightforward to observe that  
\be\label{transinint}
 M^{(0)}\left(J_{\text{i}\x\pm}^{(0)}(-x),-x\right) = M^{(0)}\left(J_{\text{i}\pm}^\text{(0)}(x),x\right)
\ee
Now, doing the contour integral over $\tilde{\a}_{1}$ by picking up the residues at the poles of the spectral function, and exploiting the invariance of each individual term under $x \to -x$, we can write the resulting integral over $x$ as 
\be \label{Iifinalx}
\mi_{\text{i}}^{(0)}=2\zeta_0(\D_i,t)\x^{2}\int_{0}^{1}dx\, \left[\F^{(0)}_+(x)+\F^{(0)}_-(x)\right]
\ee where,
\begin{align} \label{phii0}
\begin{split}
&\hspace{0.5 cm}\F^{(0)}_{\pm}(x):=\Theta^{(0)}_{\text{i}}(\tilde{\a}_{1\pm})\,M^{(0)}\left(J_{\text{i}\pm}^\text{(0)}(x), x\right)\,\left(\frac{s}{4}\right)^{J_{\text{i}\pm}^\text{(0)}(x)}\,\left[\pm\frac{(4\p)^{-2h}}{2\x^4\sqrt{1-x^2}}\right].\\
\end{split}
\end{align}
 Next, we make another variable change $x=\sqrt{1-y^2}$. The reason for doing this will be clear later. Under this change of variable one has $J^{(0)}_{i+}(-y)=J^{(0)}_{i-}(y)$. After this change of variable, we have 

\begin{shaded}
\begin{align}\label{Iifinal}
\begin{split}
\mi_{\text{i}}^{(0)}&=2\zeta_0(\D_i,t)\x^2\int_0^{1} \frac{y dy}{\sqrt{1-y^2}}\left[\F_+\left(\sqrt{1-y^2}\right)+\F_-\left(\sqrt{1-y^2}\right)\right]\,.\\
\end{split}
\end{align}
\end{shaded}
We will not attempt to write any explicit expression for this integral and will come back to it only after considering the outer integral. 

\subsubsection*{II. Outer Integral}
Next, we turn to the outer integral 
\be
\begin{split} 
\mi_{\text{o}}^{(0)}&:=
\zeta_0(\D_i,t)\left(\int_{-\infty}^{-\x^{2}}+\int_{\x^{2}}^\infty\right) d\n 
\mf_\text{outer} \\
&= \zeta_0(\D_i,t)\left(\int_{-\infty}^{-\x^{2}}+\int_{\x^{2}}^\infty\right) d\n \left[ \oint \frac{dJ}{2\p i}~M^{(0)}(J,\n)\left(\frac{s}{4}\right)^{J}\,
\left(\frac{E_{h+i\n,J}}{1-\c_0E_{h+i\n,J}}\right)\,+\,(\n\to-\n)\right].
\end{split}
\ee  

where $M(J,\n)$ is given by \eqref{appM} and the outer solution $J_{\text{o}\pm}^{(0)}(\n)$ is given by \eqref{J0out}
\be
J_{\text{o}\pm}^{(0)}(\nu)=\pm i\n+\a_{1\pm}\xi^4+\sum_{k=2}^\infty \a_k(\pm\n) \xi^{4k}\,.
\ee
the first few coefficients in the expansion \eqref{J0out} are given by,
\begin{align}
\a_2(\pm\n)&=-\frac{F_1(\pm i\n)}{2}\a_{1\pm}^2\,,\ \a_3(\pm\n)=\frac{F_2(\pm i\n)+3F_1(\pm i\n)^2}{8}\a_{1\pm}^3\,,\\
~ \a_4(\pm\n)&=-\frac{F_3(\pm i\n)+12F_1(\pm i\n)F_2(\pm i\n)+16F_1(\pm i\n)^3}{48}\a_{1\pm}^4\,.
\end{align}
With the definition
\be
F_{r}(z)=(r-1)! H_{h-1,r}+(-1)^{r-1} \left[\psi^{(r-1)}(h+z)-\psi^{(r-1)}(z)\right]\,,
\ee where $\psi^{(n)}(z)$ is the usual polygamma function and $H_{n,r}(z)$ is the \emph{generalized harmonic number of order $r$ of $n$} (see footnote \ref{harmonic}). Next, as in the inner integral, we make the change of variable $J\to\a_1$ along with the required Jacobian
\begin{align}\label{Jacoo0}  \begin{split}
\Th^{(0)}_{\text{o}}(\a_{1\pm}):=\left|\frac{\pd J_{\text{o}\pm}^{(0)}(\nu)}{\pd\a_{1\pm}}\right|=\xi^4\bigg[&1-\a_{1\pm} F_1(\pm i\n)\xi^4+\frac{3(F_2(\pm i\n)+9F_1(\pm i\n)^2}{8}\a_{1\pm}^2\xi^8\\
&\hspace{1 cm}-\frac{F_3(\pm i\n)+12F_1(\pm i\n)F_2(\pm i\n)+16F_1(\pm i\n)^3}{16}\a_{1\pm}^3\xi^{12}+\dots\bigg].\end{split}
\end{align}
Therefore, we have finally 
\begin{align}\begin{split} 
\mi_{\text{o}}^{(0)}= \zeta_0(\D_i,t)\left(\int_{-\infty}^{-\x^{2}}+\int_{\x^{2}}^\infty\right) d\n &\Bigg[\oint \frac{d\a_{1+}}{2\p i}\,\Th^{(0)}_{\text{o}}(\a_{1+})\,M^{(0)}\left(J_{\text{o}+}^{(0)}(\n),\n\right)\left(\frac{s}{4}\right)^{J_{\text{o}+}^{(0)}(\n)}
\left(\x^{-4}\frac{(4\p)^{-2 h}B(h,i\n)}{[2\a_{1+}-B(h,i\n)]}\right)\\
&\,+\,\oint \frac{d\a_{1-}}{2\p i}\,\Th^{(0)}_{\text{o}}(\a_{1-})\,M^{(0)}\left(J_{\text{o}-}^{(0)}(\n),\n\right)\left(\frac{s}{4}\right)^{J_{\text{o}-}^{(0)}(\n)}
\left(\x^{-4}\frac{(4\p)^{-2 h}B(h,-i\n)}{[2\a_{1-}-B(h,-i\n)]}\right)\Bigg].
\end{split}\end{align} 
where under $\nu \to -\n$, we have the following transformations
\be\label{transoutint}
 \a_{1+} \to \a_{1-},\qquad J_{\text{o}+}^{(0)}(-\n) \to J_{\text{o}-}^{(0)}(\n),\qquad M^{(0)}\left(J_{\text{o}+}^{(0)}(-\n),-\n\right) \to M^{(0)}\left(J_{\text{o}-}^{(0)}(\n),\n\right)
\ee 
Note that this is qualitatively different from the transformation properties in the inner integral. Now, we can do the contour integral over $\a_{1\pm}$ using Cauchy integral formula. After doing this integral and using the transformation formulae \eqref{transoutint} under $\nu \to -\nu$, we can cast the resulting $\n$ integral in the form
\be 
\mi_{\text{o}}^{(0)}= 2\zeta_0(\D_i,t)\int_{\x^2}^{\infty}d\n\,\left[\Psi^{(0)}_+(\n)+\Psi^{(0)}_-(\n)\right]
\ee 

where,
\begin{align}\label{psio0}\begin{split}
&\hspace{1.5 cm}\Psi^{(0)}_{\pm}(\n):= 
\Th^{(0)}_{\text{o}}(\a_{1\pm}) \,M^{(0)}\left(J_{\text{o}\pm}^{(0)}(\n), \n\right)\left(\frac{s}{4}\right)^{J_{\text{o}\pm}^{(0)}(\n)} \left[\frac{(4\p)^{-2h} }{2\x^4 }B(h,\pm i\n)\right],\qquad~~~\a_{1\pm}=\frac{B(h,\pm i\n)}{2}
\end{split}\end{align} 

Further, it is straightforward to observe that,  $\a_{1+}=\a_{1-}^*$ and $J_{\text{o}+}^{(0)}(\n)={J_{\text{o}-}^{(0)}}(\n)^*$. Thus, quite evidently, we have $\Psi^{(0)}_-(\n)=\Psi^{(0)}_+(\n)^*$ using which, one obtains at once
\be 
\mi_{\text{o}}^{(0)}=4\zeta_0(\D_i,t)\,\Re \left[\int_{\x^2}^\infty d\n\, \Psi^{(0)}_-(\n)\right].
\ee  
Now, we introduce the variable $y$ defined by $\n^2=(\x^4 +y^2)$. Then changing the variable of integration from $\n$ to $y$ we have, 
\be \label{Io1}
\mi_{\text{o}}^{(0)}=4\zeta_0(\D_i,t)\Re \left[\int_{0}^{\infty}dy\frac{~y}{\sqrt{\x^4+y^2}} ~ \Psi^{(0)}_-(\sqrt{\x^4+y^2}, y)\right].
\ee
\paragraph{}We observe that, although there is a similarity in structure between the expression for the inner integral \eqref{Iifinal} and this outer integral, there are still some differences.% The two expressions differ in three main aspects. \emph{Firstly}, in the inner integral the integration is over $y\in(0,1)$ while for outer integral the integration is over $y\in(0,\infty)$ which \emph{contains} the interval $(0,1)$. \emph{Secondly,} we have the combination $\sqrt{1-y^2}$ appearing in the inner integral while, $\sqrt{\xi^4+y^2}$ appears in the outer integral.
% But note that, these two can be \emph{related} by a Wick rotation $y\to iy$.    \emph{And finally,}  the relation between $\Phi^{(0)}_{\pm}$ and $\Psi^{(0)}_{-}$ is not straightforward \emph{a priori}. In the following analysis we will ponder over these issues and will see that, $\mi_{\text{o}}^{(0)}$ can actually be expressed as an integral over the interval $(0,1)$ with an integrand having \emph{exactly similar} structure as that of the inner integral. \par 

\begin{itemize}
	\item[A.] \textbf{Wick Rotation and choice of Contour}\\
	We start by Wick rotating $y$ in the integral of \eqref{Io1}. %Wick rotation really accounts to doing a closed contour integral for the sake of which we need to close the contour. 
	The direction in which we intend to close the contour is dictated by the behavior of the integrand at infinity. To do so first we will analyze large $s$ behaviour of the integrand.
	From \eqref{J0out},
	\be \label{J0}
	J_{\text{o}-}^{(0)}(\n)=-i \nu -\frac{1}{8} \xi ^4 B(h,-i \nu ) \left[-4+\xi ^4 (\psi ^{(0)}(h-i \nu )+\psi ^{(0)}(h)-\psi ^{(0)}(-i \nu )+\gamma ) B(h,-i \nu )\right]+O(\x^{12})
	\ee 
	where, now, we have used the value of $\a_{1-}$ at the pole of the spectral function, \eqref{0pole}. Next, we put $\n=\sqrt{\x^4+y^2}$ into the above expression followed by rearranging the series in powers of the coupling $\x$ to obtain the generic structure, 
	\be 
	J_{\text{o}-}^{(0)}(y)\equiv J_{-}(y)=\sum_{k=1}^\infty a_k(y) \x^{2 k}.
	\ee 
	Further, in the limit $y\to\infty$ we observe that, 
	\be 
	J_-(y)\sim -iy.
	\ee 
	Therefore,  
	\be 
	\left(\frac{s}{4}\right)^{J_-(y)}\sim \left( \frac{s}{4}\right)^{-iy},~~~~y\to\infty
	\ee 
	In the Regge limit $s\to\infty$, we close the contour for the integral of \eqref{Io1} in the lower half plane so that the integral over the infinite circular arc $C$(see below) can be dropped off. The precise contour arrangement is as shown in figure \ref{oint0mag}. From \ref{Io1} we have, 
	\begin{align}
	\begin{split}
	&\int_0^\infty dy \frac{y}{\sqrt{\xi^4+y^2}}\Psi^{(0)}_-(\sqrt{\xi^4+y^2},y)\\
	=&\int_0^{-i\infty} dy \frac{y}{\sqrt{\xi^4+y^2}}\Psi^{(0)}_-(\sqrt{\xi^4+y^2},y) -2\p i\sum_{\{y_P\}}\underset{y=y_P}{\text{Res.}}\left[\frac{y}{\sqrt{\xi^4+y^2}}\Psi^{(0)}_-(\sqrt{\xi^4+y^2},y)\right]
	\end{split}
	\end{align} 
	where $\{y_P\}$ are any poles of $y\Psi^{(0)}_-(\sqrt{\xi^4+y^2},y).$  We need to tackle contributions from these poles.
	\begin{figure}[h]
		\centering
		\begin{tikzpicture}[scale=0.5]
		\draw[-] (-6,0)--(6,0);
		\draw[-] (0,-6)--(0,6);
		\draw (7.5,0) node {$\Re(y)$};
		\draw (1.8,5.5) node {$\Im(y)$};
		\draw (5.5,-5) node {$C$};
		\draw
		[
		very thick,
		decoration={markings, mark=at position 0.5 with {\arrow[black,line width=0.5mm]{>}}},
		postaction={decorate}
		]
		(0,0)--(6,0);
		\draw
		[
		very thick,
		color=blue,
		decoration={markings, mark=at position 0.5 with {\arrow[black,line width=0.5mm]{>}}},
		postaction={decorate}
		]
		(6,0) arc[start angle=0, end angle=-90, radius=6cm];
		\draw
		[
		very thick,
		color=red,
		decoration={markings, mark=at position 0.50 with {\arrow[black,line width=0.5mm]{>}}},
		postaction={decorate}
		]
		(0,-6)--(0,0);
		\filldraw  (3,-2.5)circle[radius=2pt];
		\draw[color=magenta] (3,-2.5)circle[radius=6pt];
		\draw (3.5,-3.3) node {$y_P$};
		\end{tikzpicture}
		\caption{Contour Prescription for $\mi_{\text{o}}^{(0)}$}\label{oint0mag}
	\end{figure}\\\noindent	
	\item[B.] \textbf{Pole Contribution}\\
	Clearly, the poles that will give any  contribution towards the second piece of the above integral must have the generic structure, 
	\be 
	y_P=\Re(y_P)+i\Im(y_P),~~~~\Im(y_P)>0.
	\ee  From which it follows readily that, 
	\be 
	\text{Res.}\left[ 
	\frac{y}{\sqrt{\xi^4+y^2}}\Psi^{(0)}_-(\sqrt{\xi^4+y^2},y)
	\right]_{y=y_P}~ \sim \left(\frac{s}{4}\right)^{-\Im(y_P)+i\Re(y_P)}.
	\ee  
	In the  limit $s\to\infty$, the above residue is exponentially suppressed due to $\Im(y_P)>0$. 
	So, these pole contributions can be dropped off in this limit.  Thus we can write, 
	\be 
	\int_0^\infty dy \frac{y}{\sqrt{\xi^4+y^2}}\
	\Psi^{(0)}_-(\sqrt{\xi^4+y^2},y)= \int_0^{-i\infty} dy \frac{y}{\sqrt{\xi^4+y^2}}\Psi^{(0)}_-(\sqrt{\xi^4+y^2},y)
	\ee
	where, now, the equality is \emph{modulo exponentially suppressed pole contributions}.  Now, we can make the change of variable $y=-iY$ to obtain, 
	\be 
	\mi_{\text{o}}^{(0)}=-4\,\Re\int_0^{\infty} dY \frac{Y}{\sqrt{\xi^4-Y^2}}\, \Psi^{(0)}_-(\sqrt{\xi^4-Y^2},Y).
	\ee 
	Next, we can divide this integral into two parts as
	\begin{align}
	\begin{split} 
	&\int_0^{\infty} dY \frac{Y}{\sqrt{\xi^4-Y^2}} \Psi^{(0)}_-(\sqrt{\xi^4-Y^2},Y)\\
	=&\int_0^{\xi^2} dY \frac{Y}{\sqrt{\xi^4-Y^2}} \Psi^{(0)}_-(\sqrt{\xi^4-Y^2},Y)+\int_{\xi^2}^{\infty} dY \frac{Y}{\sqrt{\xi^4-Y^2}} \Psi^{(0)}_-(\sqrt{\xi^4-Y^2},Y).
	\end{split}
	\end{align} 
	\item[C.] \textbf{On the integral contribution from the interval $[\xi^2,\infty)$}\\
	We observe the functional dependence of $\Psi^{(0)}_{-}$ and $J_{\text{o}-}^{(0)}$ on $\n$; both are functions of $i \n$ \footnote{The $\n\sinh(\p\n)$ term in $M(J,\n)$ can be written as $-i\n\sin(i\p\n)$.}.  Further, we are interested in \emph{physical}, i.e $t\in\mathbb{R}$.  Then, one obtains straightforwardly that, $\Psi^{(0)}_-\in\mathbb{R}$ for $Y>\xi^2$\footnote{To see this use $\n=\sqrt{\x^4-Y^2}$. For $Y>\xi^2$, then $\n$ is purely imaginary. From \eqref{J0}, it readily follows then, $J_{\text{o}-}^{(0)}\in\mathbb{R}$ for $Y>\x^2$. Then, from \eqref{Mdef} \eqref{Jacoo0} and \eqref{psio0}, it follows at once that $\Psi^{(0)}_-\in\mathbb{R}$ for physical $t$ and $Y>\xi^2$. }.    But, the factor $\sqrt{\x^4-Y^2}$ is purely imaginary for $Y>\xi^2$. Therefore,  
	\be 
	\Re\left[\int_{\xi^2}^\infty dY \frac{Y}{\sqrt{\xi^4-Y^2}} \Psi^{(0)}_-(\sqrt{\xi^4-Y^2},Y)\right]=0.
	\ee \end{itemize}
\paragraph{}
Putting together all these three arguments above, we finally reach 
\be \label{Iout}
\mi_{\text{o}}^{(0)}=-4 \zeta_0(\D_i,t)\,\Re \left[\int_0^{\xi^2} dY \frac{Y}{\sqrt{\xi^4-Y^2}} \Psi^{(0)}_-(\sqrt{\xi^4-Y^2},Y)\right].
\ee 
So far we  have carried out the calculation, formally, using the outer solution. In principle, in the region $Y\in[0, \xi^2]$, the outer solution is not valid. We argue that an unique analytic continuation exists that maps $J_{\text{o}-}^{(0)}$ for $Y\in[\xi^2, \infty)$ to s $J_{\text{i}-}^{(0)}$ for $Y\in[0, \xi^2]$ and consequently $\Psi^{(0)}_{-}(\sqrt{\x^4-Y^2})$ gets analytically continued to $\Phi^{(0)}_-(\sqrt{\x^4-Y^2})$. Therefore we can  write, 
\be \label{Iofinalx}
\mi_{\text{o}}^{(0)}=-4 \zeta_0(\D_i,t)\,\int_0^{\x^2}dY \frac{Y}{\sqrt{\x^4-Y^2}} \F^{(0)}_-(\sqrt{\x^4-Y^2}),
\ee where, we have used the fact that, $\Phi^{(0)}_-(\sqrt{\x^4-Y^2})$ is real for $Y>\x^2$. In order to relate \eqref{Iofinalx} to \eqref{Iifinalx}, let us now look at the following inner integral
\be 
\zeta_0(\D_i,t)\x^{2}\int_{0}^{1}dx\, \F^{(0)}_-(x)
\ee 
With the following change of variables $x=\frac{\sqrt{\xi^4-X^2}}{\xi^2}$, this becomes 
\be 
\zeta_0(\D_i,t)\x^{2}\int_{0}^{1}dx\, \F^{(0)}_-(x) = \zeta_0(\D_i,t)\int_0^{\x^2} \frac{XdX}{\sqrt{\xi^4-X^2}}\, \F^{(0)}_-(\sqrt{\xi^4-X^2}) 
\ee 
Therefore we establish the $d$ dimensional analogue of the $d=4$ integral identities between inner and outer integral of \cite{Korchemsky:2018hnb, Chowdhury:2019hns}
\begin{shaded}
\be\label{Iofinal}
\mi_{\text{o}}^{(0)}=-4\zeta_0(\D_i,t) \,\x^{2}\int_{0}^{1}dx\, \F^{(0)}_-(x)=-4 \zeta_0(\D_i,t)\int_0^{\x^2} \frac{YdY}{\sqrt{\xi^4-Y^2}}\, \F^{(0)}_-(\sqrt{\xi^4-Y^2}) ,
\ee 
\end{shaded}

% Bit, recall that, the outer solution $J_{\text{o}+}^{(0)}(i\n)$ is valid for $|\n|>\x^2$. But, when we Wick rotated by introducing the change of variable $y=iY$, essentially, we have \emph{analytically continued} to $|\n|<\x^2$ for $Y<1$. Thus, now, we are in the domain of validity for the inner solution. In fact, as explained in \textbf{PH to PH: cite appropriate appendix}, under this analytic continuation the outer solution gets analytically continued to inner solution. To be specific, $J_{\text{o}-}^{(0)}$ gets analytically continued to $J_{\text{i}-}^{(0)}$ as a consequence of which one obtains that, $\Psi^{(0)}_{-}(\x^2\sqrt{1-Y^2})$ gets analytically continued to $\Phi^{(0)}_-(\sqrt{1-Y^2})$. 

\subsubsection*{III. Combining $\mi_{\text{i}}^{(0)}$ and $\mi_{\text{o}}^{(0)}$}
Combining the inner integral $\mi_{\text{i}}^{(0)}$, \eqref{Iifinal} and the outer integral $\mi_{\text{o}}^{(0)}$, \eqref{Iofinal},   we have
\be 
 \mi^{(0)}(\x,s,t)=2\zeta_0(\D_i,t)\x^2\int_0^1 dy \frac{y}{\sqrt{1-y^2}} [\F^{(0)}_+(\sqrt{1-y^2})-\F^{(0)}_-(\sqrt{1-y^2})].
\ee Next, using $J_{\text{i}+}^{(0)}(-y)=J_{\text{i}-}^{(0)}(y)$ (this follows from putting $\tilde{\a}_{1\pm}=\pm y$ in \eqref{J0in}) and $\F^{(0)}_+(\sqrt{1-y^2})|_{y\to -y} =\F^{(0)}_-(\sqrt{1-y^2})$, we can write

\begin{shaded}
\be \label{finalint}
 \mi^{(0)}(\x,s,t)=2\zeta_0(\D_i,t)\x^2\int_{-1}^1 dy \frac{y}{\sqrt{1-y^2}} \,\F^{(0)}_{+}(\sqrt{1-y^2})\,.
\ee 
\end{shaded}
 This is the final integral to be done. Finally let us address the argument behind the analytic continuation $J_{\text{o}-}^{(0)}$ to $J_{\text{i}-}^{(0)}$. This merits some amount of discussion since while we have not been able to prove it rigorously, we have found compelling evidence for this being so. In general for 0 magnon when $h=k,~~~ k\in {\cal Z}$, we can immediately see from the spectral function \eqref{E0} and \eqref{spec0}, that the form of the denominator will be a polynomial in $J$. This tells us that although the closed form expression for $J$ is tedious to find, but in general, since we know that the closed form solutions exist, there exist an analytic continuation which maps $J_{\text{o}-}^{(0)}$ to $J_{\text{i}-}^{(0)}$. This occurs for one magnon spectral function too \eqref{E1} and \eqref{spec1} for $h=2k$. So atleast for these special choice of dimensions, we can see that this happens. We prove it more explicitly by computing the Regge Mellin amplitudes in for $d=4$ for 0 magnon and $d=4, 8$ for 1-magnon at weak coupling using exact Regge poles a la \cite{Korchemsky:2018hnb, Chowdhury:2019hns} and match with our perturbative general $d$ prediction.

\section{Chiral fishnet Regge integral} \label{chfema1int}
In this appendix we present the argument that \eqref{chfema1k1} holds. We explicitly present the case for $k=1$ with the structural understanding that the argument can be generalised to higher $k$. The Mellin amplitude for $k=1$ is given by \eqref{chfema1k1}, 

\be
\begin{split}
	I_1=4\omega^2\int_{-1}^1 dx \left(\frac{x}{\pi^4 \sqrt{1-x^2}} \frac{(l_0(J(x),x)) M(J(x),x)}{((J-J_1^+)(J-J_1^-)(J-J_2^+)(J-J_2^-))^2}\right)|_{{\rm Res }~ J=J^1_+}\, + (s\rightarrow -s).
\end{split}
\ee 

The residue can be evaluated with the poles at \eqref{0magpolesc} but higher multiplicity. Contribution from the leading Regge trajectory is given by,

\begingroup
\allowdisplaybreaks
\begin{align}\label{appch1}
	I_1&=\sum_{J=J_\pm}\int_{-\infty}^\infty d\n~\frac{-i \kappa ^4}{1024 (J+1)^4 \nu  \left(J (J+2)+\nu ^2\right)^3} \Bigg[4 (J+1) \left(J^2+\nu ^2\right) \left(J (J+2)+\nu ^2\right) \left((J+2)^2+\nu ^2\right)\nn\\
	& \left(-\psi ^{(1)}\left(\frac{J+i \nu +2}{4}\right)+\psi ^{(1)}\left(\frac{J+i \nu +4}{4}\right)+\psi ^{(1)}\left(\frac{J-i \nu +2}{4}\right)-\psi ^{(1)}\left(\frac{J-i \nu +4}{4}\right)\right) M'(J(\nu))\nn\\
	& +M(J(\nu)) \left[-(J+1) \left(J^2+\nu ^2\right) \left(J (J+2)+\nu ^2\right) \left((J+2)^2+\nu ^2\right) \psi ^{(2)}\left(\frac{J+i \nu +2}{4}\right)+(J+1) \left(J^2+\nu ^2\right) \right.\nn\\
	&\left.\left(J (J+2)+\nu ^2\right) \left((J+2)^2+\nu ^2\right) \psi ^{(2)}\left(\frac{J+i \nu +4}{4}\right)+(J+1)\left(J^2+\nu ^2\right)\left(J (J+2)+\nu ^2\right) \left((J+2)^2+\nu ^2\right) \right.\nn\\
	&\left. \psi ^{(2)}\left(\frac{J-i \nu +2}{4}\right)-(J+1) \left(J^2+\nu ^2\right) \left(J (J+2)+\nu ^2\right) \left((J+2)^2+\nu ^2\right)\psi ^{(2)}\left(\frac{J-i \nu +4}{4}\right)\right.\nn\\
	&\left. +8 \left(J (J+2) \nu -J (J+2)+\nu ^3+\nu ^2+2 \nu \right) \left(J^2 (\nu +1)+2 J (\nu +1)+\nu  ((\nu -1) \nu +2)\right)\psi ^{(1)}\left(\frac{J+i \nu +2}{4}\right)\right.\nn\\
	&\left. -8 \left(J (J+2) \nu -J (J+2)+\nu ^3+\nu ^2+2 \nu \right) \left(J^2 (\nu +1)+2 J (\nu +1)+\nu  ((\nu -1) \nu +2)\right)\psi ^{(1)}\left(\frac{J+i \nu +4}{4}\right) \right.\nn\\
	&\left.+8 \left(J (J+2) \nu -J (J+2)+\nu ^3+\nu ^2+2 \nu \right) \left(J^2 (\nu +1)+2 J (\nu +1)+\nu  ((\nu -1) \nu +2)\right)\psi ^{(1)}\left(\frac{J-i \nu +4}{4}\right)\right.\nn\\
	&\left. -8 \left((J (J+2)+2) \nu +J (J+2)+\nu ^3-\nu ^2\right) \left((J (J+2)+2) \nu -J (J+2)+\nu ^3+\nu ^2\right) \psi ^{(1)}\left(\frac{J-i \nu +2}{4}\right)\right]\Bigg]
\end{align}
\endgroup  
$J_\pm=J^\pm_1$ given by \eqref{0magpolesc}
\be 
J^\pm=-1+\sqrt{1-\n^2\pm 2\sqrt{4\omega^4-\n^2}}
\ee
and $M(J(\n))$ is given by $M(J,\n)=\left(\frac{s}{4}\right)^J e^{i\pi J/2} M^{(c)}(J,\nu) \,\zeta_c({\D_i,t})$ where $M^{(c)}(J,\nu)$ is given by \eqref{melc}. Following \cite{Korchemsky:2018hnb} and \cite{Chowdhury:2019hns}, we can write this integral as 
\be
\begin{split}
	I_1 &=I_1^+ + I_1^-=\int^\infty_{-\infty} d\n \left(\left(\frac{s}{4}\right)^{J^+}\phi_+(\nu) +\left( \frac{s}{4}\right)^{J^-}\phi_-(\nu)\right)
\end{split}
\ee  

where
\begingroup
\allowdisplaybreaks
\begin{align}
\phi_\pm(\nu) &= -\frac{i \kappa ^4 \sinh [\pi  \nu ]\csc (J_\pm \pi ) \Gamma (J_\pm+2) \Gamma (J_\pm-i \nu +2) \Gamma \left(\frac{-J_\pm-t-i \nu +2}{2}\right) \Gamma (J_\pm+i \nu +2) \Gamma \left(\frac{-J_\pm-t+i \nu +2}{2}\right)}{2048 (J_\pm+1)^4 \pi ^9 \left(\nu ^2+J_\pm (J_\pm+2)\right)^3 \omega ^2 \Gamma (J_\pm+1) \Gamma \left(\frac{J_\pm-i \nu +2}{2}\right)^2 \Gamma \left(\frac{J_\pm+i \nu +2}{2}\right)^2}\nonumber\\
& \left(-\left(-8 \left(\nu ^3+\nu ^2+J_\pm (J_\pm+2) \nu +2 \nu -J_\pm (J_\pm+2)\right) \left((\nu +1) J_\pm^2+2 (\nu +1) J_\pm+\nu  ((\nu -1) \nu +2)\right) \right.\right.\nonumber\\
&\left.\left. \psi ^{(1)}\left(\frac{1}{4} (J_\pm+i \nu +2)\right)+8 \left(\nu ^3+\nu ^2+J (J+2) \nu +2 \nu -J_\pm (J_\pm+2)\right) \left((\nu +1) J_\pm^2+2 (\nu +1) J_\pm \right.\right.\right.\nonumber\\
&\left.\left.\left.  +\nu  ((\nu -1) \nu +2)\right) \psi ^{(1)}\left(\frac{1}{4} (J_\pm+i \nu +4)\right)+8 \left(\nu ^3-\nu ^2 +(J_\pm (J_\pm+2)+2) \nu +J_\pm (J_\pm+2)\right) \left(\nu ^3+\nu ^2 \right.\right.\right.\nonumber\\
&\left.\left.\left. +(J_\pm (J_\pm+2)+2) \nu -J_\pm (J_\pm+2)\right) \psi ^{(1)}\left(\frac{1}{4} (J_\pm-i \nu +2)\right)-8 \left(\nu ^3+\nu ^2+J_\pm (J_\pm+2) \nu +2 \nu -J_\pm (J_\pm+2)\right)\right.\right.\nonumber\\
&\left.\left. \left((\nu +1) J_\pm^2+2 (\nu +1) J_\pm+\nu  ((\nu -1) \nu +2)\right) \psi ^{(1)}\left(\frac{1}{4} (J_\pm-i \nu +4)\right)+(J_\pm+1) \left(J_\pm^2+\nu ^2\right)  \left(\nu ^2+J_\pm (J_\pm+2)\right)\right.\right.\nonumber\\
&\left.\left. \left((J_\pm+2)^2+\nu ^2\right) \psi ^{(2)}\left(\frac{1}{4} (J_\pm+i \nu +2)\right)-(J_\pm+1) \left(J_\pm^2+\nu ^2\right) \left(\nu ^2+J_\pm (J_\pm+2)\right) \left((J_\pm+2)^2+\nu ^2\right)\right.\right.\nonumber\\
&\left.\left. \psi ^{(2)}\left(\frac{1}{4} (J_\pm+i \nu +4)\right)-(J_\pm+1) \left(J_\pm^2+\nu ^2\right) \left(\nu ^2+J_\pm (J_\pm+2)\right) \left((J_\pm+2)^2+\nu ^2\right) \psi ^{(2)}\left(\frac{1}{4} (J_\pm-i \nu +2)\right)\right.\right.\nonumber\\
&\left.\left. +(J_\pm+1) \left(J_\pm^2+\nu ^2\right) \left(\nu ^2+J_\pm (J_\pm+2)\right) \left((J_\pm+2)^2+\nu ^2\right) \psi ^{(2)}\left(\frac{1}{4} (J_\pm-i \nu +4)\right)\right) \omega ^2-2 (J_\pm+1) \left(J_\pm^2+\nu ^2\right)\right.\nonumber\\
&\left. \left(J_\pm^2+2 J_\pm+\nu ^2\right) \left(J_\pm^2+4 J_\pm+\nu ^2+4\right) \left(2 \pi  \cot (J_\pm \pi ) \omega ^2+2 \psi ^{(0)}(J_\pm+1) \omega ^2-2 \psi ^{(0)}(J_\pm+2) \omega ^2\right.\right.\nonumber\\
&\left.\left. -2 \psi ^{(0)}(J_\pm+i \nu +2) \omega ^2+2 \psi ^{(0)}\left(\frac{1}{2} (J_\pm+i \nu +2)\right) \omega ^2+\psi ^{(0)}\left(\frac{1}{2} (-J_\pm-t+i \nu +2)\right) \omega ^2\right.\right.\nonumber\\
&\left.\left.-2 \psi ^{(0)}(J_\pm-i \nu +2) \omega ^2+2 \psi ^{(0)}\left(\frac{1}{2} (J_\pm-i \nu +2)\right) \omega ^2+\psi ^{(0)}\left(\frac{1}{2} (-J_\pm-t-i \nu +2)\right) \omega ^2-2 q\right)\right.\nonumber\\
&\left. \left(-\psi ^{(1)}\left(\frac{1}{4} (J_\pm+i \nu +2)\right)+\psi ^{(1)}\left(\frac{1}{4} (J_\pm+i \nu +4)\right)+\psi ^{(1)}\left(\frac{1}{4} (J_\pm-i \nu +2)\right)-\psi ^{(1)}\left(\frac{1}{4} (J_\pm-i \nu +4)\right)\right)\right) 
\end{align}
\endgroup

We can explicitly check that 
\be \label{0magnutrans}
\tilde{\phi}^\pm(-\nu) =\tilde{\phi}^\pm(\nu),\qquad J^\pm(-\nu) =J^\pm(\nu)
\ee 
 We now split the integration region in \eqref{appch1} as following,
\begin{align}\label{A.5}
\begin{split}
I_1=&\int_{-2\omega^2}^{2\omega^2}d\n~\left[\left(\frac{s}{4}\right)^{J_+}\phi_{+}(\n)+ \left(\frac{s}{4}\right)^{J_-}\phi_{-}(\n)\right]+\left(\int_{-\infty}^{-2\omega^2}d\n~\left(\frac{s}{4}\right)^{J_+}\phi_{+}(\n)+\int_{2\omega^2}^{\infty}d\n~ \left(\frac{s}{4}\right)^{J_-}\phi_{-}(\n)\right)\\
&\hspace{5 cm}+\left(\int_{-\infty}^{-2\omega^2}d\n~\left(\frac{s}{4}\right)^{J_-}\phi_{-}(\n)+\int_{2\omega^2}^{\infty}d\n~\left(\frac{s}{4}\right)^{J_+}\phi_{+}(\n)\right)\\
=&I_{1,0} +I_{1,1} +I_{1,2}
\end{split}
\end{align} 

We can use transformation properties \eqref{0magnutrans} to show
\be \label{i10}
I_{1,0}=2\int_{0}^{2\omega^2}d\n~\left[\left(\frac{s}{4}\right)^{J_+}\phi_{+}(\n)+ \left(\frac{s}{4}\right)^{J_-}\phi_{-}(\n)\right]
\ee

Following \cite{Chowdhury:2019hns}, \cite{Korchemsky:2018hnb},    we introduce the change variable $\n^2-4\omega^4=x^2$ so that,
\be 
J_{\pm}=-1+\sqrt{1-4\omega^4-x^2\pm 2ix}
\ee  
Following the argument outlined in the Appendix A.1 of \cite{Chowdhury:2019hns}, it can now be shown that $I_{1,1}$ and $I_{1,2}$ can be rewritten in the following manner 
\be \label{chbwr}
\begin{split}
I_{1,1}=I_{1,2}= 2\, \text{Re} \int_{0}^{\infty}\frac{x\,dx}{\sqrt{x^2+4\omega^4}}\left(\frac{s}{4}\right)^{J_-}\phi_{-}(\sqrt{x^2+4\omega^4})
\end{split}
\ee 
Here we took into account that $J_+$ and $J_-$ are conjugate to each other for real $x$ such that $1-4\omega^4-x^2>0$. We would like to wick rotate the contour in \eqref{chbwr}. Noting that in large $x$ limit we have,
\be 
J_-  \sim {-ix},\qquad x\rightarrow \infty
\ee 
This suggests that we would like to close the $x$-contour in the lower half of complex $x-$plane in \eqref{chbwr}. The contour that we will use is as below,
\begin{figure}[h]
	\centering
	\begin{tikzpicture}[scale=0.5]
	\draw[-] (-6,0)--(6,0);
	\draw[-] (0,-6)--(0,6);
	\draw (7.5,0) node {$\Re(x)$};
	\draw (1.8,5.5) node {$\Im(x)$};
	\draw (5.5,-5) node {$C$};
	\draw
	[
	very thick,
	decoration={markings, mark=at position 0.5 with {\arrow[black,line width=0.5mm]{>}}},
	postaction={decorate}
	]
	(0,0)--(6,0);
	\draw
	[
	very thick,
	color=blue,
	decoration={markings, mark=at position 0.5 with {\arrow[black,line width=0.5mm]{>}}},
	postaction={decorate}
	]
	(6,0) arc[start angle=0, end angle=-90, radius=6cm];
	\draw
	[
	very thick,
	color=red,
	decoration={markings, mark=at position 0.50 with {\arrow[black,line width=0.5mm]{>}}},
	postaction={decorate}
	]
	(0,-6)--(0,0);
	\filldraw  (3,-2.5)circle[radius=2pt];
	\draw[color=magenta] (3,-2.5)circle[radius=6pt];
	\draw (3.5,-3.3) node {$x_P$};
	\end{tikzpicture}
	\caption{Contour Prescription for \eqref{A.9}}
\end{figure}
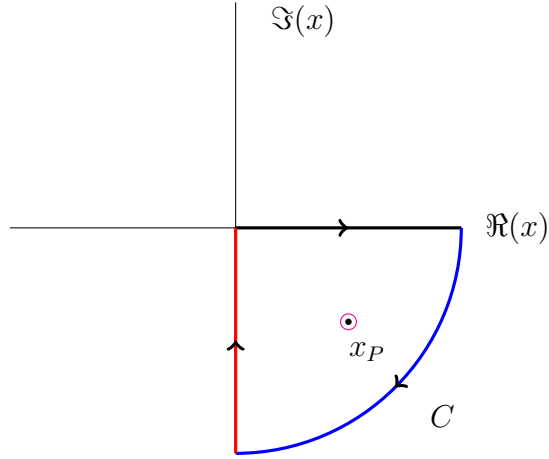\\
Now, referred to the above contour prescription, we have for \eqref{chbwr},
\be\label{A.9}
\begin{split}
	2\int_{0}^{\infty}\frac{x\,dx}{\sqrt{x^2+4\omega^4}} \left(\frac{s}{4}\right)^{J_-}\phi_{-}(\sqrt{x^2+4\omega^4})=2\int_{0}^{-i \infty}\frac{x\,dx}{\sqrt{x^2+4\omega^4}}\left(\frac{s}{4}\right)^{J_-}\phi_{-}(\sqrt{x^2+4\omega^4})-{\cal O}\left( \frac{1}{s}\right)
\end{split}
\ee

where $\lbrace x_P\rbrace$ are the  poles of $\phi_{-}(\sqrt{x^2+4\omega^4})$ in $x$ \footnote{Assuming that the poles that contribute have the generic structure,
	\be 
	x_P=\Re(x_P)-i\bar{\Im}(x_P),\qquad \bar{\Im}(x_P)>0\nonumber
	\ee, we immediately see that the residue is suppressed for $\bar{\Im}(x_P)>0$}. We introduce the \enquote{Wick Rotation} $x=-ix_E$ and finally obtain from \eqref{A.9},
\be \label{A.14}
\begin{split} 
	2\text{Re}\int_{0}^{\infty}\frac{x\,dx}{\sqrt{x^2+4\omega^4}} \left(\frac{s}{4}\right)^{J_-}\phi_{-}(\sqrt{x^2+4\omega^4})
	& \sim 2\text{Re}\int_{0}^{-i \infty}\frac{x\,dx}{\sqrt{x^2+4\omega^4}}\left(\frac{s}{4}\right)^{J_-}\phi_{-}(\sqrt{x^2+4\omega^4})\\
	&=-2\text{Re}\int_{0}^{\infty}\frac{dx_E\,x_E}{\sqrt{-x_E^2+4\omega^4}}\phi_{-}\left(\sqrt{-x_E^2+4\omega^4}\right)
\end{split} 
\ee
The
integrand has two square-root branch cuts $[-\infty,-2\omega^2)$ and $[2\omega^2,\infty)$ and deforming the contour we should not cross the cut.\\
Next we split up \eqref{A.14},
\begin{align}\label{A.15}
\begin{split}
&-2\text{Re}\int_{0}^{\infty}\frac{dx_E\,x_E}{\sqrt{-x_E^2+4\omega^4}}\left(\frac{s}{4}\right)^{J_-}\phi_{-}\left(\sqrt{-x_E^2+4\omega^4}\right)\\
&= -2\text{Re}\int_{0}^{2\omega^2}\frac{dx_E\,x_E}{\sqrt{-x_E^2+4\omega^4}}\left(\frac{s}{4}\right)^{J_-}\phi_{-}\left(\sqrt{-x_E^2+4\omega^4}\right)-2\text{Re}\int_{2\omega^2}^{\infty}\frac{dx_E\,x_E}{\sqrt{-x_E^2+4\omega^4}}\left(\frac{s}{4}\right)^{J_-}\phi_{-}\left(\sqrt{-x_E^2+4\omega^4}\right)
\end{split}
\end{align}
Note that in the term in the second line of \eqref{A.15}, the prefactor $\frac{1}{\sqrt{-x_E^2+4\omega^4}}$ is strictly imaginary in the regime of integration while the quantity $\left(\frac{s}{4}\right)^{J_-}\phi_{-}\left(\sqrt{-x_E^2+4\omega^4}\right)$ is real in the same regime. This leads to

\begin{align}
\begin{split}
-2\text{Re}\int_{0}^{\infty}\frac{dx_E\,x_E}{\sqrt{-x_E^2+4\omega^4}}\left(\frac{s}{4}\right)^{J_-}\phi_{-}\left(\sqrt{-x_E^2+4\omega^4}\right)
&\sim -2\int_{0}^{2\omega^2}\frac{dx_E\,x_E}{\sqrt{-x_E^2+4\omega^4}}\left(\frac{s}{4}\right)^{J_-}\phi_{-}\left(\sqrt{-x_E^2+4\omega^4}\right)\\
\end{split}
\end{align}

Therefore \eqref{chbwr} can be shown to be,

\be
\begin{split}
I_{1,1}=I_{1,2}= - 2\int_{0}^{2\omega^2}\frac{x\,dx}{\sqrt{-x^2+4\omega^4}}\left(\frac{s}{4}\right)^{J_-}\phi_{-}(\sqrt{-x^2+4\omega^4})
\end{split}
\ee 
Putting everything together, \eqref{A.5} becomes
\begin{eqnarray}\label{chk=1}
I_1=2\int^{2\omega^2}_{0} d\n \left(\left(\frac{s}{4}\right)^{J_+}\phi_+(\nu) -\left( \frac{s}{4}\right)^{J_-}\phi_-(\nu)\right)
\end{eqnarray}

We now make a transformation of variables $\nu = 2 \omega^2 \sqrt{1-x^2}$ and note that 
$J_+(-x)=J_-(x)$ and $\phi_+(-x)=\phi_-(x)$. The integral \eqref{chk=1}can finally brought to the form \eqref{chfema1k1} as we had promised. 
\begin{eqnarray}
I_1=4\omega^2\int^{1}_{-1} dx \frac{x}{\sqrt{1-x^2}} \left(\left(\frac{s}{4}\right)^{J_+}\phi_+(x) \right)
\end{eqnarray}

We have explicitly verified that this manipulation works for $I_2$ also and we believe this will continue to hold true for higher values of $k$ but we don't have a more physical argument to present except for structural observations.

\subsection{Explicit form for the perturbative integrals}
In this appendix, we list the explicit form for the integrals used in Section \ref{chm1}.
\begingroup
\allowdisplaybreaks
\begin{align}\label{chm1i0}
I_0=&2\int_{-1}^1 dx~\sqrt{1-x^2} e^{2 q x} \Gamma \left(1-\frac{t}{2}\right)^2\left(\frac{\omega ^2 }{128\pi ^9 x}+\frac{\omega ^4  \left(-2 q x-2 x^2 \psi ^{(0)}\left(1-\frac{t}{2}\right)+4 x^2+1\right)}{128 \pi ^9 x^2}\right. \nn\\
&\left.+\frac{\omega ^6 }{384 \pi ^9 x^3}  \left(x \left(6 q^2 x-6 q \left(2 x^2+1\right)+x \left(\pi ^2 \left(2 x^2+1\right)+6\right)\right)+3 x^2 \left(2 x \psi ^{(0)}\left(1-\frac{t}{2}\right)\right.\right.\right.\nn\\
&\left.\left.\left. \left(2 q+x \psi ^{(0)}\left(1-\frac{t}{2}\right)-4 x\right)+\left(2 x^2-1\right) \psi ^{(1)}\left(1-\frac{t}{2}\right)\right)+3\right)-\frac{\omega^8 }{384 \pi ^9 x^4} \left(2 q \left(2 q^2+\pi ^2 \right.\right.\right.\nn\\
&\left.\left.\left. +3\right) x^3-\left(6 q^2+\pi ^2+3\right) x^2+6 x^4 (2 x (q-2 x)+1) \psi ^{(0)}\left(1-\frac{t}{2}\right)^2+3 x^2 \left(2 x \left(2 x^2-1\right) (q-2 x)+1\right)\right.\right.\nn\\
&\left.\left. \psi ^{(1)}\left(1-\frac{t}{2}\right)+2 x^4 \psi ^{(0)}\left(1-\frac{t}{2}\right) \left(6 q (q-2 x)+\left(6 x^2-3\right) \psi ^{(1)}\left(1-\frac{t}{2}\right)+\pi ^2 \left(2 x^2+1\right)\right)\right.\right.\nn\\
&\left.\left. +4 \left(\pi ^2-6\right) q x^5+6 q x+4 x^6 \psi ^{(0)}\left(1-\frac{t}{2}\right)^3+\left(4 x^2-3\right) x^4 \left(\psi ^{(2)}\left(1-\frac{t}{2}\right)+12 \zeta (3)\right)-8 \pi ^2 x^6+12 x^4-3\right) \right.\nn\\
&\left. +\frac{\omega ^{10}  }{23040 \pi ^9 x^5}\left(4 \left(-10 \left(-6 \pi ^2 \left(q^2-2\right)+36 \left(q^2-1\right)+\pi ^4\right) x^6+60 q \left(2 q^2+\pi ^2+9\right) x^5-30 q \left(2 q^2+\pi ^2\right) x^3\right.\right.\right.\nn\\
&\left.\left.\left. +15 \left(6 q^2+\pi ^2\right) x^2+\left(30 \left(q^4-9\right)+15 \pi ^2 \left(2 q^2+1\right)+2 \pi ^4\right) x^4-120 \left(6+\pi ^2\right) q x^7-90 q x+38 \pi ^4 x^8+45\right)\right.\right.\nn\\
&\left.\left. +15 x^2 \left(8 x^3 \psi ^{(0)}\left(1-\frac{t}{2}\right) \left(4 q^3-6 \left(-2 q x^2+q+4 x^3-3 x\right) \psi ^{(1)}\left(1-\frac{t}{2}\right)+2 q \left(2 \left(\pi ^2-6\right) x^2+\pi ^2+3\right)\right.\right.\right.\right.\nn\\
&\left.\left.\left.\left.+x \left(4 x^2-3\right) \left(\psi ^{(2)}\left(1-\frac{t}{2}\right)+12 \zeta (3)\right)+2 \pi ^2 x \left(1-4 x^2\right)\right)+8 x^3 \psi ^{(0)}\left(1-\frac{t}{2}\right)^2 \left(\left(6 q^2+\pi ^2-6\right) x\right.\right.\right.\right.\nn\\
&\left.\left.\left.\left.-12 q x^2+6 q+\left(6 x^3-3 x\right) \psi ^{(1)}\left(1-\frac{t}{2}\right)+2 \pi ^2 x^3\right)+4 \left(x \left(6 q^2 \left(2 x^2-1\right) x+6 q \left(-4 x^4+2 x^2+1\right)\right.\right.\right.\right.\right.\nn\\
&\left.\left.\left.\left.\left. +4 \pi ^2 x^5+\left(-3-\pi ^2\right) x\right)-3\right) \psi ^{(1)}\left(1-\frac{t}{2}\right)+32 x^4 (x (q-2 x)+1) \psi ^{(0)}\left(1-\frac{t}{2}\right)^3+8 x^3 \left(q \left(4 x^2-3\right)\right.\right.\right.\right.\nn\\
&\left.\left.\left.\left.-8 x^3+7 x\right) \psi ^{(2)}\left(1-\frac{t}{2}\right)+96 x^3 \zeta (3) \left(q \left(4 x^2-3\right)-8 x^3+7 x\right)+8 x^6 \psi ^{(0)}\left(1-\frac{t}{2}\right)^4+6 \left(1-2 x^2\right)^2\right.\right.\right.\nn\\
&\left.\left.\left. x^2 \psi ^{(1)}\left(1-\frac{t}{2}\right)^2+\left(8 x^6-8 x^4+x^2\right) \psi ^{(3)}\left(1-\frac{t}{2}\right)\right)\right)\right) + {\cal O}(\omega^{12})
\end{align}
\endgroup

\begingroup
\allowdisplaybreaks
\begin{align}\label{chm1i1}
	I_1&=2\int_{-1}^1 dx~\Gamma \left(1-\frac{t}{2}\right)^2\sqrt{1-x^2}e^{2 q x}\left( \frac{3 \kappa ^4\omega ^2}{64 \pi ^9 x^3}  \zeta (3)  (x (q+x)-1)\right.\nn \\
	&\left. -\frac{\kappa ^4\omega ^4}{7680 \pi ^9 x^4}  \left(180 \zeta (3) \left(-2 x \left(-2 q^2 x+q \left(3-2 x^2\right)+2 x^3+x\right)+2 x^2 (2 x (q+x)-1) \psi ^{(0)}\left(1-\frac{t}{2}\right)+3\right)\right.\right.\nn\\
	&\left.\left. +7 \pi ^4 x^2 (2 x (q+x)-1)\right) + \frac{\kappa^4 \omega ^6}{3840 \pi ^9 x^5}\left(2 x^3 \psi ^{(0)}\left(1-\frac{t}{2}\right) \left(7 \pi ^4 x^2 (q+x)-180 \zeta (3) \left(-2 q^2 x-2 q x^2+q+2 x^3\right)\right)\right.\right.\nn\\
	&\left.\left.-7 \pi ^4 x^3 \left(-2 q^2 x-2 q x^2+q+2 x^3\right)+60 \zeta (3) \left(2 \left(9 q^2+\pi ^2+6\right) x^4+q \left(6 \left(q^2-2\right)+\pi ^2\right) x^3-\left(12 q^2+\pi ^2-3\right) x^2\right.\right.\right.\nn\\
	&\left.\left.\left.+2 \left(\pi ^2-6\right) q x^5+12 q x+2 \left(\pi ^2-6\right) x^6-6\right)+360 x^5 \zeta (3) (q+x) \psi ^{(0)}\left(1-\frac{t}{2}\right)^2\right.\right.\nn\\
	&\left.\left.+90 x^2 \left(2 \zeta (3) \left(2 q x^3-q x+2 x^4-2 x^2+1\right) \psi ^{(1)}\left(1-\frac{t}{2}\right)+5 \zeta (5) \left(4 q x^3-q x+4 x^4-2 x^2+1\right)\right)\right)\right)+{\cal O}(\kappa^4\omega^8)	
\end{align} 
\endgroup

\be\label{chm1i2}
\begin{split}
	I_2=&2\int_{-1}^1 dx~\Gamma \left(1-\frac{t}{2}\right)^2\sqrt{1-x^2}\left(\frac{9 \kappa ^8 \sqrt{1-x^2} \omega ^2 \zeta (3)^2 e^{2 q x} \left(2 \left(q^2-4\right) x^2+8 q x^3-5 q x+4 x^4+4\right)}{128 \pi ^9 x^5}\right)\\
	&+ {\cal O}(\omega^4\kappa^8)
\end{split}
\ee

%\section{Chiral fishnet Regge integral-II} \label{chfema2int}

\section{Details of various integrals }\label{intweak}
We note that in zero magnon and one magnon weak coupling case we finally are left with evaluation of the integrals of the form
\be 
\mathcal{I}_{n}(P)=\int_{-1}^{1}dx e^{P x} \sqrt{1-x^2} x^n\,,\qquad n\in\mathbb{Z}
\ee 
Now we can generate all such integrals from the basic integral by repeated applications of derivative (for non-negative $n$) or anti derivative (for negative $n$) with respect to $L$ of  the following basic integral,
\be 
\mathcal{I}_{0}(P)=\int_{-1}^{1}dx e^{P x} \sqrt{1-x^2}=\frac{\pi  I_1(P)}{P}
\ee 
where $I_\m(L)$ is Modified Bessel function of first kind.\\
\paragraph{} For non-negative $n$, we have  the following differential relation,
\be \label{npos}
\mathcal{I}_{n}(P)=\frac{d^n}{dP^n}\mathcal{I}_{0}(P),\qquad n\geq0
\ee
with $n=0$ corresponds to no differentiation.\\
For example, 
\be 
\mi_1(P)=\frac{d}{dP}\mi_0(P)=\p\frac{I_2(P)}{P}
\ee 
\paragraph{} On the other hand we note that for $n<0$ the integrand is singular at $x=0$. So in this case the integral as such does not exist. However the integral can still be given meaning in the sense of Cauchy Principal value. Thus we have the following integral under consideration,
\be 
\tilde{\mathcal{I}}_{n}(P)=\text{P.V.}\int_{-1}^{1}dx e^{P x} \sqrt{1-x^2} x^n=\lim_{\d\rightarrow0}\left[\int_{-1}^{\d}+\int_{\d}^{1}\right]dx e^{P x} \sqrt{1-x^2}x^n,\quad n\in\mathbb{Z}^-
\ee 
We can get this integral from $\mathcal{I}_0(L)$ by repeated anti derivative operation i.e, repeated indefinite integral w.r.t $L$. Thus if we define,
\be 
\ml=\int dP
\ee 
then,
\be \label{nneg}
\tilde{\mi}_n(P)=\ml^n\mi_0(P)=\int^PdP_n\int^{P_n}dP_{n-1}\ldots\int^{P_2}dL_1\,\mi_0(P_1)
\ee 
For example ,
\be 
\tilde{\mi}_{-1}(P)=\int^{P}dP_1\mi_0(P_1)=\frac{\p}{2} P \, _1F_2\left(\frac{1}{2};\frac{3}{2},2;\frac{P^2}{4}\right)
\ee 
This can be expressed in terms of modified Bessel functions and modified Struve functions as following,
\be 
\tilde{\mi}_{-1}(P)=\frac{\p}{2} (P (\pi  \pmb{L}_1(P)+2) I_0(L)-(\pi  P \pmb{L}_0(P)+2) I_1(P))
\ee 
where, $I_{\m}(z)$ is modified Bessel function of first kind and $\pmb{L}_\n(z)$ is modified Struve function. In general $\tilde\mi_{-n}(P), n>0$ can be expressed in terms of  Bessel functions and  Struve functions.
%%%%%%%%%%%%%%%%%%%%%%%%%%%%%%%%%%%%%%%%%%%%%%%%%%%
%%%%%%%%%%%%%%%%%%%%%%%%%%%%%%%%%%%%%%%%%%%%%%%%%%%
%bibliography starts here
\providecommand{\href}[2]{#2}\begingroup\raggedright\endgroup

\end{document}